\documentstyle[12pt]{book}
\input amssym.def
\input amssym

\begin{document}

\title{RELATIVISTIC PHYSICS IN ARBITRARY REFERENCE FRAMES \\
{\small Application of the theory of congruences to
description of physical effects} }

\author{Nikolai V. Mitskievich\thanks{Departamento de F{\'\i}sica,
Universidad de Guadalajara
        (on leave of absence from the Russian University of
        Peoples' Friendship, Moscow, Russia),
Apartado Postal 1-2011, C.P. 44100, Guadalajara, Jalisco, M\'EXICO.
        ~ ~ ~ ~ ~ ~ e-mail: nmitskie@udgserv.cencar.udg.mx}}
\date{~}
\maketitle

\pagenumbering{roman}
\vspace*{-.3in}
\parbox{4.5in}{
\section*{Abstract}
\addcontentsline{toc}{chapter}{Abstract}
        {\small In this paper we give a review of the most general
        approach to description  of  reference  frames,  the
        monad formalism. This approach is explicitly general covariant at
        each its step, permitting use of abstract  representation
        of tensor quantities; it is applicable also to special relativity
        theory when non-inertial effects are considered in its context;
        moreover, it involves no hypotheses whatsoever thus being
        a completely natural one. For the sake of  the
        reader's  convenience,  a  synopsis  of  tensor  calculus  in
        pseudo-Riemannian  space-time  precedes  discussion  of   the
        subject, containing many expressions  rarely  encountered  in
        the literature but essentially facilitating the  consideration.
        We give also a brief comparison of the monad formalism with the
        other approaches to description of reference frames in general
        relativity. In three special chapters we consider applications of the
        monad formalism to general relativistic  mechanics,  electromagnetic
        and gravitational fields theory.  Alongside  of  the
        general theory, which includes the monad representation of
        basic  equations
        of motion of  (charged)  particles  and  of  fields,  several
        concrete solutions are provided for  better  understanding  of
        the physical role and  practical  application  of  reference
        frames. These examples lead  often  to
        exotic consequences, but this only  clarifies  the  situation
        (e.g., cases when a rotating electrically charged fluid  does
        not produce any electric field  in  the  co-moving  reference
        frame, or when magnetic charges of a kinematic nature arise
        in a rotating frame). The topic is discussed of cases when it
        is unnecessary to introduce a reference frame, and when  such
        an introduction is essential, depending on the problems under
        consideration. Special  attention  is  dedicated  to  analogy
        between gravitation and electromagnetism, closely related to
        the dragging phenomenon.  An  analogy  is  also
        disclosed between mechanics and electromagnetic field  theory
        in  a  sense  that  in  the  Maxwell  equations  in  rotating
        reference frames appear terms of the same nature as  that  of
        the Coriolis and centrifugal forces.}
} \\

\noindent  PACS Numbers: 04.20., 04.20.Cv, 04.20.Jb, 04.40.-b, 04.40.Nr,
04.90.+e, 02.40., 03.30.+p, 03.50.De

\noindent  1991 Mathematics Subject Classification: 83-02, 83C05, 83C10,
  83C22, 53B30, 53B50, 53C50

\newpage

\chapter*{Preface}
\addcontentsline{toc}{chapter}{Preface}
During many years I have felt a necessity to have at hand a
concise, but complete enough, account of the theory of
reference frames and its applications to the field theory.
There are, of course, such publications as the books by
Yu.S. Vladimirov, by V.I. Rodichev, by O.S. Ivanitskaya,
and by A.L. Zel'manov and V.G. Agakov (all of them only in
Russian), containing expositions of such ideas, and
miscellaneous journal papers. They were however written
using languages which differed  drastically from the
traditional guidelines closely related to
intuition-nourishing symbolics, such as time-derivative,
spatial divergence and curl. It was obviously indispensable
to translate ideas of the reference frames theory into this
more physical language not losing at the same time the rigour
of our theory --- thus, without restricting to any kind of
approximations (which is done quite often in such decent
branches of science as the relativistic celestial mechanics
where concrete applications are otherwise, of course, unattainable).

{\bf Acknowledgements}

I am greatly indebted to Abram L. Zel'manov for a great many
enlightening talks on the problems of reference frames in
general relativity. He was in fact the great discoverer of
new ideas and methods in this field. My thanks are due to
Vladimir A. Fock, Anatoly E. Levashov, Olga S. Ivanitskaya,
Alexander S. Kompaneyetz, Vladimir I. Rodichev, Yakov A.
Smorodinsky who have helped me so much in different aspects.
I am thankful also to Dimitry D. Ivanenko, Alexei Z. Petrov,
and Yakov P. Terletzky of whom depended essentially my research
activities at various stages of my scientific career. All of them
were very different, I dare say, even controversial persons, with
really phantastic biographies, and I feel happy for having met
them on my personal worldline and having learned so much from their
particular ``reference frames''.

This work could not be accomplished without fruitful
collaboration with many of my students; some of them I cannot
meet more in this life; the others are my colleagues now.
I am afraid that it would be difficult to name here all of them:
Viktor N. Zaharow\dag, Jos\'e C. del Prado\dag, H\'ector A. Poblete,
Maria Ribeiro Teodoro, Alberto A. Garc{\'\i}a, Ildefonso Pulido
(who disappeared somewhere in Spain), Alexander I. Nesterov,
David A. Kalev, Jos\'e L. Cindra, Georgios A. Tsalakou,
H\'ector Vargas Rodr{\'\i}guez, and many others who were not so
closely connected with this work on reference frames.

I thank so much my friends and colleagues for encouragement,
real help, brilliant ideas and concrete collaboration:
Ernst Schmutzer, Yuri S. Vladimirov, Alexander A. Beilinson,
Bryce S. DeWitt and Cecile Morette DeWitt, Valeri D. Zakharov,
Kip S. Thorne, Jan Horsk\'y, Jerzy F. Pleba\'nski.

Reading these lists I feel that I am abusing the reader too
much, but I do really need to mention with deep gratitude all of
these persons, as well as to add my sincere acknowledgements
of patience and daily support which I always have met in
my family. As a matter of fact, I have learned from everybody,
and it is my fault not to properly understand many lessons I have
been taught in my life by them; thus all the unevenness,
inconsistency, or sheer error to be encountered in this work,
are undisputably mine.

\newpage

\contentsline {chapter}{Abstract}{i}
\contentsline {chapter}{Preface}{iii}
\contentsline {chapter}{\numberline {1}Introduction}{1}
\contentsline {section}{\numberline {1.1}A general characterization of the subject}{1}
\contentsline {section}{\numberline {1.2}A synopsis of notations of Riemannian geometry}{5}
\contentsline {chapter}{\numberline {2}Reference frames' calculus}{19}
\contentsline {section}{\numberline {2.1}The monad formalism and its place in the description of reference frames in relativistic physics}{19}
\contentsline {section}{\numberline {2.2}Reference frames algebra}{27}
\contentsline {section}{\numberline {2.3}Geometry of congruences. Acceleration, rotation, expansion and shear of a reference frame}{31}
\contentsline {section}{\numberline {2.4}Differential operations and iden\discretionary {-}{}{}ti\discretionary {-}{}{}ties of the mo\discretionary {-}{}{}nad for\discretionary {-}{}{}ma\discretionary {-}{}{}lism}{35}
\contentsline {chapter}{\numberline {3}Equations of motion of test particles}{41}
\contentsline {section}{\numberline {3.1}The electric field strength and magnetic displacement vectors}{41}
\contentsline {section}{\numberline {3.2}The energy and momentum of particles}{45}
\contentsline {section}{\numberline {3.3}Monad description of the motion of a test charged mass in gravitational and electromagnetic fields}{48}
\contentsline {section}{\numberline {3.4}Motion of photons, the red shift and Doppler effects}{50}
\contentsline {section}{\numberline {3.5}The dragging phenomenon}{57}
\contentsline {chapter}{\numberline {4}The Maxwell field equations}{73}
\contentsline {section}{\numberline {4.1}The four-dimensional Maxwell equations}{73}
\contentsline {section}{\numberline {4.2}The electromagnetic stress-energy tensor and its monad decomposition}{75}
\contentsline {section}{\numberline {4.3}Monad representation of Maxwell's equations}{78}
\contentsline {section}{\numberline {4.4}A charged fluid without electric field}{81}
\contentsline {section}{\numberline {4.5}An Einstein-Maxwell field with kinematic magnetic charges}{85}
\contentsline {chapter}{\numberline {5}The Einstein field equations}{89}
\contentsline {section}{\numberline {5.1}The four-dimensional Einstein equations}{89}
\contentsline {section}{\numberline {5.2}Monad representation of Einstein's equations}{91}
\contentsline {section}{\numberline {5.3}The geodesic deviation equation and a new level of analogy between gravitation and electromagnetism}{95}
\contentsline {section}{\numberline {5.4}New quasi-Maxwellian equations of the gravitational field}{98}
\contentsline {section}{\numberline {5.5}Remarks on classification of intrinsic gravitational fields}{101}
\contentsline {section}{\numberline {5.6}Example of the Taub--NUT field}{103}
\contentsline {section}{\numberline {5.7}Example of the spinning pencil-of-light field}{106}
\contentsline {section}{\numberline {5.8}Gravitational fields of the G{^^f6}del universe}{109}
\contentsline {chapter}{\numberline {6}Concluding remarks}{111}
\contentsline {chapter}{References}{115}
\contentsline {chapter}{Index}{127}

\newpage

\pagenumbering{arabic}
\pagestyle{headings}
\chapter{Introduction}

\section{A general characterization of the subject}

         The  concept  of  reference  frame  was  introduced  in
         physics at the early stage when its  formalization  had  just
         begun and before introduction of system  of  coordinates  and
         equations of motion. When Galileo Galilei  has  first  spoken
         about the principle of relativity of motion  (see
         Schmut\-zer and Sch\"utz (1983)),
         he already used in fact this concept. The
         more used it Newton (1962)
         having  spoken  of  the  law  of
         inertia (the First Law) and  of  the  concepts  of  time  and
         space. The concept of non-inertial reference frame  was  also
         introduced and developed  in  classical  mechanics,  together
         with the corresponding effects, the centrifugal and  Coriolis
         forces (see our discussion of  the  subject  in  (Vladimirov,
         Mitskievich and Horsk\'y 1987)). The idea of  reference  frames
         has played a fundamental role in physics;  Galilei  connected
         them closely with the thought experiment method
         (Gedankenexperiment, so important in quantum  theory),
         and  Newton  who
         considered a rotating bucket  full  of  water  has  laid  the
         foundation-stone for Mach's  principle.  In  this  paper,  we
         shall consider mainly the problem of description and  further
         applications of reference frames in relativistic physics. The
         context of general  relativity  is  essential  since  it  has
         inevitably to be used also in the Minkowski space-time if
         non-Cartesian coordinates are used. This
         approach is crucial  for  interpretation  and  evaluation  of
         physical  effects  predicted  by  four-dimensional   theories
         formulated irrespective to any concrete choice of a reference
         frame, as well as for determination of  physical  observables
         already in classical  (non-quantum)  theory, {\em i.e.} also  for
         interpretational purposes.

   We  shall  speak  primarily  of  general   relativistic
         approach to reference frames, for which a concise  review  of
         ideas, methods, and applicational results will be  given.  As
         general relativity we  understand  all  relativistic  physics
         taking account of gravitation, {\em i.e.} the space-time
         curvature.
         Reference frames are considered in  the  scope  of  classical
         (non-quantum)  physics, {\em i.e.}
         in  the  assumption  that  the
         observation and measurement procedures do not  influence  the
         behaviour of the objects being observed, in particular,  they
         do not disturb the space-time geometry.
         This means that the reference body is a {\em test} one,
         while one and the same physical system of real
         bodies and fields may be described simultaneously and equally
         well in any of the  infinite  set  of  alternative  reference
         frames. Since the reference body points are  test  particles,
         they may be accelerated arbitrarily  without  application  of
         any finite forces (hence, of any fields), --- their  masses  are
         infinitesimal too. The only limitation on their motion is due
         to the relativistic causality principle:  since  a  reference
         body represents an idealization of sets of measuring  devices
         and local  observers,  its  points'  world  lines  should  be
         time-like (lie inside the  light  cone).  It  is  not  always
         possible to describe motion  of  a  system  of  bodies  in  a
         co-moving reference frame without involving singularities (at
         least, the bodies should not mutually collide). Therefore  we
         shall assume that motion of a reference body is described  by
         a {\em congruence} of time-like world lines. This leads to a very
         fruitful employment of ideas of relativistic hydrodynamics in
         the theory of reference frames; thus  a  reference  body  may
         accelerate, rotate and undergo  variable  deformation,  which
         completely characterize  the reference frame properties. This
         mathematical  description  should  be  consistent  with   the
   assumed space-time geometry. The resulting description of arbitrary
   reference frames (including non-inertial ones) and the corresponding
         physical effects, has explicitly general covariant form.

   Dealing  with  the  reference   frame   concept   comes
         sometimes across certain  psychological  obstacles  when  one
         either confuses reference fra\-me with system  of
         coordinates (non-covariant approaches),
         or connects it to any four-dimensional basis whatsoever.  The
         both extremities, in the author's opinion, come  together  in
         the sense  of  gross  vulgarization  of  this  merely  simple
         notion; what is  much  worse,  these  approaches  are  widely
         practised, which involves frequently  a  disregard  and  even
         complete negligence of reference frames in the most parts  of
         physics. We shall revisit this problem in section  2.1  that
         precedes the description of the reference frames calculus.

   It  is  remarkable  that  all  algebraic  relations  in
         electrodynamics are the same in the standard  three-dimensional
         form and in the formalism of arbitrary  reference  frames
         in any gravitational fields. But what is more important,  the
         dynamical (differential) laws of electromagnetic field theory
         in non-inertial frames  differ  from  their  standard
         three-dimensional Maxwellian form by  additional  terms  which  are
         naturally interpreted analogously to the forces of inertia of
         the classical mechanics. Moreover, even in the flat Minkowski
         space-time, non-inertial reference frames involve  inevitably
         and rigorously a curved physical  three-space  which  in  the
         presence of a rotation is also non-holonomic in the sense that
         there exist only {\em local} three-space
         elements which do not form
         global three-hypersurfaces. Thus the  expressions
         of the forces of inertia and  their  field  theoretical
         generalizations either  do  not  change  at  all,  or  change
         inessentially when one turns from general to special  relativity
         (only the four-curvature goes). For  example,  Maxwell's
         equations retain in a non-inertial frame in special relativity
         exactly the same monad form they have in general ralativity.
         Under a monad, we understand a time-like unit vector field
         whose integral lines depict worldlines of the ``particles''
         forming a test body of reference (see below). But
         we have mentioned the role of reference frames in  interpreting
         theoretical results and  evaluating  effects  the  theory
         predicts, not in a casual remark: the very solutions  of  the
         particles and fields dynamics equations are in  general  much
         easier obtained from  their  standard  four-dimensional  form
         using simplifications due to a choice  of  coordinates  which
         corresponds to the symmetries inherent in the  problem  under
         consideration. The use of reference frames comes only later.

   Some remarks about the  structure  of  this  paper.  In
         section 1.2 we consider mathematical  (mainly
         geometrical) definitions and  relations.  This  is  necessary
         since the notations differ substantially from one publication
         to another, and moreover it is necessary to have at hand some
         exotic formulae. This paper may also be of use for colleagues
         working in other areas than gravitation theory, and a bit  of
         self-containedness would help them to get into the subject.

In chapter 2, after the  already  mentioned  section
         2.1, we give in 2.2 the basic relations of the monad formalism
         algebra, including purely spatial but nevertheless  generally
         covariant (in the four-dimensional sense)
         operations of the scalar and  vector  products.  In
         further sections of the chapter 2, the differential relations
         of the monad formalism are given: on the one hand, these
         are important characteristics common with those of  mechanics
         of continuous media; on the other hand, differential operations
         as well as the corresponding identities are considered.

   Chapters 3 and 4 deal not only with gravitation, but
   essentially with electrodynamics too.  The  first  of
         them is dealing with the relativistic mechanics of  electrically
         charged particles, the second with dynamics of the  very
         electromagnetic field. It is worth specially mentioning
         section 3.3 in which  the  equations  of  motion  of  a  test
         massive charge  are  considered  from  the  viewpoint  of  an
         arbitrary reference frame, 3.4 where an example is  given
         of  calculation  of  the  frequency  shift  effect   in   two
         characteristic cases, and 3.5 giving an opposite example, that
         of an objectively determined Killing reference frame which
         enables a standard definition of the dragging phenomenon.
         Maxwell's equations are written down in
         an arbitrary reference frame in section 4.3 where non-inertial
         terms are also considered. In section  4.4  the  reference
         frame formalism is applied to physical interpretation  of  an
         exact solution of the system  of  Einstein-Maxwell  equations
         where a non-test charged fluid does not produce  in  its  comoving
         frame any electric field whatsoever. In 4.5  kinematic
         non-inertial magnetic charge density is shown to appear in
         the space-time of  an  Einstein-Maxwell $pp$-wave  in  a  very
         natural (but rotating) frame of reference.  In section 5.2 we
         consider 3+1-splitting of Einstein's equations in  the  monad
         formalism  and  trace  some  analogies  and   dissimilarities
         between the electromagnetic and  gravitational  interactions.
         Then, in sections 5.3 and  5.4  the  next  stage  of  analogy
         between gravitation and electromagnetism  is  considered  and
         realized: now a  practically  perfect  analogy  is  attained,
         without introducing any {\em ad  hoc} hypotheses  concerning  the
         gravitational theory. Finally, in chapter  6  the  concluding
         remarks are  made, especially concerning the modern
         development  of a canonical formalism in relativistic field
         theories (including the case of gravitation), and studies
         of the gravity  quantization problem. These items are  beyond
         the  scope  of  this paper, but they are closely connected
         with  the  mathematical methods discussed in it, so that
         we feel it to be  worthwhile to mention them.

\pagestyle{myheadings}
\markboth{CHAPTER 1. INTRODUCTION}{1.2. A SYNOPSIS OF RIEMANNIAN
GEOMETRY}

\section{A synopsis of notations of Riemannian geometry}

   A great many of expressions and relations of Riemannian
         geometry  will  be  extensively  used  below.   In   existing
         literature (see for a traditional approach, {\em e.g.},
         Eisenhart (1926, 1933, 1972);
         Schouten and Struik (1935, 1938); as to a presentation of
         general relativity from the mathematical viewpoint, see
         Sachs and Wu (1977)),
         they are scattered in different parts  of  various
         publications, and the notations differ substantially from one
         source to another. This is why we give here (usually  without
         derivations, though sometimes mentioning original  publications
         or  principal  expositions)  quite  a   few   of   these
         geometrical relations. The reader may skip this  section  but
         look into it merely for clarification of some formulae. It is
         worth emphasizing that we take everywhere the speed of  light
         equal to unity  ($c=1$),  the  space-time  signature  being
         $(+ - - -)$;  the  Greek  indices   are   four-dimensional,
         running from 0 to 3, the  Einstein  summation  convention  is
         adopted, including  collective  indices.  Symmetrization  and
         antisymmetrization with  respect to  sets  of  indices  are
   denoted by the standard {\em Bach brackets}, $(\cdots)$ and $[\cdots]$
         correspondingly, with the indices  in  the  brackets.  Tetrad
   indices are written in separate parentheses, {\em e.g.},
   $F_{(\mu)(\nu)}$.  For
         important details of abstract representation  of  the  tensor
         calculus and its applications to field  theory,  see  (Israel
         1970, Ryan and Shepley 1975, Eguchi, Gilkey and Hanson  1980,
         Choquet-Bruhat, DeWitt-Morette and Dillard-Bleick  1982,  von
         Westenholz 1986).

   We admit that the tensor algebra in an arbitrary  basis
         and in abstract notations, is commonly known. Hence we  begin
         with Cartan's  forms.  To  make  notations  short,  we  shall
         sometimes employ collective indices, {\em e.g.} in the basis of a
         $p$-form ({\em cf.} Mitskievich and Merkulov (1985)):
$$
dx^a:=dx^{\alpha_1}\wedge\cdots\wedge dx^{\alpha_p}.
$$
         Here exterior (wedge) product is supposed to be an
         antisymmetrization of the tensorial product,
$$
dx^{\alpha_1}\wedge\cdots\wedge dx^{\alpha_p}:=
dx^{[\alpha_1}\otimes\cdots\otimes dx^{\alpha_p]}.
$$
  Similarly, we shall write tetrad basis of a $p$-form
  as $\theta^{(a)}$. Then a $p$-form $\alpha$ can be written as
$$
\alpha=\alpha_a dx^a=\alpha_{(a)}\theta^{(a)},
$$
         while
$$
\alpha\wedge\beta=(-1)^{pq}\beta\wedge\alpha        \eqno{(1.2.1)}
$$
where $p= \deg\alpha, ~ q= \deg\beta$
(degrees of the forms $\alpha$ and $\beta$).

The axial tensor of Levi-Civit\`a~ is defined with help of
         the corresponding symbol,
$$
E_{\alpha\beta\gamma\beta}=(-g)^{1/2}\epsilon_{\alpha\beta\gamma\beta}, ~ ~
E^{\alpha\beta\gamma\beta}=-(-g)^{-1/2}\epsilon_{\alpha\beta\gamma\beta}.
  \eqno{(1.2.2)}
$$
         In  our  four-dimensional   (essentially,   even-dimensional)
         manifold, this tensor has the following properties:
$$
E_{ag}=(-1)^{p(4-p)}E_{ga}\equiv(-1)^p E_{ga},
$$
$$
E_{ag}E^{bg}=-p!(4-p)!\delta^b_a; ~ ~ \# a=\# b=p=4-\# g.
$$
         While using here collective indices, we denote the number  of
         individual indices contained in them,  by  \#.  The  Kronecker
         symbol  with  collective  indices  is  totally  skew   by   a
         definition,
$$
\delta^b_a\equiv\delta^{\beta_1\cdots\beta_p}_{\alpha_1\cdots\alpha_p}:=
\delta^{\beta_1}_{[\alpha_1}\cdots\delta^{\beta_p}_{\alpha_p]}, ~ ~
\delta^{~ b}_{[a}\delta^{~ v}_{u]}\equiv\delta^{bv}_{au},
        \eqno{(1.2.3)}
$$
         so that $A^a\delta^b_a\equiv A^b$ (let $A^a$ be skew in all
         individual indices).
         Dual conjugation of a form is denoted by  the  Hodge  star $\ast$
         (see Eguchi, Gilkey and Hanson (1980)) before the  form.  The
         star acts on the basis,
$$
\ast dx^a:=\frac{1}{(4-p)!}{E^a}_g dx^g, ~ ~ \ast\alpha:=\alpha_a\ast dx^a.
$$
         Thus $\ast\ast\alpha=(-1)^{p+1}\alpha$ (here even-dimensional
         nature of the manifold is  essential).
         It is worth stressing that the sequence  of  the  (lower  and
         upper) indices is of great importance too, {\em e.g.} it was not
  the question of a chance when we have written ${E^a}_g$ and
  not ${E_g}^a$ (which in fact is equal to $(-1)^p {E^a}_g$,
  {\em cf.} the formula  after (1.2.2)).

   The usual dual conjugation of a bivector will  be  also
         denoted by a star, but this will be  written
         over (or  under)  the  pair  of  indices  to  which  it
         is applied, {\em e.g.}
$$
F\mbox{\footnotesize ${\kappa\lambda}\atop{\ast}$}:=
\frac{1}{2}E^{\kappa\lambda\mu\nu}F_{\mu\nu},
\eqno{(1.2.4)}
$$
  thus $F{{\ast\ast} \atop\mbox{\footnotesize $\alpha\beta$}}\equiv
  -F_{\alpha\beta}$.  It   is
         self-understood that $F_{\kappa\lambda}$ is skew
         (we called it  bivector  for
         this reason), so that the dual conjugation does not  lead  to
         any loss of information. Our choice of place for the star  is
         justified  here  by  the  convenience  in  writing  down  the
  so-called {\em crafty identities} where the stars  are  applied  to
         different pairs of indices (see {\em e.g.} (Mitskievich   and
         Merkulov 1985)):
$$
V\mbox{\footnotesize ${{\kappa\lambda}\atop\ast}
{\ast\atop{\mu\nu}}$}=-{V_{\mu\nu}}^{\kappa\lambda}
-\frac{1}{2}\delta^{\kappa\lambda}_{\mu\nu}{V_{\sigma\tau}}^{\sigma\tau}
+\delta^{\kappa\tau}_{\mu\nu}{V_{\sigma\tau}}^{\sigma\lambda}
+\delta^{\tau\lambda}_{\mu\nu}{V_{\sigma\tau}}^{\sigma\kappa}.
\eqno{(1.2.5)}
$$
         For special cases and when a contraction  is  performed,  one
obtains in electrodynamics ({\em cf.} (Wheeler 1962,  Israel  1970,
         Mitskievich and Merkulov 1985))
$$
F_{\mu\nu}F^{\lambda\nu}
-F\mbox{\footnotesize $\ast\atop{\mu\nu}$}
F\mbox{\footnotesize ${\lambda\nu}\atop\ast$}
=\frac{1}{2}\delta^\lambda_\mu F_{\sigma\tau}F^{\sigma\tau},
\eqno{(1.2.6)}
$$

$$
F\mbox{\footnotesize $\ast\atop{\mu\nu}$}F^{\lambda\nu}=
\frac{1}{4}\delta^\lambda_\mu F\mbox{\footnotesize $\ast\atop{\sigma\tau}$}
F^{\sigma\tau}.
\eqno{(1.2.7)}
$$
         For the Levi-Civit\`a axial tensor one has
$$
{E\mbox{\footnotesize ${\kappa\lambda}\atop\ast$}}_{\mu\nu}
\equiv{E^{\kappa\lambda}}\mbox{\footnotesize $\ast\atop{\mu\nu}$}
=-2\delta^{\kappa\lambda}_{\mu\nu}, \eqno{(1.2.8)}
$$
         for the Riemann-Christoffel curvature tensor,
$$
R\mbox{\footnotesize ${\ast\atop{\alpha\beta}}{{\kappa\lambda}\atop\ast}$}
=-{R_{\alpha\beta}}^{\kappa\lambda}
+R\delta^{\kappa\lambda}_{\alpha\beta}
-4R^{[\kappa}_{[\alpha}\delta^{\lambda]}_{\beta]},
        \eqno{(1.2.9)}
$$
         and for the Weyl conformal curvature tensor,
$$
{C\mbox{\footnotesize $\ast\atop{\alpha\beta}$}}^{\gamma\delta}
\equiv C_{\alpha\beta}\mbox{\footnotesize ${\gamma\delta}\atop\ast$}.
\eqno{(1.2.10)}
$$
         By the way, a repeated application of the  crafty  identities
         to a construction quadratic in  the  curvature  tensor  (with
         subsequent contractions, so that only two indices remain free
         ones), leads to the well known
         Lanczos identities,
$$
{R_{\alpha\beta\gamma}}^{\mu}R^{\alpha\beta\gamma\nu}-
\frac{1}{4}R_{\alpha\beta\gamma\delta}R^{\alpha\beta\gamma\delta}g^{\mu\nu}
-2R^{\mu\alpha\beta\nu}R_{\alpha\beta}+
R_{\alpha\beta}R^{\alpha\beta}g^{\mu\nu}-
$$
$$
-R^\mu_\alpha R^{\alpha\nu}+R~R^{\mu\nu}-\frac{1}{4}R^2g^{\mu\nu}=0
\eqno{(1.2.11)}
$$
         (see our derivation of these  identities  (Mitskievich  1969)
         which differs radically from that by Lanczos (1938) who  used
  less straightforward integral relations). Note that the  invariant
  $R_{\alpha\beta\gamma\delta}
  R\mbox{\footnotesize ${\alpha\beta}\atop\ast$}
  \mbox{\footnotesize ${\gamma\delta}\atop\ast$}$ is
         known as the Bach--Lanczos invariant, or the generalized
         Gauss-Bonnet invariant (the first of Lagrangians  of  the  Lovelock
         (1971) series reduces to it); like the electromagnetic  invariant
  $F_{\alpha\beta}F\mbox{\footnotesize ${\alpha\beta}\atop\ast$}$,
  it represents (in a four-dimensional world)  a
         pure divergence, so that it does not contribute to the  field
         equations in four dimensions. {\em Cf.} also DeWitt (1965).

   Scalar multiplication of Cartan's forms is realized with
         help of consecutive dual conjugations:
$$
\ast(dx_k\wedge\ast dx^{al})=
(-1)^{p+1}\frac{(p+q)!}{p!}{\delta^[}^l_k dx^{a]}, ~
~  \{ {{\# a=p,} \atop {\# k=q=\# l,}}
$$
         in particular,
$$
\ast(dx^\lambda\wedge\ast dx^\mu)=-g^{\lambda\mu}.
$$
         The last relation is equivalent to
$$
\ast({E^\mu}_{\pi\rho\sigma}dx^{\lambda\pi\rho\sigma})=
{E^\mu}_{\pi\rho\sigma}E^{\lambda\pi\rho\sigma}=-3!~ g^{\lambda\mu}.
$$

   The covariant differentiation axioms (see {\em e.g.} Ryan and
         Shepley (1975)) are:                  \\
(1)          $\nabla_vT$ has the same tensor valence as $T$, \,
        $v$ being a vector.\\
(2)          $\nabla_v$ is a linear operation:
$$
\nabla_v(T+U)=\nabla_vT+\nabla_vU,
$$
$$
\nabla_{au+bv}T=a\nabla_uT+b\nabla_vT.
$$
~ ~ \\
(3)          The  Leibniz property (distributive  property  when  product
         expressions  are   differentiated)   holds,   including   the
         contraction operation. \\
(4)          Action of $\nabla$ on a function (a scalar;
        the concept of scalar
         is  somewhat  different  for  the  covariant  differentiation
         denoted by a semicolon) is
         $\nabla_vf=vf$, where $v=v^\alpha\partial_\alpha$. \\
(5)          Metricity property:
        $\nabla_vg=0$ (for a further  generalization
         of the geometry, the  non-metricity  tensor  appears  on  the
         right-hand side).   \\
(6)          Zero torsion axiom:
$$
\nabla_uv-\nabla_vu=[u,v]+{\sf T}, ~ ~ {\sf T}=0.
$$
         (in a generalization of the geometry,  the torsion  tensor ${\sf T}$
         becomes different from zero).

   These  axioms  are  most  simply  realized   when   the
         connection coefficients are introduced,
$$
\nabla_{X_{(\alpha)}}X_{(\beta)}=
\Gamma^{(\gamma)}_{(\beta)(\alpha)}X_{(\gamma)}; ~ ~
\nabla_{X_{(\alpha)}}\theta^{(\gamma)}=
-\Gamma^{(\gamma)}_{(\beta)(\alpha)}\theta^{(\beta)}
$$
         (only one of these relations is independent). When the structure
         coefficients of a basis are introduced with help of
$$
[X_{(\alpha)},X_{(\beta)}]=C^{(\gamma)}_{(\alpha)(\beta)}X_{(\gamma)}, ~ ~
C^{(\gamma)}_{(\alpha)(\beta)}=
\Gamma^{(\gamma)}_{(\beta)(\alpha)}-\Gamma^{(\gamma)}_{(\alpha)(\beta)}
\eqno{(1.2.12)}
$$
         (for a holonomic, or coordinated basis,
         $C^\gamma_{\alpha\beta}= 0$; as usually, we
         write in individual  parentheses  only  the  tetrad  and  not
         coordinated basis indices), the general  solution  satisfying
         the whole set of axioms, from 1 to 6, is
$$
\Gamma^{(\gamma)}_{(\alpha)(\beta)}=
\frac{1}{2}g^{(\gamma)(\delta)}\lbrack X_{(\beta)}g_{(\alpha)(\delta)}+
X_{(\alpha)}g_{(\beta)(\delta)}-X_{(\delta)}g_{(\alpha)(\beta)}\rbrack
$$
$$
+\frac{1}{2}\lbrack C^{(\gamma)}_{(\beta)(\alpha)}+
C^{(\delta)}_{(\epsilon)(\beta)}g_{(\alpha)(\delta)}g^{(\epsilon)(\gamma)}+
C^{(\delta)}_{(\epsilon)(\alpha)}g_{(\beta)(\delta)}g^{(\epsilon)(\gamma)}
\rbrack.   \eqno{(1.2.13)}
$$
\bigskip

   In a coordinated  basis  (where  the  basis
         vectors and covectors are usually written as $\partial_\alpha$
         and $dx^\alpha$)  the
         connection coefficients reduce  to  the  Christoffel  symbols
         (which are written without parentheses before and  after each
         index, and which are symmetrical in the two lower indices).
         Orthonormal bases and those of Newman--Penrose  (see  Penrose
         and Rindler (1984a, 1984b))  are  special  cases  of  bases
   whose  vectors  have  constant  scalar  products ({\em i.e.}, the
         corresponding tetrad components of the metric are  constant).
         For such bases the connection coefficients are  skew  in  the
         upper and the first lower indices; they are sometimes  called
         Ricci rotation coefficients.

   Often  (in  the  elder  literature,  always)  covariant
         differentiation is denoted by a semicolon before  the  index.
         Then, according to the covariant differentiation  axioms  (in
         particular,  the  axiom  1,  however  paradoxical  it   would
         appear), it leads to an increase by unity of the  valence  of
         the  corresponding  tensor.  As  Trautman  (1956,  1957)  has
         remarked, such a covariant derivative can be defined as
$$
T_{a;\alpha}:=T_{a,\alpha}+T_a|^\tau_\sigma\Gamma^\sigma_{\tau\alpha}
   \eqno{(1.2.14)}
$$
  where the coefficients $T_a|^\tau_\sigma$ determine the behaviour of the
         quantity $T_a$ (which could be not merely a tensor, but  also  a
         tensor density or some more general object) under infinitesimal
         transformations of coordinates:
$$
x'^\mu=x^\mu+\epsilon\xi^\mu(x),
$$

$$
\delta T_a:=T'_a(x')-T_a(x)=\epsilon T_a|^\tau_\sigma{\xi^\sigma}_{,\tau}.
        \eqno{(1.2.15)}
$$
        It is clear that the coefficients $T_a|^\tau_\sigma$ possess a
        property
$$
(S_aT_b)|^\tau_\sigma=S_a|^\tau_\sigma T_b+S_aT_b|^\tau_\sigma
$$
        for tensor product $S\otimes T$,
        which is closely related to the Leibniz property of
        differentiation. The definition of covariant derivative
        (1.2.14) may be deduced from the concept of Lie derivative,
$$
\pounds_\xi T_a:=T_{a,\alpha}\xi^\alpha-T_a|^\tau_\sigma{\xi^\sigma}_{,\tau}
\equiv T_{a;\alpha}\xi^\alpha-T_a|^\tau_\sigma{\xi^\sigma}_{;\tau},
   \eqno{(1.2.16)}
$$
         which can be written for objects of the most  general  nature
         (leaving the limits of tensors completely) as
$$
\pounds_\xi T:=\epsilon^{-1}(T(x)-T'(x))  \eqno{(1.2.17)}
$$
({\em cf.} Yano (1955)). It is important that the both  tensors  in
the parentheses in (1.2.17), have  the  same  argument  (non-primed
         one). This leads to a possibility to extend the operation
$T_{a;\alpha}$ even to quantities of the  type  of  a  connection.
         Without going into details we would note that in  this  sense
$\Gamma^\lambda_{\mu\nu;\alpha}={R^\lambda}_{(\mu\nu)\alpha}$.

It is convenient to define the concept of curvature  in
         the Riemannian geometry by introducing the curvature operator
         (see {\em e.g.} (Ryan and Shepley 1975)),
$$
{\sf R}(u,v):=\nabla_u\nabla_v-\nabla_v\nabla_u -\nabla_{[u,v]}.
          \eqno{(1.2.18)}
$$
         This operator has the following five properties: \\
         1) It is skew in $u$ and $v$.                      \\
2) It annihilates the metric, ${\sf R}(u,v)~g= 0$. \\
3) It annihilates a scalar function, ${\sf R}(u,v)~f= 0$. \\
4) This is  a  linear  operator ({\em cf.}  the  axiom  2  of  the
         covariant differentiation). \\
         5) The  Leibniz property holds  for  it  (this  fact  is  quite
         remarkable since the curvature  operator  involves  second,
         not first order differentiation).  \\

   The curvature tensor components (which are scalars from
the point of view of the $\nabla$-differentiation) are determined as
$$
{R^{(\kappa)}}_{(\lambda)(\mu)(\nu)}:=
\theta^{(\kappa)}\cdot{\sf R}(X_{(\mu)},X_{(\nu)})X_{(\lambda)}.
     \eqno{(1.2.19)}
$$
These are the coefficients (components) of decomposition  of  the
         curvature tensor with respect of the corresponding basis, and
         in the general case they are
$$
{R^{(\alpha)}}_{(\beta)(\gamma)(\delta)}=
X_{(\gamma)}\Gamma^{(\alpha)}_{(\beta)(\delta)}-
X_{(\delta)}\Gamma^{(\alpha)}_{(\beta)(\gamma)}+
\Gamma^{(\epsilon)}_{(\beta)(\delta)}
\Gamma^{(\alpha)}_{(\epsilon)(\gamma)}-
$$
$$
\Gamma^{(\epsilon)}_{(\beta)(\gamma)}
\Gamma^{(\alpha)}_{(\epsilon)(\delta)}-
C^{(\epsilon)}_{(\gamma)(\delta)}\Gamma^{(\alpha)}_{(\beta)(\epsilon)}.
      \eqno{(1.2.20)}
$$
         It is worth remembering that  some authors use  a  definition
of curvature with the opposite sign, {\em e.g.} Penrose and Rindler
         (1984a,b) and Yano (1955)  (see  also  a  table  in  (Misner,
         Thorne and Wheeler 1973)). The  structure  of  the  curvature
         operator yields a simple property,
$$
{R^\kappa}_{\lambda\mu\nu}w^\lambda=
{w^\kappa}_{;\nu;\mu}-{w^\kappa}_{;\mu;\nu}.
        \eqno{(1.2.21)}
$$
         It is also easy to show that simple algebraic identities hold,
$$
R_{\kappa\lambda\mu\nu}=R_{[\kappa\lambda][\mu\nu]}=
R_{\mu\nu\kappa\lambda},
$$
as well as
$$
R_{\kappa[\lambda\mu\nu]}=0
={R\mbox{\footnotesize ${\nu\lambda}\atop\ast$}}_{\mu\nu}
    \eqno{(1.2.22)}
$$
(this last identity bears the name of Ricci).  Important  role  is
         also played by the differential Bianchi identity,
$$
R_{\kappa\lambda[\mu\nu;\xi]}=0
={R\mbox{\footnotesize ${\alpha\beta}\atop\ast$}}_{\kappa\lambda;\beta}.
    \eqno{(1.2.23)}
$$
         The Ricci curvature tensor we define as
  $R_{\mu\nu}\equiv R_{\nu\mu}:={R^\lambda}_{\mu\nu\lambda}$,
  {\em cf.} Eisenhart  (1926, 1933),  Landau and Lifshitz   (1971),  and
         Misner,
         Thorne and Wheeler (1973). There are also  introduced
  the scalar curvature, $R:=R^\alpha_\alpha$, and the Weyl conformal
         curvature tensor,
$$
C_{\kappa\lambda\mu\nu}:=R_{\kappa\lambda\mu\nu}+
\frac{1}{2}(g_{\kappa\mu}R_{\lambda\nu}+g_{\lambda\nu}R_{\kappa\mu}-
g_{\lambda\mu}R_{\kappa\nu}-g_{\kappa\nu}R_{\lambda\mu})
$$
\nopagebreak
$$
-\frac{1}{6}R(g_{\kappa\mu}g_{\lambda\nu}-g_{\kappa\nu}g_{\lambda\mu}).
     \eqno{(1.2.24)}
$$
         The latter (taken with one  contravariant  index,  and  in  a
         coordinated basis), does not feel  a  multiplication  of  the
         metric tensor by an arbitrary function, so it is
         {\em conformally invariant}.

   In the  formalism  of  Cartan's  exterior  differential
         forms,  see  (Israel  1970,  Ryan  and  Shepley  1975),   the
         operation of exterior differentiation is introduced,
$$
d:=\theta^{(\alpha)}\nabla_{X_{(\alpha)}}\wedge,
             \eqno{(1.2.25)}
$$
         where $\wedge$ is the discussed above wedge product (the
         operation  of exterior multiplication), so that $dd\equiv0$.
         The operation $d$ has
         its most simple representation in a coordinated basis  (since
         the Christoffel symbols are symmetric in their two lower  indices).
         It is however especially convenient to work in an orthonormalized
         or Newman--Penrose basis when the connection 1-forms
$$
{\omega^{(\alpha)}}_{(\beta)}:=
\Gamma^{(\alpha)}_{(\beta)(\gamma)}\theta^{(\gamma)}
$$
         are skew-symmetric. In the general case they are found  as  a
solution of the system of {\em Cartan's first structure equations},
$$
d\theta^{(\alpha)}=-{\omega^{(\alpha)}}_{(\beta)}\wedge\theta^{(\beta)},
       \eqno{(1.2.26)}
$$
         and relations for differential of the tetrad metric,
$$
dg_{(\alpha)(\beta)}=\omega_{(\alpha)(\beta)}+\omega_{(\beta)(\alpha)}.
$$
         These are just 24 + 40 equations for determining all  the  64
         components  of  the  connection  coefficients,  so  that  the
         solution of this set of equations should  be  unique.  It  is
         easy to accustom  to  find  the  specific  solutions  without
         performing tedious formal calculations; then the  work  takes
         surprisingly short time. Since reference frame  characteristics
         (vectors of acceleration and rotation, and rate-of-strain
         tensor) below will turn  out  to  be  a  kind  of  connection
         coefficients, this approach allows to calculate promptly also
         these physical characteristics, though for the rate-of-strain
         tensor the computation will be rather specific.

   In their turn, the curvature tensor components are also
computed easily and promptly, if one applies {\em Cartan's  second
         structure equations},
$$
{\Omega^{(\alpha)}}_{(\beta)}=d{\omega{(\alpha)}}_{(\beta)}
+{\omega^{(\alpha)}}_{(\gamma)}\wedge{\omega^{(\gamma)}}_{(\beta)}.
       \eqno{(1.2.27)}
$$
         Here the calculations are much more straightforward than  for
         the connection 1-forms, and they take less time.  It  remains
         only to read out expressions  for  the  curvature  components
         according to the definition
$$
{\Omega^{(\alpha)}}_{(\beta)}:=
\frac{1}{2}{R^{(\alpha)}}_{(\beta)(\gamma)(\delta)}\theta^{(\gamma)}
\wedge\theta^{(\delta)}.
$$
         The Ricci and Bianchi identities  read  in  the  language  of
         Cartan's forms as
$$
{\Omega^{(\alpha)}}_{(\beta)}\wedge\theta^{(\beta)}=0
            \eqno{(1.2.28)}
$$
    and
$$
d{\Omega^{(\alpha)}}_{(\beta)}=
{\Omega^{(\alpha)}}_{(\gamma)}\wedge{\omega^{(\gamma)}}_{(\beta)}-
    {\omega^{(\alpha)}}_{(\gamma)}\wedge{\Omega^{(\gamma)}}_{(\beta)}
          \eqno{(1.2.29)}
$$
         correspondingly.

   Let us give here a scheme of definition of  transports.
Denote first as $\delta_{\mbox{\tiny ${\rm T}$}}v:=v|_Q-v|_{P\rightarrow Q}$
the change of a vector (or of some other object)
under its transport  between  points $P$
         and $Q$ (see (Mitskievich, Yefremov and Nesterov  1985)).  Then
         for the parallel transport
$$
\frac{\delta_{\mbox{\tiny ${\rm P}$}}v}{d\lambda}=\nabla_u v
      \eqno{(1.2.30)}
$$
         and for the Fermi--Walker transport ({\em cf.} Synge (1960))
$$
\frac{\delta_{\mbox{\tiny ${\rm FW}$}}v}{d\lambda}=\nabla_u v+
\eta[(v\cdot\nabla_uu)u-(v\cdot u)\nabla_uu]=:
\nabla^{\mbox{\tiny ${\rm FW}$}}_u v     \eqno{(1.2.31)}
$$
  where $u^\alpha=dx^\alpha/d\lambda$ is the tangent vector
  to the transport curve
         (for the Fermi--Walker transport, it is  essential  to  choose
         canonical parameter along the transport curve in such  a  way
         that the norm $u\cdot u=\eta$ would be constant and non-zero
         --- so the
         curve  itself  should  be  non-null  for   the   Fermi--Walker
         transport). Using the definitions (1.2.14),  it  is  easy  to
         generalize the transport relations to  arbitrary  tensors  or
         tensor densities; so for the Fermi--Walker transport
$$
\frac{\delta_{\mbox{\tiny ${\rm FW}$}}T_a}{d\lambda}=
T_{a;\alpha}u^\alpha+\eta T_a|^\tau_\sigma(u_{\tau;\alpha}u^\sigma
-u_\tau{u^\sigma}_{;\alpha})u^\alpha.
              \eqno{(1.2.32)}
$$
         From (1.2.31) a  definition of  the  Fermi--Walker  connection
seems to be directly obtainable, together with the corresponding
covariant FW differentiation operation which should lead also to
         the FW-curvature ({\em cf.} (Mitskievich, Yefremov and Nesterov
         1985)):
$$
{\sf R}^{\mbox{\tiny ${\rm FW}$}}(w,v):=
\nabla^{\mbox{\tiny ${\rm FW}$}}_u
\nabla^{\mbox{\tiny ${\rm FW}$}}_v
-\nabla^{\mbox{\tiny ${\rm FW}$}}_v
\nabla^{\mbox{\tiny ${\rm FW}$}}_u -
\nabla^{\mbox{\tiny ${\rm FW}$}}_{[u,v]}.    \eqno{(1.2.33)}
$$
        The reader may find that in this generalization a difficulty
        arises: the ``derivative''
        $\nabla^{\mbox{\tiny ${\rm FW}$}}_u$ does not possess the
        linearity property. This trouble is however readily removable if
        we begin with introduction of another type of derivative, namely
        $\nabla^\tau_u$ (see below) which possesses all necessary
        properties of a covariant derivative.

   We give here the (principal and equivalent to each other)
   forms of the geodesic equation for a canonical parameter:
$$
\nabla_uu=0,        \eqno{ (1.2.34)}
$$

$$
u\wedge\ast du=0,   \eqno{(1.2.35)}
$$

$$
\frac{d^2x^\alpha}{d\lambda^2}+
\Gamma^\alpha_{\beta\gamma}\frac{dx^\beta}{d\lambda}
\frac{dx^\gamma}{d\lambda}=0,  \eqno{(1.2.36)}
$$

$$
\frac{d}{d\lambda}(g_{\alpha\beta}\frac{dx^\beta}{d\lambda})=
\frac{1}{2}g_{\sigma\tau,\alpha}\frac{dx^\sigma}{d\lambda}
\frac{dx^\tau}{d\lambda}.       \eqno{(1.2.37)}
$$
         A geodesic may be, of course, also null (light-like), but  if
         it is time-like, one has $d\lambda= ds$.
         If a non-canonical  parameter
         is used, an extra term proportional to $u$ should be  added
         in the equations (1.2.34)  -  (1.2.37),  with  a  coefficient
         being arbitrary function of the (non-canonical) parameter.

   Below we shall make use of the just mentioned generalization of the
         FW-derivative (1.2.31) along a time-like congruence
         with a tangent  vector  field $\tau$ normalized  by  unity  (the
         {\em monad}). It is convenient to denote such a derivative as
         $\nabla^\tau_u$,
$$
\nabla^\tau_uT=\nabla_uT+T_a|^\mu_\nu\left(\tau_{\mu;\alpha}\tau^\nu-
\tau_\mu{\tau^\nu}_{;\alpha}\right)u^\alpha B^a,   \eqno{(1.2.38)}
$$
  where $B^a$ is the (coordinated) basis of $T$: $T=T_aB^a$.
  Then $\nabla^{\mbox{\tiny ${\rm FW}$}}_u\equiv\nabla^u_u$.
  The definition (1.2.38) can be also written as
$$
T^{~ [\tau]}_{a;\mu}=T_{a;\mu}+T_a|^\rho_\sigma(\tau^\sigma\tau_{\rho;\mu}
-{\tau^\sigma}_{;\mu}\tau_\rho).   \eqno{(1.2.38a)}
$$
         Consider now the properties of this $\tau$-derivative when it is
         applied to a scalar function $f$, the vector $\tau$, the metric
     tensor $g$,  and the construction $b=g-\tau\otimes\tau$ (the
     three-metric or the projector in the monad formalism):
$$
\nabla^\tau_uf=uf, ~ ~ \nabla^\tau_u\tau=\nabla^\tau_ug
=\nabla^\tau_ub=0    \eqno{(1.2.39)}
$$
(action of a $\tau$-derivative on a tensor is understood in the sense of
         (1.2.38); $u$ is  considered  as  a  partial  differentiation
operator in $uf$). The corresponding $\tau$-curvature  operator  is
  denoted as ${\sf R}^\tau(w,v)$  ({\em cf.}
  (2.4.10);  see  also  (Mitskievich, Yefremov and Nesterov 1985)).

   Returning to the Lie derivative, it is worth mentioning
         the relations
$$
\pounds_\xi g_{\mu\nu}=\xi_{\mu;\nu}+\xi_{\nu;\mu},        \eqno{(1.2.40)}
$$

$$
\pounds_\xi\Gamma^\lambda_{\mu\nu}={\xi^\lambda}_{;\mu;\nu}-
\xi^\kappa{R^\lambda}_{\mu\nu\kappa}\equiv{\xi^\lambda}_{(;\mu;\nu)}-
\xi^\kappa{R^\lambda}_{(\mu\nu)\kappa}.   \eqno{(1.2.41)}
$$
         A vector field satisfying the equation
$$
\xi_{\mu;\nu}+\xi_{\nu;\mu}=0,       \eqno{(1.2.42)}
$$
         is called the Killing field. Its existence imposes
         certain  limitations  on the geometry, ---
         existence of an {\em isometry}: $\pounds_\xi g_{\mu\nu}=0$
         (see (Yano 1955, Eisenhart  1933,  Ryan  and  Shepley  1975)).
         The  Lie
         derivative plays an important role in the monad formalism, as
         we shall see below; this is why a version of  this  formalism
         well adapted to description of the canonical structure of the
         field  theory,  is  called  the ``Lie-monad   formalism'' (see
         (Mitskievich and Nesterov  1981,  Mitskievich,  Yefremov  and
         Nesterov 1985)).

   Alongside  with  exterior   differentiation $d$   which
         increases  the  degree  of  a  form  by  unity,  in  Cartan's
         formalism  there  is  employed  also   another   differential
         operation which diminishes this degree by unity,
$$
        \delta:= -\ast d\ast .    \eqno{(1.2.43)}
$$
         In fact this is the divergence operation,
$$
\delta(\alpha_{\mu h}dx^{\mu h})=p{\alpha_{\mu h}}^{;\mu}dx^h, ~ ~
p= \# h+1,      \eqno{(1.2.44)}
$$
         which does not however fulfil the Leibniz property: when applied
         to a product, {\em e.g.}, of a function and a 1-form, $fa$, it
         results in
$$
\delta(fa)=f\delta a-\ast(df\wedge\ast a),
$$
         while it acts on a function (0-form) simply annihilating  it,
         $\delta f\equiv 0$. One might call this
         a generalized  Leibniz  property.
         Of course, similarly to the identity $dd\equiv 0$, there holds also
         the identity $\delta\delta\equiv 0$.
         A combination of the both operations
         gives the de Rham operator (de-Rhamian),
$$
\triangle :=d\delta +\delta d \equiv (d+\delta)^2   \eqno{(1.2.45)}
$$
         (remember divergence of a gradient in Euclidean geometry).  A
         form annihilated by the de-Rhamian, is called  harmonic  form
         (remember harmonic functions and the  Laplace  operator).  In
         contrast to $\delta$, the de-Rhamian acts non-trivially on  0-forms,
         and in contrast to  the  exterior  differential $d$, it acts
         non-trivially on 4-forms too (also in the four-dimensional  world).
         There  are  introduced  also  the  following   classification
         elements of forms: an exact form is that which is an exterior
         derivative of another form;  a  closed  form  is  that  whose
         exterior differential  vanishes  (an  exact  form  is  always
         closed, though the converse is in general not  true);  a  co-exact
         form is a result of action by $\delta$ on  some  form  of  a
         higher (by unity) degree; a co-closed form
         is itself identically annihilated by $\delta$.
         In (topologically) simplest cases it is  possible
         to split an arbitrary form into sum of a harmonic, exact, and
         co-exact  forms  (see  Eguchi,  Gilkey  and   Hanson (1980),
         Mitskievich, Yefremov and Nesterov (1985)).

For  an arbitrary form $\alpha$,
$$
\triangle\alpha={\alpha_{a;\mu}}^{;\mu}dx^a
-p{{{R^\sigma}_\tau}^\mu}_\nu\alpha_{\mu h}|^\tau_\sigma dx^{\nu h}, ~ ~
\# a=p.   \eqno{(1.2.46)}
$$
         The number of  individual  indices  entering  the  collective
index $a$ of the components of the form $\alpha$, may be any from 0 to
  4, while the quantity $\alpha_a|^\tau_\sigma$ is  determined
  by  (1.2.15).  The expression (1.2.46) can be reduced identically to
$$
\triangle\alpha=(p!)^{-1}[{\alpha_{\mu\nu h;\lambda}}^{;\lambda}+
pR^\sigma_\mu\alpha_{\sigma\nu h}+
\frac{p(p-1)}{2}{R^{\sigma\tau}}_{\mu\nu}\alpha_{\sigma\tau h}]
dx^{\mu\nu h}     \eqno{(1.2.47)}
$$
         where $a=\mu\nu h$, $\# a=2+\# h=p$, and $p$ may be
         equal  also  to zero.

   We compare here explicitly the  results  of  action  of
         de-Rhamiam on forms of  all  degrees  in  a  four-dimensional
         world:
$$
f: ~ \triangle f= {f_{,\mu}}^{;\mu},         \eqno{(1.2.48)}
$$

$$
A=A_\alpha dx^\alpha: ~  \triangle A=({A_{\alpha;\mu}}^{;\mu}+
R^\sigma_\alpha A_\sigma)dx^\alpha,     \eqno{(1.2.49)}
$$

$$
F=\frac{1}{2}F_{\alpha\beta}dx^{\alpha\beta}: ~
\triangle F=\frac{1}{2}({F_{\alpha\beta;\mu}}^{;\mu}+
2R^\sigma_\alpha F_{\sigma\beta}+{R^{\sigma\tau}}_{\alpha\beta}
F_{\sigma\tau})dx^{\alpha\beta},    \eqno{(1.2.50)}
$$

$$
T=\frac{1}{3!}T_{\alpha\beta\gamma}dx^{\alpha\beta\gamma}: ~
\triangle T=\frac{1}{3!}({T_{\alpha\beta\gamma;\mu}}^{;\mu}+
3R^\sigma_\alpha T_{\sigma\beta\gamma}+
3{R^{\sigma\tau}}_{\alpha\beta}T_{\sigma\tau\gamma})dx^{\alpha\beta\gamma}
$$

$$
=({\tilde{A}_{\alpha;\mu}}^{;\mu}+R^\sigma_\alpha\tilde{A}_\sigma)
\ast dx^\alpha,
    \eqno{(1.2.51)}
$$

$$
V=\frac{1}{4!}V_{\alpha\beta\gamma\delta}dx^{\alpha\beta\gamma\delta}: ~
\triangle V=\frac{1}{4!}{V_{\alpha\beta\gamma\delta;\mu}}^{;\mu}
dx^{\alpha\beta\gamma\delta}={\tilde{f}_{;\mu}}^{;\mu}\ast 1.
        \eqno{(1.2.52)}
$$
         From the last two expressions it is clear that there exists a
         symmetry with respect to an exchange of a basis form to its dual
         conjugate with a simultaneous dual conjugation of  components
         of the form (compare  (1.2.48)  and  (1.2.52),  (1.2.49)  and
         (1.2.51)). Here we made use of obvious definitions
$$
\tilde{A}^\delta=
-\frac{1}{3!}T_{\alpha\beta\gamma}E^{\alpha\beta\gamma\delta}, ~ ~
\tilde{f}=
-\frac{1}{4!}V_{\alpha\beta\gamma\delta}E^{\alpha\beta\gamma\delta},
$$
         alongside with the easily checkable identity,
$$
(2R^\sigma_\alpha E_{\sigma\beta\gamma\delta}
+3{R^{\sigma\tau}}_{\alpha\beta}E_{\sigma\tau\gamma\delta})
E^{\alpha\beta\gamma\delta}\equiv 0.
$$

\newpage

\pagestyle{myheadings}
\markboth{CHAPTER 2. REFERENCE FRAMES' CALCULUS}{2.1. THE MONAD
FORMALISM}

 \chapter{Reference frames' calculus}

 \section{The monad formalism and its place in the description of
 reference frames in relativistic physics}

   It would be completely erroneous to automatically identify
   the concepts of a system of coordinates and of a  reference
 frame. Such an  identification  is  often  characteristic  for
 those physicists who  are  accustomed  to  work  in  Cartesian
 coordinates and who, moreover, try  to  reduce  (for  sake  of
 "simplicity") problems --- mainly, educational ones  ---  to
 two-dimensional diagrams dealing with  one  temporal  and  another
 spatial, coordinates. But when more than one spatial dimension
 (plus extra one, temporal dimension) are considered, a question
 arises whether it is possible to relate to different reference
 frames systems of coordinates which differ one from the  other
 by a purely spatial rotation (in the simplest  case  when  the
 both systems are Cartesian). The more such a  problem  becomes
 obvious when we pass from a  Cartesian   to  polar  system  of
 coordinates.

 Let us call  to  mind  the  Newtonian  mechanics  (see, {\em e.g.},
 Newton (1962),  Khaikin  (1947),  H. Goldstein (1965),  Schmutzer
 (1989), as well as Schmutzer and  Sch\"utz  (1983)), and  think  it
 over, in which cases there appear, for example, the forces  of
 inertia  (we  do  not  intend  to  take  part  in   scholastic
 discussions over the reality of those). What is clear is  that
 such forces are originated by  the  state  of  motion  of  the
 reference frame. But if we consider two  reference  frames  at
 rest, one with respect to the  other,---  could  any  difference
 between them be  physically  perceivable  in  course  of  some
 experiments, at least in principle? If this would be the case,
 it would be possible to experimentally determine, what kind of
 coordinate frame ({\em e.g.}, Cartesian or polar one) is used by the
 observer, or, at least, which of the relatively resting  bases
 corresponds to the latter. It is however clear that experiments
 are performed irrespective to the  choice  of  purely  spatial
 coordinates (the corresponding system of coordinates is chosen
 only at the stage  of  mathematical  description  of  physical
 effects, and the experimental devices --- even if they are  used
 for measurement of lengths or angles --- have no direct relation
 to this choice). The same relates also  to  the  choice  of  a
 purely spatial basis. Thus transcriptions of  all  expressions
 and quantities supposed to describe physical effects, have  to
 be performed invariantly with respect to the choice of spatial
 coordinates or spatial basis. This demand is not as mild as it
 may seem to be at the first sight, and it yields a set of
 geometrical identities to  be  imposed  on  the  corresponding
 congruences.

   Hence  we  see  that  a  transition  between   different
 reference  frames  should  basically  reflect  their  relative
 motion. The simplest special case of such  a  motion  is  that
 which relates two non-identical inertial reference frames both
 in Newtonian and relativistic mechanics.

   It is also obviously possible to an observer or a system
 of observers to move  together  with  a  frame  of  reference,
 locally or in a space-time region (globally).
 When an object moves together with a reference  frame,  it  is
 geometrically identified with the latter, so that if this is a
 spatially extended object, the world lines of its mass  points
 form a congruence.  Such  a  congruence  presents  a  complete
 characterization  of  the  reference  frame,  and   with   any
 reference frame a conceptual object (thus a test one)  of  the
 above type can be associated  which  is  called  the  body  of
 reference. Since it models a set  of  observers  and/or  their
 measuring  devices,  but  not  photons,  the  reference  frame
 congruence  should  be  time-like.  A  transition  to  another
 reference frame means merely a change of  such  a  congruence,
 with the  corresponding  and  automatic  change  of  the  test
 reference body. In order to avoid  mathematical  difficulties,
 one has to necessarily use congruences, since otherwise  there
 would appear a danger of either mutual  intersections  of  the
 world  lines  (which  mathematically  manifests  itself  as  a
 non-physical singularity), or in the  space-time  there  would
 arise bald spots  whose  boundaries  will  represent
 singularities too.

   The simplest way to describe  a  reference  frame  is  to
 identify the congruence of the  corresponding  reference  body
 mass-points with a congruence of the  time  coordinate  lines.
 The latters should be, of  course,  time-like,  which  is  not
 always the case for the so-called time coordinate ({\em e.g.}, for
 the Schwarzschild space-time in the curvature coordinates, the
 $t$ coordinate lines become null on the horizon  and  space-like
 inside it), thus we  come  to  a  restriction  concerning  the
 choice  of  coordinates  adequate  for  the  reference   frame
 description of physical situations. Here a strange  (from  the
 point of view  of na\"\i ve psychology)  fact arises:  it is  easy
 to see that any transformation of  the  time  coordinate  with
 fixed all other (spatial) coordinates, can at most change  the
 lines of the latters, but {\em not} the time coordinate lines, {\em i.e.}
 we remain  with  the  former  reference  frame  if  determined
 through the time coordinate description. Such  transformations
 are called the chronometric ones,  and  the  invariance  of  a
 reference frame  with  respect  to  them  gives  a  basis  for
 Zel'manov's formalism of  chronometric  invariants  (Zel'manov
 1956, 1959) being of importance primarily for  cosmology  (see
 also (Mitskievich and Zaharow 1970, Mitskievich 1969)).

   Another formalism, that of  kinemetric  invariants,  was
 also proposed and developed by Zel'manov (1973). It  is  based
 on a description of reference frames with help of families of space-like
 hypersurfaces in the space-time. We confine ourselves here to
 the corresponding reference since, as also  the  formalism of
 chronometric invariants, this is in fact a special  choice of
 the system of coordinates in the monad  formalism  extensively
 discussed below. This  latter  was  primarily  worked  out  by
 Zel'manov (1976); see also good reviews  of  all  these  three
 formalisms in Zel'manov and Agakov  (1989),  Massa  (1974a,b,c),
 and  partially  in  Shteingrad   (1974),   Mitskievich    (1972),
 Ivanitskaya, Mitskievich and Vladimirov (1985, 1986). It should
 be noted that the formalism of chronometric  invariants,  with
 some  modifications  (which  bring  it  near  to   the   monad
 formalism), was independently  developed  by  Cattaneo  (1958,
 1959,  1961,  1962) and  Schmutzer  (1968),   while   some
 applications to the study of physical  effects  are  given  in
 Schmutzer and Plebanski (1977) and Dehnen (1962). In  its  turn,
 the formalism of kinemetric invariants represents a refinement
 of the earlier formalism of Arnowitt, Deser and  Misner  (ADM)
 (see  Misner,  Thorne  and  Wheeler   (1973),   as   well   as
 Mitskievich and  Nesterov  (1981),  Mitskievich,  Yefremov  and
 Nesterov (1985)). This ADM formalism proved to be of importance
 for realizing the canonical formalism and the  quantization
 procedure in gravitation  theory.  For  applications  of  an
 approach similar to  the  monad  formalism,  to  astrophysical
 effects, see Thorne and Macdonald (1982), Macdonald and  Thorne
 (1982), Thorne, Price and Macdonald (1986). An  extensive  review
 of methods  of  description  of  reference  frames  see  in  a
 monograph  by  Vladimirov  (1982)  who  himself  has  made   a
 substantial contribution to this field, and also in a recent
 paper by Jantzen, Carini and Bini (1992).

   During many years the principal role in  description  of
 reference frames in general relativity, was played however  by
 the tetrad formalism which influenced much the  very  style  of
 thinking. In the case of an  orthonormalized  tetrad,  we  are
 dealing not only with a time-like congruence which corresponds
 to four of the sixteen components of the tetrad (the field  of
 tangent vector to the congruence), but  with  three  (spatial)
 congruences more. One does not usually  care  much  about  the
 invariance of  expressions  which  have  a  physical  meaning,
 although this  obviously  contradicts  the  conclusions  drawn
 above. If we take such a care, we come to the monad  formalism
 with auxiliary use of  an  arbitrary  triad  in  the  subspace
 orthogonal to the monad which does not practically change  our
 further conclusions. The tetrad formalism  was  considered  by
 many  authors;  for  an  orientation  in   the   corresponding
 literature, see Ivanitskaya (1969), Mitskievich (1969)  and,  in
 connection with gravitational effects, Ivanitskaya (1979).

   Tetrad bases  are  not  always  directly  applicable  to
 description of  reference  frames, {\em e.g.} in the case of a
 Newman--Penrose basis (see Penrose and Rindler  (1984a,  1984b)
 where also  the  use  of  spinors  in  constructing  bases  is
 considered). In some  calculations,  it  proved  effective  to
 combine together the Dirac  matrices  and  Cartan  forms  when
 gravitational  fields  and  reference   frames   are   studied
 (Mitskievich 1975). General methods of  the  tetrad  formalism
 function well in  the  coordinates-free  approach  to  tensor
 calculus, in particular in the  Cartan  forms  formalism  (see
 section 1.2).

   We came thus to a conclusion that it is namely the monad
 formalism which is most closely connected with description  of
 frames of reference, and this formalism is  presented  in  the
 next Chapter. A question is however  still  worthwile:  is  it
 really always indispensable to use explicitly some  definition
 of a specific reference frame when observable physical effects
 are calculated from  the  point  of  view  of  a  non-inertial
 observer? And in  what  classes  of  problems  of  theoretical
 physics, and at  which  step  of  calculation  procedure,  the
 formalisms of  description  of  reference  frames  may  be  of
 practical use?

   When local effects are  studied,  this  means  that  the
 observer may be idealized as a  single point-like test mass while he
 receives all outside information through  signals  propagating
 with  the  fundamental  velocity  (electromagnetic  and  other
 signals),  as  well  as  detecting  all  other  physical   and
 geometric factors which are  perceivable  on  his  world  line
 only.  Then  this  world  line  will  be  the  only  essential
 representative of the state of motion of the observer  in  the
 calculations, so that any extension of the reference frame  to
 the outside will be absolutely unnecessary. Moreover,  if  the
 magnitude of some  predicted  physical  effect  would  to  any
 extent depend on the concrete  way  of  this  extension,  this
 effect could not  be  considered  as  a  correctly  determined
 physical one, since there  should
 be an invariance of all observables (for our local  observer)
 with respect to an arbitrary choice  of  the  reference  frame
 outside the observer's world line, if only  this  line  would,
 with a proper  smoothness,  belong  to  the  congruence  which
 globally  describes  the  pertinent  reference  frame.  It  is
 obvious that there exists infinite and continuous multitude of
 such reference frames mutually coinciding on  a  single  world
 line. An example of a calculation of a physical effect when no
 extension of the reference frame outside a single  world  line
 of the observer should be considered, is the case  studied  in
 section 3.4, a united description  of  the  gravitational  red
 shift effect and the Doppler effect.  These  two  effects  are
 inseparable when the space-time is non-stationary,  and  their
 distinction  is  a  mere  convention.   This   conventionality
 manifests itself in the mentioned case in  the  following  two
 interpretations of the effect. When the both  time-like  world
 lines (those of the emitter of radiation and of its  detector)
 are considered to belong to one and the same congruence  of  a
 reference frame, the both emitter and detector are at rest  by
 definition, and the effect has to be interpreted as a  purely
 gravitational one. The distance between  the  two  objects  is
 however  continually  changing;  in  a  similar  situation  in
 cosmology,  the  non-stationarity  of  space-time   completely
 prevents domination of any of the alternative interpretations!
 Another approach consists of inclusion in the reference  frame
 congruence  of the observer's (detector's)  world  line  only;
 then the radiation source will be in motion  with  respect  to
 the frame, so that the effect  would  be (partly) due  to  the
 Doppler shift. An interpretation of this effect as a completely
 Doppler  one,  is  also  possible  since  the  space-time   is
 stationary; the reference frame congruence is then the Killing
 one, although it fails to be time-like in the ergosphere.  But
 outside the latter, one has to consider the  effect  as  being
 due to motions of the both emitter and detector of radiation.
 As to the time-like Killing congruence, if it is rotating, one
 usually connects it with the dragging phenomenon.
 Such an objectively determined reference frame is discussed in
 section 3.5.

   There exists however a wide class of still other problems in which
 employment of a concrete global reference frame  is  perfectly
 acceptable and desirable, while  the  predicted  in  this  way
 effects do depend  on  its  specific  choice.  These  are  the
 problems in which the  interacting  physical  factors  possess
 spatial extension, {\em e.g.} the case of charged perfect fluid
 moving in electromagnetic and gravitational fields. In section
 4.4 we consider such an exact  self-consistent  problem  which
 moreover yields a strange (at first sight) result that for the
 state of motion under  consideration,  being  consistent  with
 behaviour  of  fields,  the  non-test  charged  perfect  fluid
 produces in its co-moving reference frame only a magnetic, but
 no electric  field  whatsoever.  Hence  its  particles,  being
 undoubtedly electrically charged,  do  not  mutually  interact
 electromagnetically. We show  there  also  that  there  is  no
 paradox, but this is merely an encounter with unusual (for the
 prevailing inertial mode  of  thinking)  specific  feature  of
 electromagnetic field in non-inertial, in particular, rotating
 reference frames. We shall discuss there a similarity of  this
 situation  with  that  in   classical   mechanics   with   its
 centrifugal and Coriolis forces. The concept  of  a  reference
 frame and employment  of  formalisms  of  its  description  is
 certainly fruitful in the  relativistic  cosmology  where  the
 continuous medium filling the universe, realizes a  privileged
 ({\em really existing})
 reference frame. The fact of a non-test nature of this  medium
 does not mean that the reference frame perturbs in  this  case
 geometry, this is merely a specific --- co-moving --- frame, which
 is objectively singled out of all other  frames  by  the  very
 statement of the problem.

   We do not (and cannot)  give  here  an  unambiguous  and
 precise  answer  to  the   question   what   problems   demand
 application of  the  concept  of  reference  frame, and for what
 problems this is unnecessary. Instead we confine ourselves  to
 certain examples, since it would  be  a  scholastic  task,  to
 foresee every special case of application of  reference  frames
 in the future. As to  introduction  of  this  concept  at an
 initial stage of every calculation,  this  clearly  cannot  be
 practical, since symmetry of a  problem may more  easily  be
 connected  with  a  system  of  coordinates  and  not  with  a
 reference frame, so that the adequate choice of coordinates is
 more crucial for the most effective treatment of problems
 ({\em e.g.}, when Einstein--Maxwell fields are derived).
 When however one  turns  to  interpretation  of  the  obtained
 solutions, introduction of reference  frames  becomes  quite
 opportune, and it may play a decisive role. These considerations
 bring us to a better understanding of the  fact  that  the
 first of all complete and practical methods of description  of
 reference frames was  Zel'manov's  formalism  of  chronometric
 invariants (the quotient system of coordinates co-moving with  a
 frame of reference), and this was primarily done in problems
 of  relativistic  cosmology.  In  the  same  context,  certain
 problems  connected  with  the  Noether  theorem   were   also
 considered  by Mitskievich  and  Mujica  (1967) and  Mitskievich  and
 Ribeiro Teodoro (1969).

   The  concept  of  reference  frame  should  be  probably
 essential also  at  starting  points  of  development  of  new
 theoretical ideas.  As  an  example  the  idea  of  zero-point
 radiation may  be mentioned  which  seems  to  be  still  at  a
 relatively early stage of its development. We only recall that
 the zero-point radiation possesses such a spectrum that  under
 Lorentz transformation, it retains its form  (while  a  black-body
 radiation would alter its temperature), but  a transition
 to a uniformly accelerated frame yields  in  addition  to  the
 former zero-point radiation, also  a  black-body  thermal  one
 (Boyer 1980). The reader should not be perplexed by  the  fact
 that the integral energy of the zero-point radiation  diverges
 (it is  usually  treated  in  the  spirit  of  renormalization
 ideas). These items are closely related to the concept  of
 the Rindler vacuum, and they accord with the quantum theory of
 black holes and the black hole evaporation (Thorne, Price  and
 Macdonald 1986). In quantum physics the concept  of  reference
 frame has however not yet found its adequate realization.  At
 the same time,  its  possible  application  areas   are  quite
 far-reaching: this is the  theory  of  quantum  fields  itself
 (Gorbatsievich 1985) as well as motivation and fundamentals of
 the field quantization (the  ADM  formalism to be once more mentioned
 (Misner, Thorne and Wheeler 1973), alongside of a  new  development
 in this area --- Ashtekar's canonical  formalism  (Ashtekar
 1988), some application of which will be sketched out  in  the
 concluding chapter of this paper).

   As it was motivated above,  we  have  chosen  the  monad
 approach for description of frames of reference,  while  monad
 means a fixed unit time-like vector field which  is  naturally
 interpreted as a field of four-velocities of  local  observers
 equipped  with  all  necessary  measurement  devices.  In  the
 space-time region where such a reference frame is realised, we
 may therefore consider a congruence of integral curves of this
 vector field, {\em i.e.} of the world-lines of particles forming a
 reference body. Every local observer represents
 such a test  particle.  The  classical  (non-quantum)  physics
 approach admits that acts of measurement do not influence  the
 objects and processes which are subjected to the  measurement,
 in full conformity with our treatment of  local  observers  as
 test particles, --- in  other  words,  in  conformity  with  the
 identification of a reference frame and a time-like congruence
 which, of course, should be geometrically  admissible  in  the
 space-time region under consideration. Then  a  transition  to
 another reference frame should be equivalent to a  new  choice
 of such an admissible congruence, without any  change  of  the
 four-geometry of the world itself.

   Rodichev (1972, 1974) has undertaken an attempt to  overpass
 the limits of this  approach  to  reference  frames,
 but we consider his attempt as  a  premature  and
 technically  wrong  one,  though  it  is   worth   making   an
 acquaintance with his concepts which were expressed  with
 much inspiration.  These  concepts  might  become  useful  for
 far-reaching generalizations in  the  future,  such  as  those
 connected with quantum  gravity.  As  a  matter  of  fact,  in
 quantum  theory  it  is  useless  to  try  to  introduce  test
 particles in the classical meaning, and  moreover,  the  local
 observers cannot be considered there without perturbations  of
 the  four-geometry,  as  well  as  of  all  involved  physical
 objects. Thus one has to consider a radical  revision  of  the
 concept of reference frame when passing  to  quantum  physics.
 Some ideas of such a generalization were outlined by Fock (1971)
 (see also (Schmutzer 1975)),
 and some (rather obscure) remarks  were  made  by
 Brillouin (1970), but probably the most powerful  insight  can
 be found in the early paper by Bohr and Rosenfeld (1933)  (see
 also Regge (1958) and Anderson (1958a,b; 1959)).  We  intend  here,
 however,  to  confine  ourselves  to  classical   relativistic
 physics.

 \pagestyle{headings}

 \section{Reference frames algebra}

   The  first  steps in  the  development  of  the   monad
 formalism were made  already  in  the  forties  and  fifties
 by Eckart (1940) and Lief  (1951)  on  the  basis  of  hydrodynamical
 analogies. In our general relativistic approach,  this  means
 that in the region under consideration, a time-like four-vector monad
 field $\tau$ is introduced in  addition  to  the  four-dimensional
 metric $g$, with a normalization condition $\tau\cdot\tau=+1$.
 Then a symmetric tensor,
$$
 b=g-\tau\otimes\tau,     \eqno{(2.2.1)}
$$
 will be automatically the projector onto the  local  subspace
 orthogonal to $\tau$:
$$
\begin{array}{rr}
b(\tau,\cdot)=0, ~ ~  & b^\lambda_\lambda=3,~   \\
\det b_{\mu\nu}=0, ~ ~  & b_{\mu\nu}b^{\lambda\nu}=b^\lambda_\mu.
\end{array}
  \mbox{\Huge $\rbrace$}         \eqno{(2.2.2)}
$$
 As we shall see now, the tensor $b$ plays  simultaneously  the
 role of a metric  in  this  subspace  (the  three-dimensional
 physical space of the reference frame).  The  three-space  is
 then non-holonomic in general (namely, when the $\tau$-congruence
 is rotating).

   Consider  now  an  arbitrary  four-(co)vector $q$;   its
 pro\-jec\-tion on\-to the mo\-nad is a sca\-lar,
$$
\stackrel{(\tau)}{q}:=q\cdot\tau,    \eqno{(2.2.3)}
$$
 and the projection onto  the  three-space  of  the  reference
 frame, a four-di\-men\-si\-o\-nal (co)vector,
$$
\stackrel{(3)}{q}:=b(q,\cdot),      \eqno{(2.2.4)}
$$
 which is by a definition orthogonal  to  the  monad  and  not
 changing by a repeated projection. Hence,
$$
q=\stackrel{(\tau)}{q}\tau+\stackrel{(3)}{q}.    \eqno{(2.2.5)}
$$
 Applying the same procedures to another  (co)vector, $p$,  one
 can write now
$$
p\cdot q\equiv g(p,q)=\stackrel{(\tau)}{p}\stackrel{(\tau)}{q}+
\stackrel{(3)}{p}\cdot\stackrel{(3)}{q}=
:\stackrel{(\tau)}{p}\stackrel{(\tau)}{q}-
\stackrel{(3)}{p}\bullet \stackrel{(3)}{q},   \eqno{(2.2.6)}
$$
 where  we  have  introduced  the  notation $\bullet$ for  a
 three-dimensional scalar product relative to the reference frame $\tau$.
 The sign before this three-scalar product is chosen
 according to signature of the tensor $b$, which can be written
 symbolically as
$$
    {\rm sign}\;  b= (0,-,-,-),      \eqno{(2.2.7)}
$$
 it guarantees non-negativity  of  such  a  three-scalar
 square of any vector. Now observe that
$$
p\bullet q=\ast[(\tau\wedge p)\wedge\ast(\tau\wedge q)] \eqno{(2.2.8)}
$$
 where the (co)vectors $p$ and $q$ are considered to be  arbitrary
 (four-di\-men\-si\-o\-nal), and their pro\-jec\-tion  on\-to  the  physical
 three-space of the reference frame  is  provided automatically
 by the projecting properties of the three-metric.

   Let us apply this simple result to the squared interval
 using notations $dt$ and $d\!\stackrel{(3)}{x}$ for
 the elements of physical time
 and three-spatial displacement  (relative  to  the  reference
 frame under consideration) correspondingly, according to  the
 definitions  of  projections,  (2.2.3)  and  (2.2.4).   These
 elements are, of course, not total differentials ({\em i.e.}, they
 are non-holonomic). Thus
$$
ds^2=dt^2-dl^2, ~ ~ dl^2:=d\stackrel{(3)}{x}\cdot d\stackrel{(3)}{x}
\equiv-b(dx,dx),  \eqno{(2.2.9)}
$$
 and for propagation of a light pulse (signal),
$$
ds^2=(1-v^2)dt^2=0 ~ ~ \Longrightarrow ~ ~ v=1,  \eqno{(2.2.10)}
$$
 where $v=dl/dt$ is the absolute value of the  three-velocity.
 This result is a standard assertion of  the  universality  of
 the speed of light (but now in an arbitrary  reference  frame
 and in any gravitational fields). This  conclusion  does  not
 contradict  the  experimental  facts  of  the  time-delay  of
 electromagnetic signals when  propagated  near  large  masses
 ({\em e.g.}, that of the Sun),  since  in  such  experiments, the
 distances were  calculated  in  the  same  sense  as  in  our
 approach, but the physical time intervals were measured by  a
 clock on the Earth, while in  the  expression  (2.2.10),  the
 local  clocks  are  supposed  to  be  used  which  should  be
 localized continuously along the path  of  the  signal  (this
 means, of course, merely the corresponding re-calculation, as
 it was the case for the distances!).

   It is clear that the four-velocity of any object can be
 easily expressed through the  monad  vector  and  the
 three-velocity of this object,
$$
u=\frac{dt}{ds}(\tau+v), ~ v=b(\frac{dx}{dt},\cdot),  \eqno{(2.2.11)}
$$
 while
$$
\frac{dt}{ds}= \stackrel{(\tau)}{u}=(1-v^2)^{-1/2}.  \eqno{(2.2.12)}
$$

   Multy-index quantities (components of tensors)  are  to
 be projected in each of their indices onto either the  monad,
 or the three-space orthogonal to  the  latter,  so  that  the
 whole family of objects of various tensor ranks arises  (some
 examples can be found below).

   Let us introduce the three-spatial  (and  hence  three-index)
   axial Levi-Civit\`a tensor,
$$
e_{\lambda\mu\nu}:=\tau^\kappa E_{\kappa\lambda\mu\nu}.  \eqno{(2.2.13)}
$$
Practically, we shall not use this objects, but its existence
 is obvious, and  considering  it  helps  to  understand  more
 profoundly some expressions  below.  Similarly  to
 (1.2.3) we introduce also  the  skew  tensor $b^a_b$,  with
 antisymmetrization in all individual indices forming each  of
 the collective ones. Then
$$
\begin{array}{ll}
e_{\lambda\mu\nu}e^{\alpha\beta\gamma}=
 -0!3!b_{\lambda\mu\nu}^{\alpha\beta\gamma}; ~ & ~
e_{\alpha\mu\nu}e^{\alpha\beta\gamma}=
 -1!2!b_{\mu\nu}^{\beta\gamma}; \\
e_{\alpha\beta\nu}e^{\alpha\beta\gamma}=
 -2!1!b_\nu^\gamma; ~ & ~
 e_{\alpha\beta\gamma}e^{\alpha\beta\gamma}= -3!0!.
\end{array}
  \mbox{\Huge $\rbrace$}   \eqno{(2.2.14)}
$$
As it was the case in the flat  three-dimensional  world  for
 the Levi-Civit\`a symbol, the axial tensor  (2.2.13)  makes  it
 possible to relate rank two skew tensors to the corresponding
 vectors which also belong  to  the  three-space  of  the
 reference frame. This fact becomes, as usually, the basis for
 definition of  the  vector  product  (and,  later,  the  curl
 operation, (2.4.1)). Its symbolic expression reads
$$
p\times q:=\ast(p\wedge\tau\wedge q).   \eqno{(2.2.15)}
$$
 As in the scalar product (2.2.8), one may  take  here  simply
 the {\em four}-vectors $p$ and $q$ without  specially  projecting  them
 onto the  three-space  of  the  reference  frame,  while  the
 resulting vector product will depend on the properly  projected
 vectors, and it belongs itself to this  three-space.  The
 corresponding expression for a mixed triple product is
$$
n\bullet(p\times q)=\ast(\tau\wedge n\wedge p\wedge q). \eqno{(2.2.16)}
$$

   It is convenient to introduce 1-form quantities
$$
b^{(\alpha)}=\theta^{(\alpha)}-\tau^{(\alpha)}\tau, ~ ~
b_{(\alpha)}=X_{(\alpha)}-\tau_{(\alpha)}\tau  \eqno{(2.2.17)}
$$
 (the coordinated basis could be also used).  Now  the  three-metric
 itself arises as a three-scalar product,
$$
\ast(b^{(\alpha)}\wedge\ast b^{(\beta)})\equiv
\ast(b^{(\beta)}\wedge\ast b^{(\alpha)})=
b^{(\alpha)}\bullet b^{(\beta)}=-b^{(\alpha)(\beta)}   \eqno{(2.2.18)}
$$
 while we have, naturally,
$$
\ast(b^{(\alpha)}\wedge\ast\theta^{(\beta)})\equiv
\ast(\theta^{(\alpha)}\wedge\ast b^{(\beta)})=
-b^{(\alpha)(\beta)}.    \eqno{(2.2.19)}
$$

   Other important formulae involving $\tau$ and $b^{(\alpha)}$ are
$$
dx^{\alpha\beta}=\tau^\alpha\tau\wedge b^\beta-
\tau^\beta\tau\wedge b^\alpha+b^\alpha\wedge b^\beta  \eqno{(2.2.20)}
$$
 and
$$
dx^{\alpha\beta}=\tau^\alpha\tau\wedge b^\beta-
\tau^\beta\tau\wedge b^\alpha-{e^{\alpha\beta}}_\gamma\ast(\tau\wedge
b^\gamma).    \eqno{(2.2.21)}
$$
Transition from the first to the second relation  means  that
 there holds identity
$$
b^\alpha\wedge b^\beta=-{e^{\alpha\beta}}_\gamma\ast(\tau\wedge b^\gamma)
  \eqno{(2.2.22)}
$$
which can be easily checked.

   The monad formalism was  first  introduced  by
 Zel'manov (1976). This monad description of reference  frames
 is connected with the formalism  of  chronometric invariants
 (which belongs also to Zel'ma\-nov (1956)) by a  special  choice
 of the system of coordinates.
 It is important that the
 both approaches do not  involve  any  restrictions  upon  the
 choice of a reference frame, except for the standard  limitations
 due to geometry and physics in general, but  the  monad
 formalism allows in addition  also  an  arbitrary  choice  of
 systems of coordinates which are then by no  means  connected
 to the reference frame.

        Zel'manov's formalism of chronometric invariants is in fact
 the mo\-n\-ad formalism written in coordinates co-moving with the
 monad, {\em i.e.} such that $\tau^\mu={\delta^\mu_0}/\sqrt{g_{00}}$.
 This means that in this case the monad congruence coincides with
 the time coordinate lines. This coincidence does not change under
 arbitrary transformations of the time coordinate (so that this may
 arbitrarily mix with the spatial coordinates), but the new spatial
 coordinates should be (any) functions of the old {\em spatial}
 coordinates only. This special choice of $\tau$ substituted into
 all algebraic and differential relations of the monad formalism,
 yields the complete structure of Zel'\-ma\-nov's formalism of
 chro\-no\-me\-t\-r\-ic in\-va\-ri\-ants. The latter may seem to be
 intuitively
 more manageable than the monad formalism is, but in the reality
 relations involving chro\-no\-me\-t\-r\-ic in\-va\-ri\-ants
 are more complicated;
 moreover, in the monad formalism one may attain additional
 simplifications by choosing special systems of coordinates which
 are by no means constrained by an already fixed reference frame.
 This is why we shall not consider below the
 formalism of chro\-no\-me\-t\-r\-ic in\-va\-ri\-ants in more detail.

\pagestyle{myheadings}
\markboth{CHAPTER 2. REFERENCE FRAMES' CALCULUS}{2.3. GEOMETRY OF
CONGRUENCES}

 \section{Geometry of congruences. Acceleration, rotation,
 expansion and shear of a reference frame}

   As we have seen above, when we discussed in section 2.1
 the setting of reference frames formulation, the crucial role
 in their description has to be played by a study of congruences
 of  the  world  lines  of  particles  forming  bodies  of
 reference, {\em i.e.}  the  physical  time  lines  for  the  chosen
 reference frame. The congruence concept is essential  because
 for the sake of regularity of the mathematical description of
 the frame, these lines have not to  mutually  intersect,  and
 they  must  cover  completely  the  space-time  region  under
 consideration, so that at every world point one has  to  find
 one and only one line passing through it.  Exactly  the  same
 approach is used for  description  of  a  regular  continuous
 media, {\em i.e.}  in  hydrodynamics  of  a  compressible   fluid.
 Therefore it is worth using here the  standard  formalism  of
 hydrodynamics (relativistic: Lichnerowicz (1955), Taub (1978),
 non-relativistic: S. Goldstein (1976)). In a line
 with these ideas we introduce below the concepts of acceleration,
 rotation and rate-of-strain for a frame of reference.

   Differential operations on a manifold can (and for  our
 purposes have to) be projected onto the physical time
 direction of a frame of reference and its local  three-space  with
 help of the projectors $\tau$ and $b$, as it was the case for tensor
 quantities, in particular,  exterior  forms.  It  is  however
 important to keep in mind that when quantities lying  in  the
 three-space of a reference frame  are  being  differentiated,
 new terms arise which are non-orthogonal to the monad  field.
 Hence in order to obtain after such a differentiation  purely
 three-spatial quantities, one has to perform projections  not
 only in the differentiation indices, but also in those  which
 stem from the  very  quantities  being  differentiated:  in
 general, neither the three-projector, nor  the  monad  vector
 are constant relative to the differentiation  operations,  if
 the latters are not reformulated specially for this  end.  We
 mean here the operations with the properties of the  type  of
 (1.2.39), which are to be used for this  new  differentiation.
 Another operation which is evidently destined  for  taking  a
 derivative with respect to the physical time of  a  reference
 frame, is the Lie derivative with respect to $\tau$; but  it  does
 not commute either with the covariant  components  of  $\tau$,  or
 with the three-projector (both covariant  or  contravariant).
 Moreover, the Lie derivative, to a greater  extent  than  the
 nabla operator, depends on the choice of norm of  the  vector
 with respect to which the  differentiation  is  performed  (a
 scalar cannot be brought from this vector since it enters the
 Lie derivative under the differentiation sign, see  (1.2.16).
 We  have  made  use  of  this  "deficiency"  when  discussing
 hypersurface geometry in (Mitskievich, Yefremov and  Nesterov
 1985) --- the integrating factor method works here in the  case
 of normal congruences only.

   The formalism we treat below, is well fitted  not  only
 for the reference frames description, but  also,  to  a  full
 extent, in studies of relativistic hydrodynamics.
   In an analogy with four-dimensional  hydrodynamics,  we
 may write the covariant gradient of monad vector as
$$
\tau_{\mu;\nu}=\tau_\nu G_\mu+A_{\nu\mu}+ D_{\nu\mu},   \eqno{(2.3.1)}
$$
 where $G$ is the {\em acceleration (co)vector} of reference frame, $A$
 the {\em angular velocity tensor} or {\em rotation}
 of the frame,  and $D$ the
 {\em rate-of-strain tensor} (the sum $A+D$ represents  the  purely
 spatial part of the gradient, which is further separated into
 symmetric and skew  parts,  $D$ and $A$). Let us consider  now  these
 quantities more closely.

   Acceleration is the most obvious of  these  constructions.
   It is  clear  that  it  should  be  equal  to  $\nabla_\tau\tau$;  a
 substitution of the expression (2.3.1) brings this to
$$
\nabla_\tau\tau={\tau^\lambda}_{;\mu}\tau^\mu\partial_\lambda=G.
    \eqno{(2.3.2)}
$$
 Since the norm of  $\tau$  is  equal  to unity  by  a  definition,
 $\tau^\mu\tau_{\mu;\nu}\equiv0$,  and  all  other  terms
 lie  completely  in  the
 three-space of the reference frame. From  the  definition  of
 Lie derivative and (2.3.1), we know that
$$
\pounds_\tau\tau_\mu=\tau_{\mu;\nu}\tau^\nu+
{\tau^\nu}_{;\mu}\tau_\nu=G_\mu     \eqno{(2.3.3)}
$$
 which yields
$$
\pounds_\tau b_{\mu\nu}=\pounds_\tau g_{\mu\nu}-\tau_\mu G_\nu-
\tau_\nu G_\mu=2D_{\mu\nu}   \eqno{(2.3.4)}
$$
 (see also (2.3.1)). This conforms to the understanding of the
 rate-of-strain tensor as a measure of  evolving  the  spatial
 scales, these being determined by the three-metric $b$. Finally,
 rotation of the $\tau$-congruence is determined as a covector
$$
\omega:=\frac{1}{2}\ast(\tau\wedge d\tau)=\ast(\tau\wedge A), ~ ~
A:=\frac{1}{2}A_{\mu\nu}dx^{\mu\nu}.   \eqno{(2.3.5)}
$$
 (from here one-to-one correspondence is obvious  between  the
 bivector $A$ and axial (co)vector $\omega$, the both bearing
 one  and  the
 same information about rotation of the reference frame). Hence
$$
A=-\ast(\tau\wedge\omega), ~ ~ G=-\ast(\tau\wedge\ast d\tau).
    \eqno{(2.3.6)}
$$
   The rate-of-strain tensor (2.3.4) splits naturally into
 its trace (the scalar of expansion, or dilatation)
$$
\Theta:=\frac{1}{2}{\tau^\alpha}_{;\alpha}=\frac{1}{2}D^\alpha_\alpha
\equiv\frac{1}{2}D_{\mu\nu}b^{\mu\nu}  \eqno{(2.3.7)}
$$
 and the traceless part (the shear tensor),
$$
\sigma_{\mu\nu}:=D_{\mu\nu}-\frac{2}{3}\Theta b_{\mu\nu},
     \eqno{(2.3.8)}
$$
 whose square
$$
\sigma^2\equiv\sigma_{\mu\nu}\sigma^{\mu\nu}\equiv\sigma_{\mu\nu}D^{\mu\nu}
=D_{\mu\nu}D^{\mu\nu}-\frac{4}{3}\Theta^2  \eqno{(2.3.9)}
$$
 invariantly characterizes the presence of shear by  evolution
 of the three-space of a reference frame. These concepts  were
 borrowed by the reference frames  theory  from  hydrodynamics
 where they play an important role.  They  all  are  important
 also in theory of null (light-like) congruences often used in
 the classification of gravitational fields by principal  null
 directions  (the  Petrov  types)  and  generation  of   exact
 Einstein--Maxwell solutions (see Kramer, Stephani, MacCallum
 and Herlt (1980)). When a
 null complex (Newman--Penrose) tetrad is considered,  some  of
 the above relations change their form.

   Consider now exterior differential of the monad covector,
$$
d\tau=\tau\wedge G+2A   \eqno{(2.3.10)}
$$
 which can be rewritten as
$$
d\tau=-G\wedge\tau+2\ast(\omega\wedge\tau)   \eqno{(2.3.11)}
$$
 ({\em cf.} (3.1.16) for the electromagnetic field case). This relation
 is a part of Cartan's first structure equations adapted to
 a monad basis which can be supplemented (without leaving  the
 coordinated basis approach) by quantities similar to (2.2.17),
 a covector
$$
b^\mu=dx^\mu-\tau^\mu\tau,  \eqno{(2.3.12a)}
$$
and vector
$$
b_\mu=\partial_\mu-\tau_\mu\tau.   \eqno{(2.3.12a)}
$$
 One has to keep in mind that the free index $\mu$ appearing  in
 these expressions, does not contribute to the tensor  properties
 of the involved quantities from the point of view of the
 coordinate-free methods (the operations $\nabla$ and exterior
 differentiation), but it spoils the properties of  the  components
 of these quantities so that Christoffel  symbols  enter  such
 expressions, {\em e.g.}
$$
db^\mu=-2\tau^\mu A+[\tau^\mu G_\nu-({D_\nu}^\mu+{A_\nu}^\mu)
+\Gamma^\mu_{\lambda\nu}\tau^\lambda]b^\nu\wedge\tau  \eqno{(2.3.13)}
$$
 Here these symbols play role of scalar  functions  (from  the
 point of view of exterior forms formalism). One  can  however
 get rid of them by introduction of a differential operation
$$
\stackrel{(b)}{d}:=b^\alpha\nabla_b\wedge\equiv b^\alpha
\nabla_{\partial_\alpha}\wedge\equiv-\tau\nabla_\tau\wedge.
    \eqno{(2.3.14)}
$$
 Making use of the relation
$$
\tau\wedge\nabla_\tau b^\mu=(\Gamma^\mu_{\nu\lambda}\tau^\lambda b^\nu+
\tau^\mu G)\wedge\tau,   \eqno{(2.3.15)}
$$
 we obtain finally
$$
\stackrel{(b)}{d}b^\mu=-2\tau^\mu A-({D_\nu}^\mu+
{A_\nu}^\mu)b^\nu\wedge\tau   \eqno{(2.3.16)}
$$
    Here ({\em cf.} (2.3.5)) it is possible to write
$A=(1/2)A_{\mu\nu}b^\mu\wedge b^\nu$.
 The price one has to pay for this simplification, is
 incompleteness of the system of structure  equations  (2.3.10)  and
 (2.3.16), which however does not prevent obtaining from  them
 complete information about the quantities $G$, $D$ and $A$. Let  us
 simultaneously write down relations analogous to (2.3.2):
$$
\nabla_{b_\mu}\tau=({D_\mu}^\nu+{A_\mu}^\nu)b_\nu,  \eqno{(2.3.17)}
$$
$$
\nabla_\tau b_\mu=-\tau_\mu G-G_\mu\tau+\Gamma^\lambda_{\mu\nu}
\tau^\nu b_\lambda,    \eqno{(2.3.18)}
$$
$$
\nabla_{b_\mu}b_\nu=(\Gamma^\kappa_{\nu\lambda}b_\kappa-
\tau_\nu\tau^{\kappa}_{;\lambda}b_\kappa-\tau_{\nu;\lambda}\tau)
b^\lambda_\mu.   \eqno{(2.3.19)}
$$

\markboth{CHAPTER 2. REFERENCE FRAMES' CALCULUS}{2.4. DIFFERENTIAL
OPERATIONS IN MONAD FORMALISM}

\section{Differential  operations  and  iden\-ti\-ties  of  the  mo\-nad
 for\-ma\-lism}

   When a reference frame is not taken  into  account  and
 only the four-di\-men\-si\-o\-nal na\-ture of the universe matters, the
 role of curl is played by exterior derivative,  and  role  of
 divergence by the operation $\delta$ (1.2.43). With  help  of  these
 operations the four-dimensional  de-Rhamian  is  then  built
 which represents a generalization of the Laplacian
 (d'Alembertian) operator to the  theory  of  differential  forms.  In
 order to introduce analogous concepts in the scope  of  monad
 description of reference frames, it is sufficient  to  recall
 their definitions  in  three-dimensional  Euclidean  geometry
 where, {\em e.g.} curl is considered as a vector  product  in  which
 the first factor is the nabla operator. Here we shall base on
 the vector product expression (2.2.15), but the monad  vector
 $\tau$ will be not influenced by the exterior differentiation,
$$
{\rm curl}\stackrel{(3)}{a}=\ast(\tau\wedge d\stackrel{(3)}{a}).
         \eqno{(2.4.1)}
$$
 In this definition, curl is acting on  a  covector  which  is
 already lying in the  three-space  of  the  reference  frame.
 Remember that in the definition (2.2.15) it  was  unnecessary
 to consider any of the covectors taking part  in  the  vector
 multiplication, to be projected onto  the  three-space:  they
 were projected automatically by virtue of  (2.2.15).  But  in
 the case of curl this kind of definition leads  to  inclusion
 of an additional term compensating the part proportional to $\tau$
 of the covector being differentiated. It is easy to see  that
 then
$$
{\rm curl}~ a=\ast(\tau\wedge da)-2\stackrel{(\tau)}{a}\omega.
    \eqno{(2.4.2)}
$$
 It might seem that in this expression the overall sign should
 be chosen in the opposite sense ({\em cf.}  Mitskievich,  Yefremov
 and Nesterov (1985)), but the expression (2.4.2) is  perfectly
 correct: as a matter of fact, the  Euclidean  expression  for
 curl being a contravariant vector, it contains the
 differentiation operation (nabla) which is treated  there  as  if  it
 yielded directly a contravariant index.  It  is  however  the
 simplest to compare the curl components (2.4.1) brought to
 the contravariant level, and its components in  the  standard
 Euclidean vector calculus.

   Let us turn now to the three-divergence operation.  The
 natural way to define it, is to change in $\delta$,  (1.2.43),  the
 exterior differential $d$ to $\stackrel{(b)}{d}$ , (2.3.14). The latter  formula
 gives also the connection between  the  both  operations,  so that
$$
{\rm div}\stackrel{(3)}{a}:=-\ast\stackrel{(b)}{d}\ast\stackrel{(3)}{a}
=\delta\stackrel{(3)}{a}+\ast(\tau\wedge\ast\nabla_\tau\stackrel{(3)}{a})
       \eqno{(2.4.3)}
$$
 and finally
$$
{\rm div}\stackrel{(3)}{a}:=\delta\stackrel{(3)}{a}
-G\bullet\!\stackrel{(3)}{a}.     \eqno{(2.4.4)}
$$
 The extension of this three-divergence operation to covectors
 which have not been beforehand projected onto three-space  of
 the reference frame, is achieved by  a  compensation  of  the
 redundant terms, as in the curl case. The final expression of
 the  three-dimensional  divergence  operation  in  the  monad
 language (the point of view of a reference frame) is
$$
{\rm div}~ a:=\delta a-G\bullet a-2\stackrel{(\tau)}{a}\Theta
-\tau d\stackrel{(\tau)}{a}.   \eqno{(2.4.5)}
$$
   It is worth mentioning that when  the  reference  frame
 congruence rotates  (the  three-space  is  non-holonomic),  the
 second-order operation div curl does not vanish  identically,
 but it yields
$$
{\rm div~ curl}\stackrel{(3)}{a}=\omega\bullet\pounds_\tau
\stackrel{(3)}{a}.      \eqno{(2.4.6)}
$$
 This can be easily checked with help of relation (2.3.6)  and
 the definition of Lie  derivative. We leave to  the
 reader to compare this result with  the  identity  (13.9)  in
 Zel'manov and  Agakov  (1989);  our  next  task  is  here  to
 consider  other   important   differential   identities   for
 acceleration,  rotation,  and   rate-of-strain   tensor,   in
 particular those involving also curvature.

   The first of them is a scalar identity most easily  obtained
   when one applies operator $\delta$ to the rotation vector
   $\omega$:
$$
\delta\omega={\omega^\alpha}_{;\alpha}=2\omega\bullet G.
    \eqno{(2.4.7)}
$$
 This identity means that in the derivatives of $\omega$ which  enter
 $\delta\omega$, the second derivatives of monad $\tau$ cancel out.
 The same is
 true for the magnetic field vector $B$ which is  a part of $dA$
 ($A$ being the electromagnetic four-potential  covector),  and
 the second derivatives of $A$, {\em cf.} (4.3.11).

   Consider now curl applied  to  the  acceleration  of  a
 reference frame. The resulting covector  should  lie  in  the
 three-space, hence all terms proportional to $\tau$  must  vanish.
 But first we have
$$
{\rm curl}~ G=-2\ast[G\wedge\ast(\tau\wedge\omega)]-
\ast\d(\tau\wedge G)    \eqno{(2.4.8)}
$$
 where the first term is collinear to $\tau$, so that  it  will  be
 exactly compensated, while the last term should be rearranged
 in such a way that it would not lead  to  a  vicious  circle.
 To this end  we  use  here  relations $d(\tau\wedge G)=-2dA$ (see
 (2.3.10)),
 $$
 \pounds_\tau\omega=(\omega_{\beta;\alpha}\tau^\alpha+
 \omega_\alpha{\tau^\alpha}_{;\beta})dx^\beta
 $$
 and
 $$
 \tau_{\alpha;\beta}+\tau_{\beta;\alpha}=G_\alpha\tau_\beta+
 G_\beta\tau_\alpha+2D_{\alpha\beta},
 $$
 from where
$$
\frac{1}{2}{\rm curl}~ G=-({\omega^\alpha}_{;\alpha}-2G\bullet\omega)\tau
+\pounds_\tau\omega-2\omega^\alpha D_{\alpha\beta}dx^\beta+
2\Theta\omega.    \eqno{(2.4.9)}
$$
 Here the first right-hand side term (the parentheses)  vanishes
 by virtue of (2.4.7), as we have expected.

   In order to find  identities  involving  curvature,  we already
 introduced alongside  with  the  FW-curvature  (1.2.33),  also
 $\tau$-curvature making use of the operator (1.2.38) having remarkably
 better properties that those of (1.2.32). Thus
$$
\stackrel{[\tau]}{{\sf R}}(w,v):=\nabla^\tau_w\nabla^\tau_v -
\nabla^\tau_v\nabla^\tau_w-\nabla^\tau_{[w,v]}.   \eqno{(2.4.10)}
$$
 This $\tau$-curvature has two pairs of skew indices, but there  is
 no analogue of Ricci identities for it, so that  one  has  to
 take a linear combination  of  its  components  in  order  to
 obtain a three-space tensor with properties of  the  standard
 Riemann--Christoffel  tensor  ({\em cf.} a different approach in
 Zel'manov and Agakov (1989)):
$$
r_{\alpha\beta\gamma\delta}=
\frac{1}{6} (2\stackrel{[\tau]}{R}_{\alpha\beta\gamma\delta}
+2\stackrel{[\tau]}{R}_{\gamma\delta\alpha\beta}+
\stackrel{[\tau]}{R}_{\gamma\beta\alpha\delta}
+\stackrel{[\tau]}{R}_{\alpha\delta\gamma\beta}
+\stackrel{[\tau]}{R}_{\alpha\gamma\beta\delta}
+\stackrel{[\tau]}{R}_{\beta\delta\alpha\gamma}).
$$
 Such  a  combination  automatically  satisfies   Ricci   type
 iden\-ti\-ties if the build\-ing-ma\-te\-rial object (here,
 $\stackrel{[\tau]}{R}$) is skew
 in its first and last pairs of indices; this antisymmetry  is
 also inherited by $r_{\alpha\beta\gamma\delta}$,
 yielding therefore symmetry under a
 collective   permutation  of $[\alpha\beta]$ and $[\gamma\delta]$,
 so   that $r_{\beta\gamma}:=r_{\alpha\beta\gamma\delta}b^{\alpha\delta}$
 (of the type of Ricci tensor) becomes symmetric.

   Since the operator (2.4.10) identically  annuls  $\tau$,  we
 apply it to the basis vector $b_\lambda$ closely related  to  the
 monad. In contrast to $\tau$, we have here not a pure zero, but  a
 non-trivial expression because of the ``scalar'' nature of the
 free index of $b_\lambda$ with respect to coordinate-free  operations.
 Hence
\newpage
$$
 \stackrel{[\tau]}{{\sf R}}(b_\mu,b_\nu)b_\lambda=
 [{R^\alpha}_{\beta\gamma\delta}b^\kappa_\alpha
 b^\beta_\lambda b^\gamma_\mu b^\delta_\nu+
 ({D_\mu}^\kappa+{A_\mu}^\kappa)(D_{\nu\lambda}+A_{\nu\lambda})
$$
$$
 -(D_{\mu\lambda}+A_{\mu\lambda})({D_\nu}^\kappa+{A_\nu}^\kappa)]b_\kappa.
     \eqno{(2.4.11)}
$$
 For the same reasons, the identity connected with  $\tau$  can  be
 represented as
$$
b^\mu\cdot[{\sf R}(b_\lambda,b_\mu)\tau]\equiv
R_{\beta\gamma}\tau^\beta b^\gamma_\lambda=2\Theta_{,\beta}b^\beta_\lambda
-({D_\beta}^\mu+{A_\beta}^\mu)_{;\alpha}b^\alpha_\mu b^\beta_\lambda
-2G_\beta A_{\beta\lambda}.   \eqno{(2.4.12)}
$$
 We  call  identities  (2.4.11)  the  old  generalized   Gauss
 equations and identities  (2.4.12)  the  generalized  Codazzi
 equations (see Mitskievich, Yefremov  and  Nesterov  (1985);
 for the standard theory of hypersurfaces which is not based
 on the reference frame approach, see Eisenhart (1926, 1972));
 the generalization is  meant  in  the  sense  that  they  are
 written in non-holonomic local three-submanifolds of rotating
 reference frames. These equations turn into the  usual  Gauss
 and Codazzi equations correspondingly when $\omega=0$ ({\em i.e.},
 for a non-rotating congruence we come to the usual theory of
 hypersurfaces). Since $\stackrel{[\tau]}{R}$ does not possess
 the standard properties
 of curvature tensor, it is however  necessary  to  make  the
 next step of  generalization  using $r_{\alpha\beta\gamma\delta}$.
 Then  the  new
 generalized Gauss equations for the three-curvature read
$$
r_{\kappa\lambda\mu\nu}=R_{\alpha\beta\gamma\delta}b^\kappa_\alpha
b^\beta_\lambda b^\gamma_\mu b^\delta_\nu+D_{\kappa\mu}D_{\lambda\nu}
-D_{\kappa\nu}D_{\lambda\mu}+A_{\kappa\lambda}A_{\mu\nu},
   \eqno{(2.4.13)}
$$
 so that for the corresponding Ricci (three-)tensor we have
$$
r_{\lambda\mu}\equiv r_{\kappa\lambda\mu\nu}b^{\kappa\nu}=
R_{\alpha\beta\gamma\delta}
b^\beta_\lambda b^\gamma_\mu b^{\alpha\delta}+D_\mu^\nu D_{\lambda\nu}
-D_\nu^\nu D_{\lambda\mu}+A_\lambda^\nu A_{\mu\nu}
$$
$$
=R_{\alpha\beta\gamma\delta}
b^\beta_\lambda b^\gamma_\mu b^{\alpha\delta}+D_\mu^\nu D_{\lambda\nu}
-D_\nu^\nu D_{\lambda\mu}-\omega_\lambda\omega_\mu -
\omega\bullet\omega ~ b_{\lambda\mu},      \eqno{(2.4.14)}
$$
 where we have used simple identities ~ $A_{\kappa\mu}A_{\lambda\nu}
 -A_{\kappa\nu}A_{\lambda\mu}\equiv A_{\mu\nu}A_{\kappa\lambda}$ ~ and ~
${A_\lambda}^\nu A_{\mu\nu}\equiv \omega_\lambda\omega_\mu
+\omega\bullet\omega ~ b_{\lambda\mu}$.

   The identities we have just obtained play a fundamental
 role in the 3+1-splitting of Einsteins equations and in
 formulation of the Cauchy problem in general relativity; they are
 also important for a comparison of the latters with Maxwell's
 equations which we  discuss  in  section  5.5.  It  is  worth
 mentioning here another non-trivial fact: The definitions and
 identities given above imply that in the  case  of  a  normal
 congruence (no rotation) there exists an integrating  factor,
 {\em i.e.} such a function $N$ that $d(N^{-1}\tau)= 0$.
 In  a  region  with
 simple topological properties, the 1-form $\zeta=N^{-1}\tau$ is  not
 only closed but even an exact one, so that $\zeta=dt$.  If  a
 quantity $\xi=N\tau$ is  simultaneously  introduced  (which
 usually appears as a contravariant vector), a simple reciprocal
 norm is realized, $\zeta\cdot\xi=1$. Moreover, alongside  with  the
 trivial identity $\pounds_\xi\xi^\mu\equiv0$, a new
 important  identity  arises,
 $\pounds_\xi\zeta_\mu= 0$. The projecting tensor
 (three-metric) possesses  now
 the components $b^\alpha_\beta=\delta^\alpha_\beta-\xi^\alpha\zeta_\beta$
 and it is therefore constant
 relative to the Lie differentiation with respect to the vector
 field $\xi$: $\pounds_\xi b^\alpha_\beta\equiv 0$.

 It is worth mentioning that similar  ideas
 (although in spaces of other variables  than  the  space-time
 ones) can be successfully formulated using Cartan's  exterior
 forms, also in thermodynamics (Torres del Castillo 1992)

\newpage

\chapter{Equations of motion of test particles}         

\pagestyle{myheadings}
\markboth{CHAPTER 3. EQUATIONS OF MOTION}{3.1. ELECTRIC AND MAGNETIC
FIELD VECTORS}

\section{The electric field strength and
magnetic displacement vectors}  

        When the geodesic equation (1.3.29) is written in form
$$
{u^\mu}_{;\alpha}u^\alpha=0,     \eqno{(3.1.1)}
$$
its left-hand side is interpreted as an absolute derivative of the
four-velocity with respect to the  proper  time  of  the  particle
(partial derivative in special relativity if Cartesian  coordinates
are used). For a charged test particle the equation of motion
can be written then in a general covariant form as
$$
m{u^\mu}_{;\alpha}u^\alpha=eF^{\mu\alpha}u_\alpha   \eqno{(3.1.2)}
$$
({\em cf.}, {\em e.g.}, Landau and Lifshitz  (1971)).
A  comparison  of  this
equation with the forms of geodesic equation (3.1.1) and  (1.3.30)
equivalent to each other, shows that the 3-form equation,
$$
u\wedge\ast d(mu+eA)=0,         \eqno{(3.1.3a)}
$$
or its dual conjugate (1-form equation),
$$
\ast(u\wedge\ast d(mu+eA))=0,         \eqno{(3.1.3b)}
$$
are equivalent to (3.1.2). Here
$$
A:=a_\mu dx^\mu         \eqno{(3.1.4)}
$$
is 1-form of the electromagnetic four-potential, which should  not
be confused with the reference frame rotation  tensor  (2-form) $A$
(2.3.5); anyhow, we shall use more frequently the rotation  vector
$\omega$ (see also (2.3.5)). 2-form of the electromagnetic field strength
is then
$$
F:=dA=\frac{1}{2}F_{\mu\nu}dx^{\mu\nu}, \eqno{(3.1.5)}
$$
while $m$ and $e$ in (3.1.3) are the rest mass and electric charge  of
the test particle correspondingly, $u$ being its four-velocity.

        Thus we come to a conclusion that the complete expression for
the Lorentz force in the right-hand side of (3.1.2), can be represented as
$$
{\cal F}={\cal F}_\mu dx^\mu=eF_{\mu\alpha}u^\alpha dx^\mu
=e\ast(u\wedge\ast dA).  \eqno{(3.1.6)}
$$

Consider now a question: how could one approach logically  to  the
introduction of the concept of strength of a  physical  field?  As
one of the most simple cases studied in the theory  of  relativity
in great detail, electrodynamics represents an ideal example of  a
formulation of such an approach. On its basis, one can insist that
for a field acting on a corresponding test charge in some state of
motion with a certain force, the  field  strength  is  a  quantity
which enters the expression of the force ({\em pro} unit  charge  value)
and which does not include characteristics of the  motion  of  the
test charge. Thus the field strength describes a field existing in
space-time independently of the charges used for its detection and
measurement. In electrodynamics, this is the very tensor $F$;  in  a
more complicated case of gravitational field, one has to introduce
a concept of the {\em relative field strength},  since  in  the geodesic
equation  there is no gravitational force term whatsoever, so that
it is necessary to consider  the geodesic deviation equation  (see
(5.1.1)). Then the tensor rank of the relative strength becomes  4
and not 2, as it was the case for the field strength  in  electrodynamics.

        We know however from the special relativistic electrodynamics
the connection existing between the four- and
three-dimensional Lorentz force. This connection can be  expressed  from
the point of view of the reference frame $\tau$ as
$$
\stackrel{(3)}{\cal F}=\stackrel{(\tau)}{u}e(E+v\times B),
\eqno{(3.1.7a)}
$$
while
$$
\stackrel{(\tau)}{\cal F}=\stackrel{(\tau)}{u}eE\bullet v
\eqno{(3.1.7b)}
$$
({\em cf}. the decomposition  (2.2.5)  and  the  three-projection  of  a
vector, (2.2.4), as well as definition of the three-velocity of  a
particle, (2.2.11)). We rewrite now the first of  these  relations
using the definition (2.2.15) of the vector product,
$$
\stackrel{(\tau)}{u}E+\ast(u\wedge\tau\wedge B)=\ast(\tau\wedge\ast F)
+\ast(v\wedge\ast F)-[\tau\cdot\ast(v\wedge\ast F)]\tau,
 \eqno{(3.1.8)}
$$
where the right-hand side is also projected onto  the  three-space
orthogonal to $\tau$-congruence. This expression can also be written as
$$
E+v\times B=F_{\alpha\beta}(\tau^\beta+v^\beta)b^\alpha.
  \eqno{(3.1.9)}
$$
This relation holds for particles moving arbitrarily with  respect
to the reference frame, {\em i.e.} for any vector $v$ with  a norm  less
than unity. Then one can write immediately (Mitskievich and  Kalev
1975)
$$
E=F_{\alpha\beta}\tau^\beta dx^\alpha   \eqno{(3.1.10)}
$$
or equivalently
$$
E=\ast(\tau\wedge\ast F)        \eqno{(3.1.11)}
$$
({\em cf}. (2.3.6)). From the arbitrariness  of $v$ the  remaining  part
yields
$$
E_{\kappa\lambda\mu\nu}\tau^\kappa B^\lambda b^\nu=
F_{\nu\lambda}b^\lambda_\mu b^\nu,      \eqno{(3.1.12)}
$$
or better
$$
E_{\kappa\lambda\mu\nu}\tau^\kappa B^\lambda b^\mu\wedge b^\nu=
-F_{\mu\nu}b^\mu\wedge b^\nu.  \eqno{(3.1.13)}
$$
Here the left-hand side is equal to $2\ast(\tau\wedge B)$.
But since $\ast(\tau\wedge\ast(\tau\wedge B))\equiv-B$,
we have after exterior multiplication of (3.1.13) by $\tau$ from
the left and a subsequent dual conjugation,
$$
B=\ast(\tau\wedge F),   \eqno{(3.1.14)}
$$
an analogue of the expression (3.1.11) ({\em cf}.  (2.3.5)).  Then  as  an
analogue of (3.1.10), we have  expression  with  the  opposite  sign
(Mitskievich and Kalev 1975):
$$
B=-F^{\;\ast}_{\alpha\beta}\tau^\beta dx^\alpha.    \eqno{(3.1.15)}
$$
Thus we have obtained very simple and applicable for all reference
frames and arbitrary gravitational fields definitions of the electric
field strength and magnetic displacement vectors, $E$ and $B$. We
remind that they follow unambiguously from the standard  form  for
the Lorentz force, {\em i.e.} from  the  electric  and  magnetic  fields
interpretation in the spirit of traditional theory.
The  definitions (3.1.11) and (3.1.15) yield automatically a decomposition
$$
dA=F=E\wedge\tau+\ast(B\wedge\tau),     \eqno{(3.1.16)}
$$
and vice versa ({\em cf.} (2.3.11)).

        In order to illustrate the simplicity of application of these
definitions to concrete problems, we calculate now the electromagnetic
field invariants through the three-space  vectors $E$  and $B$
using the contracted crafty identities  (1.2.6)  and  (1.2.7).  To
this end let us multiply each of the identities by
$\tau_\lambda\tau^\mu$ and make
use of (3.1.10) and (3.1.15):
$$
F_{\sigma\tau}F^{\sigma\tau}=2(B\bullet B-E\bullet E),
 \eqno{(3.1.17)}
$$
$$
F^{\;\ast}_{\sigma\tau}F^{\sigma\tau}=4E\bullet B,
 \eqno{(3.1.18)}
$$
where the definition of three-scalar product $\bullet$ (2.2.6),
(2.2.8) is also taken into account.

        Let us build, with help  of  (3.1.16),  self-dual  quantities
while first writing the expression dual conjugate to (3.1.16),
$$
\ast F=-B\wedge\tau+\ast(E\wedge\tau).  \eqno{(3.1.19)}
$$
This relation  as well as its counterpart (3.1.16),  can  be  more
easily related to the component representation of the electric and
magnetic fields (3.1.10) and (3.1.15)  if  one  takes  account  of
identities (2.2.20)  and (2.2.22).

        We denote self-dual and anti-self-dual quantities using signs
$+$ or $-$, correspondingly, over the root letter:
$$
\stackrel{\pm}{F}:=F\mp i\ast F=\stackrel{\pm}{E}\wedge\tau
\mp i\ast\left(\stackrel{\pm}{E}\wedge\tau\right). \eqno{(3.1.20)}
$$
It is easy to see that
$$
\stackrel{\pm}{E}:=E\pm iB.  \eqno{(3.1.21)}
$$
The quantities $\stackrel{\pm}{F}$ possess the properties
$$
\ast\stackrel{\pm}{F}=\pm i\stackrel{\pm}{F}.   \eqno{(3.1.22)}
$$
Then the contracted crafty identities yield
$$
\stackrel{\pm ~}{F_{\mu\nu}}\stackrel{\pm ~}{F^{\mu\nu}}=
-4i\stackrel{\pm}{E}\bullet\stackrel{\pm}{E}, ~ ~
\stackrel{\pm ~}{F_{\mu\nu}}\stackrel{\mp ~}{F^{\mu\nu}}\equiv0,
 \eqno{(3.1.23)}
$$
the both relations following from (1.2.6) as well as from  (1.2.7)
since these mutually coincide in this case  due  to  the  equality
(3.1.22). We see that an (anti-)self-dual electromagnetic field is
equivalent to a simultaneous existence of the  both  electric  and
magnetic fields, though differing one from the other by a  complex
factor $(\mp i)$, so that if for the initial 2-form $F$ the correspondence
$F\Leftrightarrow(E,B)$ holds, for $\stackrel{\pm}{F}$ we have
$$
\stackrel{\pm}{F} ~ \Leftrightarrow\left(\stackrel{\pm}{E}, ~ ~
\mp i\stackrel{\pm}{E}\right).  \eqno{(3.1.24)}
$$

\markboth{CHAPTER 3. EQUATIONS OF MOTION}{3.2. THE ENERGY AND MOMENTUM
OF PARTICLES}

\section{The energy and momentum of particles}   

        We consider in this section general properties of the
energy-momentum four-vector of mass points from the viewpoint of the theory
of reference frames, as well as general problems of description
of media consisting of such points which mutually interact  {\em via}
pressure and gravitational field.

        The general covariant energy-momentum vector of a mass  point
is decomposed into the energy scalar and momentum three-vector  in
a standard way (Mitskievich, Yefremov and Nesterov 1985)
$$
p=mu={\cal E}\tau+{\cal P}, ~ ~ {\cal P}\cdot\tau=0,    \eqno{(3.2.1)}
$$
while
$$
{\cal E}:=p\cdot\tau=\stackrel{(\tau)}{u}m, ~ ~
{\cal P}:=\stackrel{3}{p}={\cal E}v,    \eqno{(3.2.2)}
$$
so that
$$
p\cdot p={\cal E}^2-{\cal P}\bullet{\cal P}=m^2, ~ ~ {\cal E}\geq m\geq0.
        \eqno{(3.2.3)}
$$
The scalar mass of a particle, often called its rest mass, is constant
for elementary particles only (till a  decay;  moreover,  if
the lifetime of a particle is short, the constancy of $m$ or universality
of this quantity for different particles is limited by  the
uncertainty relation for time and energy. In the case of a
macroscopic particle or a body, one cannot demand that  its  rest  mass
should be constant ({\em e.g.}, for a radiating star  or  a  body  which
throws away or evaporates its parts). Note  that  in  general  the
idea of a fixed rest mass is alien to classical (non-quantum) physics,
and it comes from observation of stable elementary particles
for which this constancy has an exclusively quantum nature.

        If a medium consisting of  particles  which  form  a  perfect
(possibly, electrically charged) fluid, is considered, this can be
characterized by the energy-momentum (or: stress-energy) tensor,
$$
T^{\rm pf}=(\mu+p)u\otimes u-pg,        \eqno{(3.2.4)}
$$
and the four-current density,
$$
j=\rho u        \eqno{(3.2.5)}
$$
(not a vector density, but a vector!). Here $\mu$ and $\rho$ are  invariant
(taken in the co-moving reference frame of the  medium)  densities
of the energy (mass) and charge of the fluid, and $p$ is its pressure
(not to be confused with the four-momentum in (3.2.1):  we  did
not change the somewhat contradictory standard notations which  in
fact do not meet together in a close context). In such a co-moving
frame the energy-momentum tensor takes the form
$$
T^{\rm pf}=\mu u\otimes u-pw,   \eqno{(3.2.6)}
$$
which justifies the choice of structure in (3.2.4); here the
four-velocity $u$ of the fluid plays the role of monad,
and $w=g-u\otimes u$
is  the corresponding three-metric  tensor  (signature of $w$ is
$(0,-1,-1,-1)$). Then in an arbitrary reference frame the
energy-momentum tensor of a perfect fluid is
$$
T^{\rm pf}=\left[\stackrel{(\tau)}{u}^2\mu+
\left(\stackrel{(\tau)}{u}^2-1\right)p\right]\tau\otimes\tau+
$$
$$
+\stackrel{(\tau)}{u}^2(\mu+p)(\tau\otimes v+v\otimes\tau+
v\otimes v)-p\; b     \eqno{(3.2.7)}
$$
where scalar coefficient in the first term represents  the  energy
(mass) density in the reference frame $\tau$, the  coefficient  in  the
second term expresses the proportionality of density of the energy
flow (the three-momentum density) of the fluid, to the three-velocity
of the fluid in this reference frame; the last term describes
the isotropic part of pressure (and stresses), while the  next  to
the last one, characterizes its anisotropic part which is  due  to
motion of the fluid with respect to the  reference  frame.  It  is
interesting that invariant quantities, the energy density and the
pressure, taken in the co-moving frame of the fluid, both  contribute
to the energy density, its flow density, and anisotropic part
of the pressure, taken in an arbitrary reference frame, while  the
fluid undoubtedly is a Pascal  one.  By  evaluating  the  involved
quantities one has to take into account the fact that the  relativistic
factor $\stackrel{(\tau)}{u}$ is always not smaller than unity
$\left(\stackrel{(\tau)}{u}=dt/ds=(1-v^2)^{-1/2}\geq1\right)$.
Expression for the non-invariant densities of
the charge and three-current are much simpler,
$$
j=\stackrel{(\tau)}{u}\rho(\tau+v).     \eqno{(3.2.8)}
$$
It is worth emphasizing that the expressions "invariant" and
"noninvariant" are both used for scalar quantities  depending  not  on
the choice of a system of coordinates, but  on  the  choice  of  a
reference  frame  only,  while  speaking  on  "three-current"  and
"three-velocity", etc, we mean in  fact  four-vectors,  which  are
lying in the local submanifold orthogonal to the monad $\tau$. One  has
to remember that no transformations of coordinates  can  influence
the choice of reference frame, while in one and the same system of
coordinates, there could be considered simultaneously as many
different reference frames as one would desire. Exactly in
this sense
the  energy  density  (even  if  non-invariant  under  transitions
between different choices of $\tau$) should always be  a  scalar  under
arbitrary four-dimensional coordinates transformations, though  it
might describe energy density of the fluid particles  not  in  the
state of rest. We speak here in such a detail on these  elementary
facts because they are in an acute contradiction with  the  vulgar
and wide-spread understanding of  reference frames as  exclusively
systems of coordinates.

        Now let us touch upon the problem of  dynamics  of  a  fluid,
which is expressed by vanishing  of  the  four-divergence  of  its
energy-momentum tensor, ${T^{{\rm pf}\mu\nu}}_{;\nu}=0$.
This means, of  course,  that
we consider now an electrically neutral fluid, otherwise it  would
exchange energy-momentum with the electromagnetic field,  so  that
we had to take the latter into account. Moreover, we shall  consider
here the fluid's behaviour only in  a  co-moving  frame, $u=\tau$.  From
(3.2.6) we have
$$
  \mu_{,\alpha}\tau^\alpha\equiv\pounds_\tau\mu=-2(\mu+p)\Theta,
  \eqno{(3.2.10)}
$$
$$
p_{,\alpha}b^\alpha_\beta\equiv
\left(\stackrel{(b)}{\rm grad}p\right)_\beta=(\mu+p)G_\beta.
        \eqno{(3.2.11)}                                          
$$
We see that a non-homogeneity of the  pressure  produces  a  force
acting on particles of the fluid and thus making their motion
non-geodesic. But if the pressure  is  homogeneous  in  the  co-moving
frame of the fluid, its particles do move geodesically ({\em cf.} (Synge
1960)), as it is the case  for  homogeneous  cosmological  models,
otherwise the sum of the pressure and  energy  density  should  be
equal to zero. We shall not give here any  further  discussion  of
these problems, and remark only that the  last  of  the  mentioned
possibilities is closely connected with the properties of  the  de
Sitter universe.

\markboth{CHAPTER 3. EQUATIONS OF MOTION}{3.3. MONAD DESCRIPTION OF
THE MOTION}

\section{Monad description of the motion of a test charged mass in
gravitational and electromagnetic fields}        

        We consider here from the viewpoint of a reference frame $\tau$
the splitting of eq. (3.1.4) which describes the motion of a  test
charged mass point. Its non-gravitational part (3.1.5) has already
met a comprehensive treatment in section 3.1 from where  we  shall
borrow the corresponding expressions; now it remains  to  consider
splitting of the term
$$
\ast(u\wedge\ast dp)=\stackrel{(\tau)}{u}\ast[(\tau+v)\wedge\ast
d({\cal E}\tau+{\cal P})]       \eqno{(3.3.1)}
$$
in which it was suitable to pass to the dynamical quantities,  the
energy ${\cal E}$ and three-momentum ${\cal P}$ characterizing
the particle and already introduced in section 3.2.

        It is easy to see that the four terms into which  splits  the
right-hand side of (3.3.1), read
$$
\ast (\tau\wedge\ast d({\cal E}\tau))=d{\cal E}-
(\pounds_\tau{\cal E})\tau-{\cal E}G,      \eqno{(3.3.2)}
$$
$$
\ast(v\wedge\ast d({\cal E}\tau))=2{\cal E}~v\times\omega-
({\cal E}~v\bullet G-v\bullet d{\cal E})\tau,   \eqno{(3.3.3)}
$$
$$
\ast(\tau\wedge\ast d{\cal P})=-\pounds_\tau{\cal P}    \eqno{(3.3.4)}
$$
(this is true only for a three-vector orthogonal to $\tau$, here,
${\cal P}$), and
$$
\ast(v\wedge\ast d{\cal P})=-(v\bullet\pounds_\tau{\cal P})\tau
+v^\mu({\cal P}_{\mu,\lambda}-{\cal P}_{\lambda,\mu})b^\lambda.
 \eqno{(3.3.5)}
$$
Differentiation of the identity (3.2.3) for $m =$ constant, yields
$$
\pounds_\tau({\cal E}^2)=-\pounds_\tau(g^{\mu\nu}{\cal P}_\mu{\cal P}_\nu)
=-2{\cal P}^\mu\pounds_\tau{\cal P}_\mu-{\cal P}_\mu{\cal P}_\nu
\pounds g^{\mu\nu},     \eqno{(3.3.6)}
$$
where $\pounds_\tau g^{\mu\nu}=-(2D^{\mu\nu}+\tau^\mu G^\nu
+\tau^\nu G^\mu)$ and
$v^\mu\pounds_\tau{\cal P}_\mu=-v\bullet\pounds_\tau{\cal P}$,
hence
$$
\pounds_\tau{\cal E}=v\bullet\pounds_\tau{\cal P}+
{\cal E}D^{\mu\nu}v_\mu v_\nu.  \eqno{(3.3.7)}
$$
Note also that
$$
\pounds_\tau{\cal E}-v\bullet d{\cal E}=(m/{\cal E})\pounds_u{\cal E}.
 \eqno{(3.3.8)}
$$

        As a result we obtain the splitted equations of motion  of  a
particle having $m =$ constant: two equivalent forms for the  scalar
equation (the component along $\tau$),
$$
\pounds_\tau{\cal E}-v\bullet d{\cal E}={\cal E}v_\mu v_\nu D^{\mu\nu}
+v\bullet(eE-{\cal E}G) \eqno{(3.3.9)}
$$
and
$$
\pounds_u{\cal E}=({\cal E}/m)[{\cal E}v_\mu v_\nu D^{\mu\nu}
+v\bullet(eE-{\cal E}G)], \eqno{(3.3.10)}
$$
and the vector equation (projection onto the  three-space  of  the
reference frame),
$$
\pounds_\tau{\cal P}+\nabla^\tau_v{\cal P}=
e(E+v\times B)+{\cal E}(-G+2v\times\omega),     \eqno{(3.3.11)}
$$
{\em cf.} (Mitskievich, Yefremov and Nesterov 1985). In the  last  case,
we do not employ the Lie derivative with respect to the
four-velocity of the particle, since being applied to (co)vectors
(in  contrast to scalars),  this derivative brings the quantity out of the
three-dimensional subspace. In equation (3.3.11) the  differentiation
operation $\nabla^\tau_v$, (1.2.34), is used, which takes  the  reference
frame explicitly into account. The corresponding term can also  be
written as $\nabla^\tau_v{\cal P}={\cal P}_{\mu;\nu}v^\nu b^\mu$.

        In the resulting equations, a  far-reaching  analogy  between
gravitation and electromagnetism can  be  clearly  traced  in  the
aspect of the action of these fields on test particles.  Below  we
shall further discuss this analogy (see chapters 4 and 5).

        The equations of motion (3.3.9) to (3.3.11)  of  course  hold
also in special relativity when no gravitational field (curvature)
is present; then the terms containing $G$ and $\omega$ represent  the
non-inertial frame effects. The very last term  in  (3.3.11)  is  then
clearly the Coriolis force; the centrifugal force is hidden in the
next to the last term proportional to $G$, if
acceleration of the reference frame is expressed  through $\omega$
and radius vector in the case of a pure (rigid) rotation.
When gravitation is
present, a mixture of gravitational and  inertial  forces  arises,
these forces being unseparable one from the  other  (remember  the
equivalence of gravity and acceleration, usually considered  at  a
more na\"\i ve level).

\markboth{CHAPTER 3. EQUATIONS OF MOTION}{3.4. RED SHIFT AND DOPPLER
EFFECTS}

\section{Motion of photons, the red shift and Doppler effects}      

        Some concrete physical effects can be however calculated  and
described in such a way that they  are  expressed  through  scalar
quantities being functions of states of motion of the  participating
particles only, including, if  necessary,  quanta  which  move
with the fundamental speed. While the  massive  objects  mentioned
above may be used (if their world lines do not intersect) to build
a reference frame in which  these  participating  objects  are  at
rest, the massless quanta cannot of course be used in constructing
such  reference  frames,  since  null  lines  are  alien  to the
latters. Therefore one has not to unnecessarily  universalize  the
trend towards description of all physical effects using  reference
frames. When they enter naturally into  an  experiment  scheme  or
simplify essentially its description  or  calculations,  they  are
surely advisable, but a mutual adjustment of setting an experiment
and constructing a reference frame, would be of an appallingly low
standard, if this would be done for the end of realization of  the
idea of universal application of  reference  frames  only.  As  an
example of such a situation we consider here a rigorous description
of one of the most important relativistic effects,  that  of
frequency shift of a signal, which in fact unifies the gravitational
red shift and Doppler shift effects (in general,  their
unambiguous separation is impossible). Some examples of alternative
separation cases can be easily invented, but, if the space-time
under consideration is non-stationary, there exists  no  criterion
for choosing a preferential interpretation. Therefore we shall not
be captivated by such a scholastic attitude, and we shall consider
here an example of use of a local reference frame, up to a numerical
evaluation of the frequency shift effect. Such local reference
frames should be chosen individually for every  specific  problem,
and there exists no general formalism  of  their  application,  in
contrast with global reference frames for which the  monad  formalism
excellently works.

        Four decades ago, Schr\"odinger (1956) has  proposed
an invariant procedure for determination of  the  frequency  shift
effect (in the description of the  cosmological  red  shift),  but
this procedure remained practically unnoticed until  Brill  (1972)
has given in his short report a rigorous  deduction  of  Schr\"odinger's
formula using coordinates-free approach, the same as applied
in our paper. While giving here a deduction  of  the  Schr\"odinger--Brill
formula, we propose to the reader to draw by her/himself the
corresponding figure, so simple the idea is  and  so  advantageous
this would be for its best understanding. Let the emitter and  detector
of signals have time-like world lines 1 and  2  correspondingly,
while the signals propagate along null geodesics  from  the
emitter to detector (thus the light  cone has its
apex on emitter's world line, and the generatrix intersects
the detectors's world line). Then two signals sent from  the
emitter with an interval of the proper time between them, equal to
some standard period, reach the detector being divided by  another
interval of the proper time (now, along  detector's  world  line),
different from the initial standard period.  If  the  detector  is
provided with a frequency standard moving  together  with  it  and
reproducing the standard period, now from  the  viewpoint  of  the
proper time of the detector, it is possible to compare  these  two
time intervals, {\em i.e} to measure
the frequency shift by  the  transmission
of the signal from emitter to detector. This is  a  fundamental
principle of the (general as well  as  special)  relativity
theory that the period (or, of course, frequency) of signals  expressed
through the proper time,  is one and the same,  independent
of the choice of the world line of the generator of these  signals
on which the proper time is to be measured. By the way,  the  same
idea of relativity (and the equivalence principle),  makes  it  in
principle impossible to separate Doppler shift  and  gravitational
red sift effects. By definition, the red shift  is  expressed {\em via}
the formula
$$
z=(\nu_{\rm em}-\nu_{\rm det})/\nu_{\rm em}=1-ds_1/ds_2 \eqno{(3.4.1)}
$$
where for a computational convenience, we introduced infinitesimal
proper time intervals on the corresponding world lines, 1 for  the
emitter and 2 for detector, while $ds_2$ is the interval between  two
signals which reach the detector, having initially --- at their
departure moments on the world line of the emitter ---  some  standard
proper time interval $ds_1$  between them. It is clear that this  formula
simultaneously describes the  both  usual  gravitational  red
shift effect and the Doppler shift (due to motion of  the  emitter
as well as the detector). In general one cannot speak on a relative
velocity between detector and emitter,  if  the  space-time  is
curved: there will be no unambiguous way to compare the four-velocities
since the comparison would depend crucially on  the  choice
of the transportation path, so  that  only  in  the  case  of  the
Minkowski space-time (absence of any genuine gravitational  field)
such an absolute comparison of motions is possible: this is why we
are so much accustomed to the independent existence of the Doppler
shift in special relativity. Thus our task now is to determine the
connection between $ds_1$ and $ds_2$ for  one  and  the  same  pair  of
signals.

        Let us complement the time-like world lines 1 and 2 by
introducing some intermediary time-like lines (the concrete choice will
not affect the results) which, as the previous lines 1 and 2, have
not necessarily to be geodesics. Thus we have built  a  family  of
lines which may be parametrized holonomically when a  parameter $\sigma$
will vary along each of them (the other parameter, $\lambda$, is  introduced
a little later, and it enumerates the individual lines  continuously).
Then the tangent vector field of this family of lines is
$q=\partial_\sigma$. The other family of lines
will be that of the null  lines
of the signals propagating between the initial world lines 1 and 2.
These null lines are, of course, geodesics, with a canonical parameter
$\lambda$ and tangent vector field $k=\partial_\lambda$, so that
$$
\nabla_kk=0, ~ ~ ~ k\cdot k=0.  \eqno{(3.4.2)}
$$
Let the parameter $\lambda$ be holonomic too, thus the condition to have two
holonomic parameters (which may be used as coordinates) can be  expressed as
$$
\nabla_qk-\nabla_kq\equiv[q,k]=0        \eqno{(3.4.3)}
$$
({\em cf.} the sixth axiom of the covariant differentiation).  Note,  on
the one hand, that the scalar product $q\cdot k$ is constant under transport
along the null lines. In order to prove this fact, one has to differentiate
$q\cdot k$ with respect to the parameter $\lambda$:
$$
\partial_\lambda(q\cdot k)\equiv\nabla_k(q\cdot k)=k\cdot\nabla_kq
=k\cdot\nabla_qk\equiv\frac{1}{2}\nabla_q(k\cdot k)\equiv0.
        \eqno{(3.4.4)}
$$
On the other hand, this is not the vector field $q$ (introduced  for
the first family of lines) which is physically meaningful, but the
four-velocity vectors $u$ on the world lines 1 and 2, or equivalently,
the proper time intervals along these lines. Therefore  it  is
worth writing the conserved (by the  null  transport  between  the
lines 1 and 2) quantity as
$$
q\cdot k=(ds/d\sigma)u\cdot k.  \eqno{(3.4.5)}
$$
The holonomicity property of $\sigma,~\lambda$ means in particular that the
difference $d\sigma$ is one and the same on the both lines 1 and 2, if it is
taken for one  and  the  same  pair  of  signals;  then  evidently
$(u\cdot k~ ds)_1=(u\cdot k~ds)_2$. We have thus finally obtained  the
Schr\"odinger--Brill formula (Schr\"odinger 1956; Brill 1972; Mitskievich
and Nesterov 1991)
$$
z=1-(u\cdot k)_2/(u\cdot k)_1   \eqno{(3.4.6)}
$$
which holds both in special and general relativity  and  describes
Doppler frequency shift in the Minkowski world,  as  well  as  the
gravitational red shift effect in  the  Schwarzschild  space-time,
and the cosmological red shift in the Friedmann world, to  mention
a few of the principal cases.

        We now apply this general formula to two cases, the  cosmological
red shift and a free fall of the detector onto a black hole.
In the first case the metric is
$$
d s^2=a^2(\eta)[d\eta^2-d\chi^2-b^2(\chi)(d\theta^2+
\sin^2\theta d\phi^2)].  \eqno{(3.4.7)}
$$
It is most simple to use the coordinates co-moving with the matter
o universe (with which we also connect the both  emitter  and
detector of radiation). Let the detector rest in the spatial origin
($\chi_2=0$; this is justified by  homogeneity  of  the  Friedmann
universe), so that signals propagate radially. Their  equation  of
motion, $ds^2=0$, is integrated trivially: $\eta_2=\eta_1+\chi_1$.
Here $\eta_1$ and $\eta_2$ are cosmological times
of transmission and reception  of  a
signal, the difference between them being equal to  the  cosmological
coordinate "distance" of the  radiation  source  from  the  origin
where the observer resides (this $\chi_1$ is at the same time integration
constant in full accordance with the  use  of  a  co-moving
frame in which all the matter of the universe is  at  rest).  Then
$u = a^{-1}\partial_\eta$, $k=d\eta+d\chi$
(it is easy to check that $k$ forms a
geodesic field). A substitution into (3.4.6) yields $z=a_2/a_1$. If
expansion of $a(\eta_1)$ about the observation moment $\eta_2$ is performed
(with the corresponding assumption that the two moments are
sufficiently close one to the other), we
obtain $a_1=a_2-(da/d\eta)_2\chi_1$.
Since the physical distance between  the  observer  and  radiation
source is equal to $l=a\chi$,  this  formula  yields  the  standard
Hubble expression
$$
H=a^{-2}da/d\eta=a^{-1}da/dt    \eqno{(3.4.8)}
$$
(where transition to the physical time of the observer  is  also
performed according to the relation $ad\eta=dt$). In this  example
a system of coordinates (co-moving with the reference frame  whose
language is used in our discussion) was employed in which the both
radiation source and observer were continuously at rest, moving at
the same time geodesically. Nevertheless, the obtained  effect  is
treated usually as a Doppler shift. It is closely  connected  with
the velocity of galaxies' parting away (the expansion of the
universe); however, in our frame the matter is  everywhere  at  rest,
but the scales are changing (the rate-of-strain tensor is non-zero
for the reference frame under consideration). But it is clear that
there can be no physical sense in  arguing  into  one  or  another
interpretation of the nature of this red shift effect, the
gravitational or Doppler one.

        The second case is related to a Gedankenexperiment concerning
a situation when over a black hole, a mother  craft  is  hovering,
and a probe dives from it with zero initial velocity into a free
fall, while standard signals are continuously sent to it from  the
mother craft (Mitskievich and Nesterov 1991). The  problem  is  to
find out the frequency shift of  these  signals  received  by  the
probe. We consider the general case of a black hole when it possesses
a mass $M$, Kerr parameter $a$, and electric charge $Q$.  Determine
first of all the tangent vectors $u$ and $k$ in the Boyer--Lindquist
coordinates. The first integrals of geodesic equation in the Kerr--Newman
field are (see Mitskievich and Nesterov (1991))
$$
\Delta\Sigma\dot t=EA,
$$
$$
\Sigma^2\dot r^2=E^2(r^2+a^2)-\Delta(K+\eta r^2)=:F^2,
$$
$$
\Sigma^2\dot\theta^2=K-\eta a^2\cos^2\theta-E^2a^2\sin^2\theta,
$$
$$
\Delta\Sigma\dot\phi=aE(2Mr-Q^2)
$$
where $\Delta=r^2+a^2+Q^2-2Mr$, $A=(r^2+a^2)^2-\Delta a^2\sin^2\theta$,
$\Sigma=r^2+a^2\cos^2\theta$, $E$ and $K$ being the first
integrals of energy and the
total angular momentum (determined with help of the  Killing  tensor,
{\em cf.} (Marck 1983)),
$\eta=0$ or 1 depending on null or time-like lines are considered;
the speed of light and  Newtonian gravitational constant  are
assumed to be equal to unity. Let now $\dot\theta=0$, so that the motion is
radial up to such dragging effects (in $\phi$ direction)
which cannot be excluded at all radii simultaneously.

        Although the world line 1 is non-geodesic, we  choose  it  so
that it corresponds to all other motions, {\em i.e.}
$u_1=\dot t~\partial_t+\dot\phi~\partial_\phi$.
Here the expressions from above are taken for $\dot t$ and $\dot\phi$
while $\eta=1$,
while normalization of the four-velocity yields a  special  choice
of the constant $E$ (the Carter integral $K$ is  already  fixed  by  the
demand $\dot\theta=0$). The same value of $E$ is inherited
by the probe since the world line 1 gives in fact the initial data
of its motion. For the probe we have then
$u_2=\dot t~\partial_t+\dot r~\partial_r+ \dot\phi~\partial_\phi$, with
the same remarks which were made with respect to $u_1$.
Since the angle $\theta$ does
not change along the world lines, the coordinate $r$ is  essential
only (all quantities do not explicitly depend on $t$ and $\phi$), so that
it is not necessary to determine the  intersection points  of  the
lines under consideration. In the common (initial)  point  of  the
world lines 1 and 2 (only the latter is a geodesic) the both lines
have a common tangent vector. As to the null geodesics, their tangent
vectors
are $k=\dot t_0\partial_t+\dot r_0\partial_r+\dot\phi_0\partial_\phi$,
and  since  all
components are proportional to one and the same constant $E_0$, this
can be equalized without loss of generality to the already used $E$.
Hence $\dot t_0=\dot t, ~ \dot\phi_0=\dot\phi, ~ \dot\theta_0=\dot\theta=0$,
and the only remaining non-trivial relation is
$$
\Sigma^2\dot r_0^2=E^2[(r^2+a^2)^2-\Delta a^2\sin^2\theta].
$$
Thus $\Sigma^2\dot r^2_0=\Sigma^2\dot r^2+\Delta\Sigma$,
and $k=u_2+(\dot r_0-\dot r)\partial_r$. The radius at
which the signal comes to the world  line  2,  we  denote  as  the
moving coordinate $r$. Since $\dot r_1=0$, $(k\cdot u)_1=1$,
and in general $k\cdot u=1+\dot r(\dot r_0-\dot r)g_{rr}$.
It is easy  to find  that $\dot r=-F/\Sigma$ and
$\dot r_0=-EA^{1/2}/\Sigma$,
where the choice of sign  (minus) is determined by
the direction of motion (the fall). Moreover,
since $AE^2=F^2+\Delta\Sigma$,
we obtain for the red shift measured by the probe falling onto the
Kerr--Newman black hole, finally
$$
z=(F^2/\Sigma\Delta)[(1-\Sigma\Delta/F^2)^{1/2}].       \eqno{(3.4.9)}
$$

        Since at the horizons (the both event and Cauchy ones) $\Delta=0$,
we see that the limiting value of the red shift is $z=\frac{1}{2}$ when
the probe crosses a horizon. It is remarkable that  this  quantity
does not depend on anything whatsoever --- on mass,  Kerr  parameter
and charge of the black hole, and  on  the  radius  at  which  the
mother craft is hovering (with a drift in the azimuthal direction)
and the polar angle value to which all the motions correspond. The
frequency shift occurs namely to the red, {\em i.e.} it can  be  treated
as if the Doppler shift would dominate over the gravitational  (in
this case, violet) one. Thus, using the red frequency shift of the
standard signals coming from the mother craft, it is  possible  on
the probe to determine the moment of crossing the horizon  of  any
black hole. In this example, in  contrast  with  the  cosmological
case, it is impossible to introduce a frame which is simultaneously
co-moving with respect to the both source and receiver  of  the
signals: at the initial moment their world lines coincide, so they
cannot belong to one and the same congruence. This  is  however  a
meaningful example of application of a bilocal approach to reference
frames, complementary with respect to the monad formalism.

\pagestyle{myheadings}
\markboth{CHAPTER 3. EQUATIONS OF MOTION}{3.5. THE DRAGGING PHENOMENON}
\section{The dragging phenomenon}   

        The dragging phenomenon embraces a large variety of
effects predicted for test particles moving in gravitational fields
of rotating (and similar to them) systems of material objects, and
it probably has a very profound and general physical nature. It is
usually characterized as ``dragging of local inertial frames'',
being considered --- in the absolute majority of cases --- in
space-times admitting a time-like Killing vector field $\xi$
(see, {\em e.g.}, Misner, Thorne and Wheeler (1973)).
It is clear that in such space-times a privileged reference frame
is easily introduced, namely that which coincides with the Killing
congruence. Rotation of this reference frame describes then the
dragging effects. We shall see that this rotation (essential stationarity
of the space-time) is a necessary prerequisite for existence of dragging,
though usually in discussions of the dragging phenomenon nothing is
mentioned concerning reference frames being used,
maybe some traditional words
about the local reference frames only (but without any elements of a
sensible formalization of the reference frames theory). In this section,
we attempt to give an introduction to consideration of the dragging
phenomenon in the realm of this theory, as well as additional hints
pointing at some prospects and needs in generalizing the very concept
of dragging.

        In a space-time admitting time-like Killing vector, there exists
a preferred reference frame with
$$
\tau=\frac{\xi}{\sqrt{\xi\cdot\xi}}.    \eqno{(3.5.1)}
$$
Such a Killingian reference frame can be interpreted as a co-moving
one with the gravitational field (or with the physical system
itself which produces this field), so that the frame is in the state of
acceleration and rotation together with the space-time being considered
(no deformation can obviously be present in this case).
If this space-time is not static, but stationary one, the dragging
phenomenon occurs. It is connected with existence in such a
reference frame, of a gravitomagnetic field --- the rotation
vector $\omega$ ({\em cf.} Nordtvedt (1988); Jantzen, Carini and Bini 1992).
If we change to another monad, $\tilde\tau$, this vector
field may be of course transformed away, but since the preferred
Killingian frame is supposed to be in a rotation, $\omega$ has to be
considered as possessing an absolute meaning. We speak then on a
rotating Killingian reference frame,
whose rotation yields gravitomag\-netic
effects, and this is precisely the dragging phenomenon. Since there
usually exists another (now, space-like) Killing vector field $\eta$
which may be considered as complementary to $\xi$, a combination of
these two fields (with constant coefficients) also represents a Killing
vector field, the concrete choice of coefficients determining a
hypersurface on which the corresponding rotation vector vanishes
locally (this analysis may be most characteristic for studies of
the ergosphere region, {\em e.g.}, of the Kerr space-time, and it
corresponds to the local stationarity property of that space-time;
{\em cf.} Hawking and Ellis (1973), p. 167. In the case of a
pencil-of-light space-time, local stationarity
is in fact a global property).

        It is worth giving here a concrete, but general enough example
of how works the existence (or absence) of the rotation vector, from
the point of view of a coordinated basis connected with $\tau$.
The Killing equation (1.2.42) can be rewritten as
$$
\pounds_\xi g_{\mu\nu}\equiv g_{\mu\nu,\alpha}\xi^\alpha-
g_{\alpha\nu}{\xi^\alpha}_{,\mu}-g_{\mu\alpha}{\xi^\alpha}_{,\nu}=0.
        \eqno{(3.5.2)}
$$
One may choose coordinate (say, $t$) lines coinciding with the
integral curves of the field $\xi$, $\xi\cdot\xi>0$,
together with a corresponding
parametrization along these lines, so that in the new system of
coordinates, $\xi=\partial_t\equiv\delta^\mu_t\partial_\mu$. Then
the last equation takes form $g_{\mu\nu,t}=0$ ({\em i.e.},
all metric coefficients
are independent of the Killingian coordinate $t$). If there exist
several independent Killing vector fields, the independence of all
components $g_{\mu\nu}$ of all corresponding new (Killingian)
coordinates may be achieved, only if these Killing vectors mutually
commute. If not, the constancy of metric coefficients is realized
in different systems of coordinates only (not unifiable into one and
the same system). We confine ourselves to a consideration of
one single Killing vector field, $\xi$, and we will show that
a non-rotating Killingian congruence corresponds to a static space-time,
and {\em vice versa}. In the corresponding Killingian system of coordinates,
$$
ds^2=g_{tt}dt^2+2g_{ti}dtdx^i+g_{ij}dx^idx^j,
$$
$i,j=1,2,3$ if $t=x^0$; all $g_{\mu\nu}$ are independent of $t$.
Equivalently,
$$
ds^2=g_{tt}\left(dt+\frac{g_{ti}}{g_{tt}}dx^i\right)^2+
\left(g_{ij}-\frac{g_{ti}g_{tj}}{g_{tt}}\right)dx^idx^j
        \eqno{(3.5.3)}
$$
or
$$
ds^2=(\tau_\mu dx^\mu)^2+b_{\mu\nu}dx^\mu dx^\nu
$$
where
$$
\tau_\mu=\frac{g_{t\mu}}{\sqrt{g_{tt}}}, ~ ~
b_{\mu\nu}=g_{\mu\nu}-\tau_\mu\tau_\nu.
$$
(The first of these last expressions represents a monad in
coordinates co-moving with the reference frame (Zelmanov's
chronometric invariants formalism), while the second one simply
duplicates (2.2.1).) A complete separation of $t$ and $x^i$ in
(3.5.3) --- orthogonalization of $t$ axis with respect to all other
(here, spatial) coordinates --- occurs when the expression
$\zeta:=\tau/\sqrt{g_{tt}}=dt+(g_{ti}/g_{tt})dx^i$ is a total
differential. This case
is known as the static field case (a simultaneous orthogonality
of $t$ with respect to other axes {\em and} $t$-independence of
all $g_{\mu\nu}$). This corresponds to fulfilment of the condition
$$
\left(\frac{g_{ti}}{g_{tt}}\right)_{,j}-
\left(\frac{g_{tj}}{g_{tt}}\right)_{,i}=0,
$$
the same as vanishing of $A_{\mu\nu}$ (or, equivalently, of $\omega$).
If $\omega\neq0$, the space-time is a stationary (but not static) one.

        Thus we have come to the following chain of conclusions:
It is {\em always} possible to choose the coordinates in such a way that
$\xi=\partial_t$; then $\xi\cdot\xi\equiv g_{tt}$, so that
$\tau=\xi/\sqrt{g_{tt}}$. Now, $\zeta:=\tau/\sqrt{g_{tt}}=\xi/g_{tt}
=\xi/(\xi\cdot\xi)$, but it is possible to introduce a new time
coordinate, $\tilde t$, so that $\zeta=d\tilde t$, {\em if and only if}
$\omega=0$.

        We shall consider first a very simple case of dragging,
that which occurs in the equatorial plane of the Kerr space-time.
The Kerr metric in the Boyer--Lindquist coordinates reads
$$
ds^2=\left(1-\frac{2\gamma mr}{r^2+a^2\cos^2\vartheta}\right)dt^2-
\frac{r^2+a^2\cos^2\vartheta}{r^2-2\gamma mr+a^2}dr^2-
(r^2+a^2\cos^2\vartheta)d\vartheta^2
$$
$$
-\left(r^2+a^2+\frac{2\gamma ma^2r\sin^2\vartheta}
{r^2+a^2\cos^2\vartheta}\right)\sin^2\vartheta d\phi^2+
2 ~ \frac{2\gamma mar\sin^2\vartheta}{r^2+a^2\cos^2\vartheta}d\phi dt,
        \eqno{(3.5.4)}
$$
so that
$$
\sqrt{-g}=(r^2+a^2\cos^2\vartheta)\sin\vartheta.
$$
For a circular equatorial orbit, radial component of the geodesic
equation
$$
\frac{d}{ds}\left(g_{r\nu}\frac{dx^\nu}{ds}\right)=0=
\frac{1}{2}g_{\alpha\beta,r}\frac{dx^\alpha}{ds}\frac{dx^\beta}{ds}
        \eqno{(3.5.5)}
$$
gives enough information for solving the problem. Here
$dx^\mu/ds=(dt/ds)\delta^\mu_t+(d\phi/ds)\delta^\mu_\phi$
($r$ and $\vartheta$ do not change in the course of motion).
There are two roots of the eq. (3.5.5),
$$
\dot\phi_\pm=\left(a\pm\sqrt{\frac{r^3}{\gamma m}}\right)^{-1}
        \eqno{(3.5.6)}
$$
($\dot\phi:=d\phi/dt$), so that
$$
T_\pm:=2\pi\mid\dot\phi_\pm\mid^{-1}
=2\pi\left(\sqrt{\frac{r^3}{\gamma m}}\pm a\right)=T_N\pm\Delta T
        \eqno{(3.5.7)}
$$
(we consider here $a$ being a smaller term than $\sqrt{r^3/(\gamma m)}$
in a physically meaningful region of the motion). The main part of the
period has here the standard Newtonian value $T_N$,
while the dragging term simply simply is additive.
It is rather remarkable that it does not depend
on the orbit's radius, mass of the attracting centre, and even on the
gravitational constant, thus being suggestible from purely dimensional
considerations. At the same time, it is worth stressing that this result
is exact, and not approximate one (see Mitskievich and Pulido (1970),
Mitskievich (1976), Mitskievich (1990); for an approximate approach,
see Mitskievich (1979)). The fact that we have used here
the coordinate (not proper) time and (coordinate) angle $\phi$,
does not make our
results non-covariant, since these coordinates are the Killingian ones,
thus invariantly reflecting geometric properties of the Kerr space-time.
It is clear that the middle point between the two test particles moving
in the opposite directions, then itself rotates with the angular velocity
$$
\dot\phi_m= \frac{\gamma ma}{\gamma ma^2-r^3}        \eqno{(3.5.8)}
$$
(in (Mitskievich and Pulido 1970), this point was interpreted as
``the meeting
point'' of the particles), and a simple calculation shows that
(Mitskievich 1976) if for usual celestial bodies the effect is quite
small, for super-dense objects (such as pulsars) it is very large: see
the Table in which we have taken the circular orbit's radius equal to
$r=3^{1/3}R$ where R is radius of the real attracting centre, $D$ the
meeting point drift in seconds of arc per century
(the Schwarzschild precession $\Delta$ in the same units being
also given as illustrative data, the quantity $r$ being then the large
semi-axis of an elliptic orbit).

~~~~ \\
\begin{tabular}{|p{0.8in}|c|c|c|c|c|c|} \hline
Central body & $m,~g$ & $R,~cm$ & $L,~\frac{g\cdot cm^2}{s}$ & $T_N,~s$
        & $D$ & $\Delta$ \\
\hline\hline
{\footnotesize Sun} & $2\cdot10^{33}$ & $7\cdot10^{10}$ &
        $1.3\cdot10^{49}$ & $2\cdot10^4$ & 600 & $10^6$ \\
{\footnotesize  Earth} & $6\cdot10^{27}$ & $6\cdot10^8$ &
        $7\cdot10^{40}$ & $9\cdot10^3$ & 4.4 & 670 \\
{\footnotesize Jupiter} & $2\cdot10^{30}$ & $7\cdot10^9$ &
        $7\cdot10^{45}$ & $2\cdot10^4$ & 280 & $10^4$ \\
{\footnotesize Rapidly rotating} & ~ & ~ & ~ & ~ & ~ &  \\
{\footnotesize class B star} &
        $3\cdot10^{34}$ & $3.5\cdot10^{11}$ & $2\cdot10^{53}$ & $5\cdot10^4$
        & $7.6\cdot10^4$ & $10^6$ \\
{\footnotesize Pulsar at a} & ~ & ~ & ~ & ~ & ~ &  \\
{\footnotesize breaking point} &
        $10^{33}$ & $10^6$ & $4\cdot10^{48}$ & $10^{-3}$  &
        93 $\frac{rad}{sec}$ & 750 $\frac{rad}{sec}$  \\
\hline
\end{tabular}
\bigskip

A closely analogous, but exotic dragging effect is present in the NUT
field --- the field of a gravitational dyon (Mitskievich 1981,
Mitskievich 1983). Here a gravitoelectric field of the Schwarzschild
type is accompanied by a gravitomagnetic field, similar to the
magnetic field of a magnetic monopole. If one considers in this field
circular orbits of test masses, one finds that the field centre cannot
be in the plane of such an orbit, but it is shifted in the direction
perpendicular to the plane, by a distance proportional to the value
of the orbital angular momentum and to the NUT parameter $l$, the
direction of the shift being the same as direction of the angular
momentum $L$ for $l>0$, and opposite to this direction for $l<0$. The
effect is described by the exact formula
\[
\cos\vartheta=-2E\frac{l}{L}.
\]

Another analogous dragging effect is that of a Kerr gravitational
lens when two light rays passing by the Kerr centre in its equatorial
plane with the same impact parameter but on opposite sides of the
centre, are deviated differently by the gravitational field of the
rotating source, and they are focused off the ``straight'' line
going through the centre parallel to the initial common direction
of the rays (a transverse shift of the focus of these rays).
The magnitude of this shift from the above ``straight'' line is
{\em equal} to the Kerr parameter (if measured in the units of distance),
this being however an approximate result which becomes more precise
for large values of the impact parameter (Mitskievich and Gupta 1980,
Mitskievich and Uldin 1983, Mitskievich and Cindra 1988).

        Now let us consider a stationary case of the pencil-of-light
space-time:
$$
ds^2=dt^2-d\rho^2-dz^2-\rho^2d\phi^2+
8\gamma\epsilon\ln(\sigma\rho)(dt-dz)^2.        \eqno{(3.5.9)}
$$
Here $\epsilon$ is the energy (or, equivalently, momentum
$z$-component) linear density, while $\sigma$ serves simply for
making the argument of logarithm dimensionless (in a non-stationary case,
both $\epsilon$ and $\sigma$ are treated as arbitrary functions of
the retarded time $t-z$).

The already mentioned stationarity property of the metric (3.5.9)
is non-trivial. There always exist Killing vectors $\xi=
\partial/\partial t$ and $\eta=\partial/\partial z$ (as well as
$\partial/\partial\phi$, of course). The squares of $\xi$ and $\eta$
are not sign-defenite, so that $t$ is not always a timelike
coordinate, and $z$ not always a spacelike one. In the region
$-1<8\gamma\epsilon\ln\sigma\rho<+1$ the coordinates are
respectively time- and spacelike. When $8\gamma\epsilon\ln\sigma\rho<-1$
the both coordinates are spacelike, and when
$8\gamma\epsilon\ln\sigma\rho>+1$, they are timelike. But at any point
one may take linear combinations (with constant coefficients) of
these two Killing vectors, yielding two new Killing vectors being
correspondingly timelike and spacelike. Moreover, the band
$-1<8\gamma\epsilon\ln\sigma\rho<+1$ may be then shifted to every
position (retaining its finite width) if the following coordinate
transformation is used:
\[
t=(1-L)t'+Lz', ~ ~ z=-Lt'+(1+L)z', ~ ~
L=-4\gamma\epsilon\ln(\sigma'/\sigma),
\]
while
\[
8\gamma\epsilon\ln(\sigma\rho) ~ \rightarrow ~
8\gamma\epsilon\ln(\sigma'\rho).
\]

From the geodesic equation it is easy to conclude about existence in
the pencil-of-light field of the phenomenon of dragging; at the same
time, the timelike Killing vector congruence (in the above-mentioned
band) rotates, so that in the privileged frame there exists a
gravitomagnetic field (alongside with a gravitoelectric one). With
$\lambda$ as a canonical parameter, the first integrals of motion
are
\[
(1+8\gamma\epsilon\ln(\sigma\rho))\frac{dt}{d\lambda}=\alpha, ~ ~
(1-8\gamma\epsilon\ln(\sigma\rho))\frac{dz}{d\lambda}=\beta,
\]
\[
\rho^2\frac{d\phi}{d\lambda}=\mu,
\]
$\alpha$, $\beta$ and $\mu$ being the integration constants. For $\rho$
we have the equation
\[
\frac{d^2\rho}{d\lambda^2}=-\frac{4\gamma\epsilon}{\rho}
(\alpha-\beta)^2+\frac{\mu^2}{\rho^2}
\]
(it can be easily integrated to a first integral, but this reduces
to the squared interval (3.5.9)). In the right-hand side, the first
term describes gravitoelectric attraction to z axis, and the second
one, the centrifugal repulsion. The dragging effect is in this case
compensated by the (conserved) $z$-component of linear momentum
of the test particle.

If the test particle even did momentarily move without changing its
$z$-coordinate, being however not on a circular orbit in a plane
perpendicular to $z$ axis, this particle is dragged in the next
moment along this axis, the sign of the dragging being determined
by the expression
\[
(\alpha-\beta)8\gamma\epsilon\frac{\Delta\rho}{\rho_0}
\]
where $\rho_0$ is the value of $\rho$ when, momentarily,
$dz/d\lambda=0$, and $\Delta\rho$ is the deviation of $\rho$ from
$\rho_0$.

It is worth mentioning that a photon (lightlike geodesic) moving
parallel to $z$ axis in its positive direction, does not interact
with the pencil-of-light field at all (see for a general discussion
Mitskievich (1989)), but a photon moving in the
opposite sense will fall subsequently onto the pencil of light
($z$ axis). This fact of no interaction holds not only for test
objects in the pencil-of-light field, but it is also true for
self-consistent problems with light-like moving objects in general
relativity. Thus we come to exact additivity of gravitational fields
of parallel (not anti-parallel!) pencils of light; this being an
exact generalization of the known approximate result obtained by
Tolman, Ehrenfest and Podolsky (1931).

        Another case of the pencil-of-light field involves not only
luminal (null) motion of the field's source, but its rotation too,
this rotation being considered as an analogue of polarization of an
infinitesimally thin beam of light. The corresponding metric reads
$$
ds^2=2d(t-z)[d(t+z)+\left(4\gamma\epsilon\ln(\sigma\rho)-a\phi\right)
d(t-z)]-d\rho^2-\rho^2d\phi^2   \eqno{(3.5.10)}
$$
which may be modified to
$$
ds^2=2d(t-z)[d(t+z)+4\gamma\epsilon\ln(\sigma\rho)d(t-z)
+(t-z)ad\phi]-d\rho^2-\rho^2d\phi^2     \eqno{(3.5.11)}
$$
where $a$ is a new constant corresponding to the ``polarization''
(in a more general case considered from another point of view in
chapter 5, instead of the product $(t-z)a$ there stands an arbitrary
function $g(t-z)$; then the space-time becomes however non-stationary).
The metric (3.5.10) describes a stationary gravitational field,
thus permitting
application of the standard approach to the dragging phenomenon.
Here, as well as in the alternative version (3.5.11), we see that
the rotational
motion when coexisting with the luminal one, inevitably leads to
a {\em dragging of dragging} effect, that is, to appearance
of the product term $dzd\phi$, in addition to $dt d\phi$, $dt dz$ being
already present by the virtue of the pencil-of-light nature of the
initial metric. This is not a merely accidental presence of all three
terms in a concrete solution, but {\em inevitability} of the presence
of them all, if any two of them are already present in a solution.

        The simplest approach to consideration of dragging in the
pencil-of-light space-time consists of a calculation of the (proper)
time derivatives of coordinates of a free test object. We mean here
not covariant, but simply partial derivatives which manifest the
tendency of varying the coordinates (or their derivatives, if only
the corresponding higher order derivatives become non-vanishing). Thus,
considering as an initial condition an instantaneous state of ``rest'',
we see that the geodesic equation yields (through the second order
derivatives) a (non-covariant) acceleration towards the pencil-of-light
source (in the $-\rho$ direction); these (and possibly the third order)
derivatives show, moreover, that {\em a tendency} of dragging
exists, acting in the $\phi$ (polarization) and $z$
(orientation of the light-like motion of the pencil of light) directions.
We encounter then, as manifestations of dragging in this special kind
of field, the {\em germs} of acceleration (see Mitskievich and
Kumaradtya (1989)).

Let us discuss this problem quantitatively using a more general
form of the {\em spinning} pencil of light (SPL) metric,
\[
ds^2=2dv\left(du+k(v)\ln\sigma\rho dv+g(v)d\phi\right)-d\rho^2-\rho^2d\phi^2
\]
(a possible dependence $\sigma(v)$ is inessential); here the variables
$v$ and $u$ correspond to $t-z$ and $t+z$ in (3.5.10)).

Since the metric coefficients are independent of $u$ anf $\phi$,
one immediately arrives at two first integrals of motion,
\[
g_{vu}\frac{dv}{d\lambda}=\alpha>0, ~ ~
g_{v\phi}\frac{dv}{d\lambda}+g_{\phi\phi}\frac{d\phi}{d\lambda}=-\beta.
\]
The further exact integration of the geodesic equation is in general
impossible, but we are not interested in obtaining approximate
solutions ({\em e.g.}, using perturbative procedures), so another way
to deal with the dragging phenomenon should be chosen. We shall
consider only {\em germs} --- the tendency of test particles motion,
{\em i.e.}, the first non-vanishing higher order derivatives of
the particles'
spatial coordinates when the initial state of motion is given. This
initial motion is most naturally chosen as a state of momentary rest
(if the test particle has non-zero rest mass, thus $d\lambda=ds$).
Then
\[
\left(\frac{d\rho}{ds}\right)_0=\left(\frac{d\phi}{ds}\right)_0=
\left(\frac{dz}{ds}\right)_0=0.
\]
We may now express $z$ in analogy with the Peres wave:
$z=(1/2)(u-v)$ ({\em cf.} also (3.5.10)). Then
\[
\left(\frac{dv}{ds}\right)_0=\left(\frac{du}{ds}\right)_0.
\]
The remaining components of the geodesic equation yield
$$
\left(\frac{d^2\rho}{ds^2}\right)_0=-\frac{k\alpha^2}{\rho},
        \eqno{(3.5.12)}
$$
$$
\left(\frac{d^2z}{ds^2}\right)_0=-\frac{g{\dot g}\alpha^2}{2\rho^2}
-\frac{\alpha^2k\ln\sigma\rho}{2}, \eqno{(3.5.13)}
$$
$$
\left(\frac{d^2\phi}{ds^2}\right)_0=-\frac{{\dot g}\alpha^2}{\rho^2},
        \eqno{(3.5.14)}
$$
where the right-hand side quantities are also taken at the initial
moment (in the sense of $(...)_0$).
Since $k=8\gamma\epsilon_0>0$ (the linear energy density of the SPL
source must be positive), (3.5.12) describes attraction of the test
particle to the SPL. The noncovariant accelerations in the directions
of $z$ and $\phi$, (5.3.13) and (5.3.14), may be either positive or
negative depending on values and signs of $g$, $\dot g$, and $k$, as
well as on $\rho$.
The sign of dragging in the $\phi$ direction coincides with
that of $\dot g$, and this means that not the function $g$, but its
first derivative is directly related to the angular momentum of the
SPL which is responsible for dragging in this direction. Moreover, if
$g$ is constant, it can be transformed away by merely introducing
$\tilde u=u+g\phi$, so that only $\dot g$ can have a physical meaning.

As to the motion of a lightlike particle, its initial state should be
now chosen in a different way, since the former initial conditions
are incompatible with $ds^2=0$. So we shall take only two of these
conditions,
\[
\left(\frac{d\rho}{d\lambda}\right)_0=
\left(\frac{d\phi}{d\lambda}\right)_0=0.
\]
Inserting them into $ds^2$, we have
\[
\left(\frac{dv}{d\lambda}\right)_0\left(\frac{du}{d\lambda}+
k\ln\sigma\rho\frac{dv}{d\lambda}\right)_0=0,
\]
which leads to two possibilities:
\[
\left(\frac{dv}{d\lambda}\right)_0=0
\]
and
$$
\left(\frac{du}{d\lambda}+
k\ln\sigma\rho\frac{dv}{d\lambda}\right)_0=0.  \eqno{(3.5.15)}
$$
The first one corresponds to absence of interaction between parallelly
moving lightlike objects (in fact, also not test ones). The alternative case
(3.5.15) does not admit $(dv/d\lambda)_0=0$ since such a case would
correspond to a world point (an event) and not a world line. Combining
the condition (3.5.15) with the geodesic equation (including the first
integrals of motion), we come to the relations
\[
\left(\frac{d^2\rho}{d\lambda^2}\right)_0=-\frac{k\alpha^2}{\rho},
\]
\[
\left(\frac{d^2\phi}{d\lambda^2}\right)_0=\frac{{\dot g}\alpha^2}{\rho^2},
\]
\[
\left(\frac{d^2u}{d\lambda^2}\right)_0=
-\frac{g{\dot g}\alpha^2}{\rho^2}-k\alpha^2\ln\sigma\rho.
\]
These accelerations are in fact the same as in the case of a massive
test particle, though we have here $u$ instead of $z$. We see that
a photon moving parallel to a SPL does not interact with it (in
particular, it does not feel dragging in the $\phi$ direction),
while a photon moving antiparallel to it, both falls onto the SPL and
starts to rotate in the $\phi$ direction.

        Let us now apply definition of the rotation covector
$\omega$ (2.3.5) to the Killingian frame in the Kerr space-time.
In fact, this is the only object which could be connected with the
quantitative manifestations of the dragging phenomenon.
Since in a stationary space-time (as also in a static one)
$\xi=\partial_t$, (3.5.1) reads
$$
\tau=\frac{g_{\mu t}}{\sqrt{g_{tt}}}dx^\mu;
$$
thus (2.3.5) yields
$$
\omega=\frac{g_{\kappa t}}{2g_{tt}}g_{\mu t,\lambda}
{E^{\kappa\lambda\mu}}_\nu dx^\nu.
$$
First, we observe that $\kappa,~\mu=t,~\phi$ and $\lambda=r,~\vartheta$.
Hence, $\nu=\vartheta,~r$ (complementary to $\lambda$). Then it is
clear that on the equator, $\omega$ is directed along
$\vartheta$, {\em i.e.}
the rotation occurs in the ($\pm$) $\phi$ direction. This is quite natural
and corresponds to a similar rotation of the source of Kerr's field.
The general exact expression is actually
$$
\omega=\frac{1}{2}E^{tr\vartheta\phi}\left[\left(g_{t\phi,\vartheta}-
\frac{g_{t\phi}}{g_{tt}}g_{tt,\vartheta}\right)\partial_r-
\left(g_{t\phi,r}-\frac{g_{t\phi}}{g_{tt}}g_{tt,r}\right)
\partial_\vartheta\right]        \eqno{(3.5.16)}
$$
(contravariant vector representation).

        On the equator ($\vartheta=\pi/2$), the rotation vector
(3.5.16) takes the form
$$
\omega_{\rm eq}=\frac{\gamma ma/r^4}{1-2\gamma m/r}\partial_\vartheta,
~ ~ \sqrt{{\omega_{\rm eq}}^2}=\frac{\gamma ma}{r^2(r-2\gamma m)}.
        \eqno{(3.5.17)}
$$
We have to compare this absolute value (which is a {\em scalar}) with
(3.5.8) (which is neither a scalar nor component of a vector).
They coincide (up to the sign) when the field is weak ({\em i.e.}
$r\gg\gamma m$, $r\gg a$). Why this coincidence realizes for a weak field
only? The answer could be that (1) we compare quantities with different
geometric natures (see our remarks in the parentheses above) which may become
similar in the vicinity of flat space-time only; (2) the very description
of the dragging effect as a mean value of two rotation frequencies is
highly tentative, and could serve well in the weak field approximation,
but not in the general case.

In this comparatively short account on dragging we can but mention
many other aspects of dragging, though they
can be easily related to the reference frames' theory. These include,
first of all, effects of change of orientation of test bodies
possessing multipole (beginning with the dipole) moments (thus
without spherical symmetry, of scalar property). This is, in
particular, the Schiff effect (precession of a gyroscope). Effects
of this type are sometimes classified as non-dragging ones, since
they are connected with a local rotation of test particles
without necessarily a translational motion, but we cannot consider
this distinction as essential. See, {\em e.g.},
Sakina and Chiba (1980), Marck (1983), Tsoubelis, Economou and
Stoghianidis (1987),
Tsoubelis and Economou (1988). This motion includes precession-like
manifestations not only of dragging, but also of the tidal forces
{\em etc} (see also Ashby and Shahid-Saless (1990)). The mentioned
papers contain additional bibliography on the subject.

A specific character of these effects
includes arbitrariness of the gyroscope world line which can be
assigned artificially (usually, one takes a geodesic which is a good
approximation for a freely moving gyroscope). As another example
one may consider a gyroscope suspended at some fixed point over the
surface of the Earth.

To this type of dragging belong effects involving non-scalar
(in particular, rotating) test bodies
which do not change their orientation (relative to the directions
determined by symmetries of the gravitational field) in course of the
motion. Thus no precession-type effects are now considered. For
example, a spinning particle can hover over a Kerr source pole if
\[
M(z^2-a^2)=-2aS\frac{3z^2-a^2}{z^2+a^2}\left(1+\frac{2m|z|}{z^2+a^2}\right).
\]
Here $z$ is position of the test particle over the pole and $S$, spin of
the particle. This effect represents the spin-spin interaction
(Epikhin, Pulido and Mitskievich 1972). An example of the spin-orbital interaction
can be given for a circular motion in the Schwarzschild field of a pair
of spinning test particles with antiparallel spins orthogonal to the
plane of their common orbit (see the same publication). Their periods
of revolution are approximately given by the formula
\[
T_\pm=2\pi\left(\sqrt{\frac{r^3}{\gamma m}}\mp\frac{3S}{2mc^2}\right);
\]
hence, these two particles will chase one another with different linear
velocities: the difference,
\[
\Delta v=\frac{3\gamma m S}{Mc^2r^2},
\]
leads to a drift
\[
\Delta l=6\pi\sqrt{\frac{\gamma m}{rc^2}}\frac{S}{Mc}
\]
of one particle with respect to the other per one revolution. Then
in order to attain a drift of some 1 {\AA}ngstr\"om per one revolution
in a low orbit about the earth, the gyroscopes should spin with
an angular velocity
\[
\Omega\approx10^5R^{-2} ~ sec^{-1},
\]
where $R$ (cm) is the characteristic radius of inertia of the
gyroscope. Some similar dragging (or quasi-dragging) effects, but
in a quantum-mechanical description, were discussed in the monograph
(Mitskievich 1969). Some other problems related to reference frames,
gravitation and dragging in quantum physics, see in (Schmutzer 1975).

        It is worth finally noticing still another important manifestation
of the dragging phenomenon
(not merely in mechanics but essentially in the electromagnetic
field theory) which is connected with results of DeWitt (1966)
supporting the conclusion about a unity of gravotimagnetic and
gravitoelectric effects in general relativity, which should be then
both considered to belong  to the dragging phenomenon. These
results consist of two parts:

        (1) When a conductor is placed into a superposition of electric
and gravitoelectric (Schwarzschild type) fields, or is considered (in the
presence of an electric field) in a translationally accelerated reference
frame, inside such a conductor there vanishes not the electric field,
but a certain combination of that and of the gravitoelectric field,
$$
eE-mG=0,        \eqno{(3.5.18)}
$$
$e$ and $m$ being the charge and mass of electron (inside the conductor,
a ``free'' electron gas has to be considered). Hence,
in the conductor there must exist an electric field exactly corresponding
to the acceleration of the monad congruence which describes the co-moving
frame of this conductor.

        (2) When a superconductor is placed into a superposition of a
magnetic and gravitomagnetic fields (one may consider, {\em e.g.}, the
rotating Earth's field --- say, the Lense--Thirring
or Kerr one), or the
superconductor belongs to a rotating reference frame, inside this
superconductor there vanishes not the magnetic field, but its combination
with the gravitomagnetic field,
$$
eB+2m\omega=0.  \eqno{(3.5.19)}
$$
So it is usually said that in a rotating superconductor a corresponding
magnetic field is generated.

        We propose to write in the case of a superconductor a general
covariant equation
$$
d(mu+eA)=0   \eqno{(3.5.20)}
$$
(this is a general covariant form of eqs. (3.5.18) and (3.5.19);
{\em cf.} also
(3.1.3)). One may characterize this effect as {\em dragging of
electric and magnetic fields by the corresponding gravitational ---
or inertial --- fields}, {\em cf.} below (in section 4.3) kinematic
``source'' terms which appear in Maxwell's equations in non-inertial
frames, and which are analogous to the Coriolis and centrifugal inertial
forces terms in the equations of classical mechanics.

        It is clear that this new type of dragging makes it easier
to measure gravitational fields, since it reduces gravitational
measuring problems to the corresponding electromagnetic ones.
Some proposals along these lines
were made, {\em e.g.}, by Papini (1966, 1969), although no general
covariant exact formulation was ever proposed (see also a consideration
of the superconductivity theory in general relativity by Meier and
Sali\'e (1979) and Sali\'e (1986)).

        It is also obvious that our approach makes it possible as well to
consider non-stationary gravitational fields (and non-uniformly accelerated
and rotating frames). Thus we predict that when a gravitational wave
overrides a superconductor sample, corresponding electromagnetic
oscillations are
to be observable inside it, the measuring problem being reduced
to that of detecting electromagnetic oscillations inside a superconductor.
This is the case when no time-like Killing congruence could be present,
which in the general opinion is however
an indispensable premise for consideration
of the dragging phenomenon itself. So, losing the symmetry in the sense
of time, do we really lose the very thread of our argument?!

        It was however so easy to lose this tiny thread which does exist
in highly idealised cases only. Adding merely a ``grain'' of impurity to
the pure symmetry (here: stationarity), we extinguish the latter completely.
But if a {\em physical effect} --- or {\em phenomenon} --- is so very
vulnerable, it either does not exist at all, or its formulation is to be
crucially revised. For example, in the Kerr field we encounter a great
multitude of dragging effects; so, if from some distant source an
infinitesimally weak gravitational wave comes to this Kerr field region,
could this lead to a total and instantaneous breakdown of all such effects?
Is it not much more likely that our {\em definition} of the dragging
phenomenon (however natural it might seem to be) is in fact doomed to
a radical revision? A symmetry is usually a good tool which makes it
possible to dig up some principal results in the quickest and most elegant
way, but we have not to reject a possibility of generalizing these results
beyond the limits of that symmetry approach, especially if the latter is so
critically unstable. Description of the dragging phenomenon becomes very
vague without the use of ideas of symmetry, because we have then no
preferred reference frame --- a very typical situation in general
relativity. It is however quite plausible that the effects considered
above in a stationary case, should hold also for fields essentially
dependent on time. Thus in a cavity inside superconducting medium, there
should be generated electromagnetic oscillations when a gravitational
radiation pulse would override that region, so that the oscillations
should imitate that pulse. Such a feature (one may speak on a tendency)
is characteristic for gravitation itself (Mitskievich 1983). Somebody
could insist that the notion of dragging is foreign to such effects,
and probably it is not so important what name to give to the phenomenon
the boundary of which is so diffuse. We are inclined nevertheless to
speak about generalized manifestations of dragging, --- let it be,
for example, a combination of gravitoelectric and gravitomagnetic
fields and corresponding effects. In fact, to {\em every} electromagnetic
effect should correspond a certain gravitational effect (though not
{\em vice versa}), which one may call a gravitoelectromagnetic effect,
and a class of such effects pertains, in the limiting case of stationary
space-times, to the dragging phenomenon area.

\newpage

\pagestyle{headings}
\chapter{The Maxwell field equations}                 

\section{The four-dimensional Maxwell equations}        

            To the end  of  the  future  applications  of  Maxwell's
 equations, we give here  different  representations  of  these
 equations in the four-dimensional  form  (without  considering
 specific reference frames), with some relevant comments.

                We recall first the structure of the Lagrangian  density
 (see Landau and Lifshitz 1971; Synge 1965; Mitskievich 1969)
$$
\frak L\;_{\rm em}=-\frac{\sqrt{-g}}{16\pi}F_{\mu\nu}F^{\mu\nu}
        \eqno{(4.1.1)}
$$
which  yields  Maxwell's  equations   through   the   standard
 variational principle. This Lagrangian density can be reshaped
 by singling out a divergence term:
$$
F_{\mu\nu}F^{\mu\nu}=(2A_\nu F^{\mu\nu})_{;\mu}-
2A_\nu{F^{\mu\nu}}_{;\mu}=\ast d(A\wedge\ast F)-A\cdot\delta F.
        \eqno{(4.1.2)}
$$
 The  divergence  (as  usually)  does  not  contribute  to  the
 Lagrange--Euler equations, but the variational principle should
 now be  regarded  as  applied  to  a  Lagrange  density  which
 includes higher (second) order derivatives. The marvel of  this
 reshaping is that the corresponding  action  integral  can  be
 varied  according  to  the  Palatini  method  known   in   the
 gravitation  theory  (Palatini  1919), {\em i.e.} when the  both
 electromagnetic four-potential and  field  tensor  are  varied
 independently.  Then  the  variation  with  respect   to   the
 four-potential yields the Maxwell  equations,  and  that  with
 respect to the field tensor, the standard expression for  this
 tensor as a four-dimensional curl of the  four-potential  ({\em cf.}
 the situation in the gravitational field theory). As  it  is  in
 the gravitational  case,  one  has  herewith  to  assume  that
 Lagrangian densities of the other fields do  not  contain  the
 electromagnetic field tensor, so  that  if  it  entered  those
 Lagrangians before the reshaping (4.1.2) (which  would  be  by
 itself exotic enough), then after passing to (4.1.2), one  has
 to use for it in  the  Lagrangians  four-curl  of $A_\alpha$ in  its
 explicit form. Meanwhile the electromagnetic field  Lagrangian
 does not explicitly contain in this formulation any derivatives
 of the four-potential. Thus the Palatini approach ascribes
 to the expression of the electromagnetic field tensor (2-form)
$$
F=dA   \eqno{(4.1.3)}
$$
 a dynamical sense.

                Now  Maxwell's  equations  can  be  written  in  several
 equivalent forms:
$$
\left.\begin{array}{ll}\delta F=4\pi j, & dF=0; \\
{F^{\mu\nu}}_{;\nu}=-4\pi j^\mu, & {F^{\mu\nu}_{~ \ast}}_{;\nu}=0; \\
(\sqrt{-g}F^{\mu\nu})_{,\nu}=-4\pi\sqrt{-g}j^\mu,
& (\sqrt{-g}F_{~ \ast}^{\mu\nu})_{,\nu}=0; \\
d\ast F=-4\pi\ast j, & \delta\ast F= 0.
\end{array}\right\}      \eqno{(4.1.4)}
$$
 Here the left-hand column represents variants of  the  Maxwell
 equations proper (let us denote  all  of  them  as  (4.1.4a)),
 while  the  right-hand  one,  (4.1.4b),  exspresses  identical
 relations if the  connection  between $F$ and $A$ was  imposed
 initially ($F=dA$),
 otherwise  this  is  an  independent   system   of
 equations. It is worth mentioning that asserting $F$ to  be  a
 closed form, is not in general equivalent to consider it as an
 exact form; hence the theory is ramifying at this point.

                If one has assumed (4.1.3), only the equations  (4.1.4a)
 have to be considered, these taking the form
$$
\Delta A-d\delta A=4\pi j       \eqno{(4.1.5)}
$$
 where it is convenient to impose the Lorenz gauge condition,
$$
\delta A=0.     \eqno{(4.1.6)}
$$
 It is worth reminding that  this  condition  is  due  to  L.V.
 Lorenz of Copenhagen, Denmark, and  not  to  H.A.  Lorentz  of
 Leyden, Holland (see Penrose and Rindler 1983). Now the de-Rhamian
 $\Delta$ in the left-hand side includes the  Ricci  curvature
 tensor, see (1.2.49).

\pagestyle{myheadings}
\markboth{CHAPTER 4. THE MAXWELL FIELD EQUATIONS}{4.2. ELECTROMAGNETIC
STRESS-ENERGY TENSOR}

\section{The electromagnetic stress-energy  tensor  and  its  monad
decomposition}          

                The Noether theorem (see Noether 1918, Mitskievich 1969,
 Mitskievich, Yefremov and Nesterov 1985) yields definition
 of the energy-momentum tensor density:
$$
\frak T\;^{\mu\nu}\equiv\sqrt{-g}T^{\mu\nu}=
-2\frac{\delta\frak L\;}{\delta g_{\mu\nu}}.    \eqno{(4.2.1)}
$$
 We substitute here the Lagrangian density (4.1.1) using in  it
 a tensor density of the weight $\frac{1}{2}$,
$\gamma^{\sigma\tau}=(-g)^{1/4}g^{\sigma\tau}$:
$$
\frak L\;_{\rm em}=-\frac{1}{16\pi}F_{\sigma\tau}F_{\alpha\beta}
\gamma^{\sigma\alpha}\gamma^{\tau\beta}.
$$
 An intermediate step will be
$$
\frak T\;_{\rm em}^{\mu\nu}=\frac{1}{4\pi}F_{\sigma\tau}
F_{\alpha\beta}\gamma^{\sigma\alpha}
\frac{\partial\gamma^{\tau\beta}}{\partial g_{\mu\nu}}
$$
 where
$$
\frac{\partial\gamma^{\tau\beta}}{\partial g_{\mu\nu}}=
-(-g)^{1/4}\left(g^{\mu(\tau}g^{\beta)\nu}-
\frac{1}{4}g^{\mu\nu}g^{\tau\beta}\right).
$$
 Remark that
$g_{\mu\nu}\frac{\partial\gamma^{\tau\beta}}{\partial g_{\mu\nu}}\equiv0$
which results in vanishing of trace of the tensor $T_{\rm em}$.
Finally  we  come  to  the  standard expression
$$
{T_{\rm em}}^\nu_\mu=-\frac{1}{4\pi}\left(F_{\mu\lambda}F^{\nu\lambda}-
\frac{1}{4}\delta^\nu_\mu F_{\kappa\lambda}F^{\kappa\lambda}\right),
        \eqno{(4.2.2)}
$$
 which can be written in a more symmetric form with help of the
 "crafty identities" (1.2.6):
$$
{T_{\rm em}}^\nu_\mu=-\frac{1}{8\pi}(F_{\mu\lambda}F^{\nu\lambda}+
F^{\; \ast}_{\mu\lambda}F_{~ \ast}^{\nu\lambda}).        \eqno{(4.2.3)}
$$
 Due to (3.1.20), this expression is equivalent to
$$
{T_{\rm em}}^\nu_\mu=-
\frac{1}{8\pi}\left(\stackrel{+~~}{F_{\mu\lambda}}
\stackrel{-~~}{F^{\nu\lambda}}+
\stackrel{-~~}{F^{~ \ast}_{\mu\lambda}}
\stackrel{+~~}{F_{~ \ast}^{\nu\lambda}}\right)\equiv
-\frac{1}{4\pi}\stackrel{+~~}{F_{\mu\lambda}}
\stackrel{-~~}{F^{\nu\lambda}}
        \eqno{(4.2.4)}
$$
 which immediately yields ${T_{\rm em}}^\nu_\nu\equiv0$.

The following identities ({\em cf.} Wheeler 1962) hold for the
 electromagnetic stress-energy tensor:
$$
{T_{\rm em}}^\nu_\mu{T_{\rm em}}^\lambda_\nu=(8\pi)^{-2}
[(B^2-E^2)-4(E\bullet B)^2]\delta^\lambda_mu,   \eqno{(4.2.5)}
$$
 or equivalently
$$
{T_{\rm em}}^\nu_\mu{T_{\rm em}}^\lambda_\nu=(8\pi)^{-2}
[(B^2+E^2)-4(E\times B)^2]\delta^\lambda_mu,   \eqno{(4.2.6)}
$$

                In order to express  the  electromagnetic  stress-energy
tensor  through  the  observables,   let  us  consider  1-form
${T_{\rm em}}^\nu_\mu\tau_\nu dx^\mu$ (the electromagnetic stresses
are to be  considered separately). Then, making use  of  (3.1.9)
and  (3.1.14),  we obtain
$$
{T_{\rm em}}^\nu_\mu\tau_\nu dx^\mu=(8\pi)^{-1}
(F_{\mu\lambda}E^\lambda-F^{\;\ast}_{\mu\lambda}B^\lambda)dx^\mu
     \eqno{(4.2.7)}
$$
 Since
$$
F_{\mu\lambda}E^\lambda dx^\mu=\ast(E\wedge\ast F) ~ ~ \mbox{and} ~ ~
F^{\;\ast}_{\mu\lambda}B^\lambda dx^\mu=-\ast(B\wedge F),
 \eqno{(4.2.8)}
$$
 we have
$$
{T_{\rm em}}^\nu_\mu\tau_\nu dx^\mu=(8\pi)^{-1}
\ast(E\wedge\ast F+B\wedge F).
       \eqno{(4.2.9)}
$$
 After substituting here the  decomposition  (3.1.15)  and  its
 dual conjugate (3.1.18) and easily identifying the scalar  and
 vector products (see (2.2.15)), we obtain finally
$$
{T_{\rm em}}^\nu_\mu\tau_\nu dx^\mu=(8\pi)^{-1}
[(E^2+B^2)\tau+2E\times B].
    \eqno{(4.2.10)}
$$
 It is clear that the component along the physical time of  the
 reference frame ($\tau$), is the  electromagnetic  energy  density,
 while the part lying  in  the  three-space  of  the  reference
 frame, the electromagnetic energy flux density  (the  Poynting
 vector), or the numerically identical with it (since $c=1$)
 electromagnetic  momentum  density.  We  see  that  all  these
 quantities are expressed through  the  three-dimensional  (but
 nevertheless  generally   covariant)   electromagnetic   field
 observables  (the  electric  field  strength $E$ and  magnetic
 displacement $B$) in the very same way as they are expressed in the
 Maxwell theory in flat Minkowski space-time. This compels
 us to conclude that on the algebraic level, the  electrodynamics
 theory has one and the same formulation in both  special
 and general relativity, neither the gravitational field,  nor
 the  non-inertiality  of  a  reference   frame   giving   any
 essential contribution in its structure.

                As to the  purely  spatial  part  of  the  stress-energy
 tensor, we have to consider it not as an exterior form, but as
 a tensor,
$$
Stress~={T_{\rm em}}^\nu_\mu~b^\mu\otimes b_\nu,    \eqno{(4.2.11)}
$$
 where $b^\mu$ and $b_\mu$ are semi-coordinated spatial  bases  (2.2.12),
 this tensor being (for convenience) of a mixed variance  (once
 covariant and once contravariant). A combination  of  (4.2.11)
 and (4.2.3) yields
$$
Stress~ =(8\pi)^{-1}(F_{\mu\lambda}b^\mu\otimes F^{\nu\lambda}b_\nu
+F^{\;\ast}_{\mu\lambda}b^\mu\otimes F_{~ \ast}^{\nu\lambda}b_\nu).
        \eqno{(4.2.12)}
$$
 In order to translate this to the reference frame language, we
 use a simple identity
 ${\rm Something}^\lambda\equiv{\rm Something}^\kappa\delta_\kappa^\lambda$,
 with a subsequent insertion of
 $\delta_\kappa^\lambda=\tau^\lambda\tau_\kappa+b^\lambda_\kappa$,
{\em cf.} (2.2.1). The terms  with
 $\tau$ yield the field vectors $E$ and $B$ {\em via}  (3.1.9)  and  (3.1.14),
 while the terms with $b^\lambda_\kappa$ are expressed through
 the Levi-Civit\`a axial  tensor.
 We use here the expression (3.1.11) as well as
 its natural counterpart which follows from the  similarity  of
 expressions (3.1.15) and (3.1.18). Further the second identity
 in (2.2.14) yields the final expression for the $Stress$:
$$
Stress~=(8\pi)^{-1}[(E^2+B^2)b^\mu\otimes b_\mu
-2(E\otimes E+B\otimes B)],   \eqno{(4.2.13)}
$$
 the same as in the Minkowski space-time ({\em cf.}  Synge  1965,  p.
 323).

                On the basis of invariants (3.1.16)  and  (3.1.17),  the
 standard classification  of  electromagnetic  fields  follows,
 consisting of the three types, electric ($F_{\mu\nu}F^{\mu\nu}<0$), magnetic
 ($F_{\mu\nu}F^{\mu\nu}>0$), and null ($F_{\mu\nu}F^{\mu\nu}=0$) ones,
 as it was the case for
 the Minkowski space-time. Like it was the case also in the flat  world,
 such electromagnetic fields do not in general reduce to purely
 electric or purely magnetic fields (they are  in  general  not
 pure fields) in no specific reference frame. However, also  as
 in the Minkowski world, the pure field condition is  expressed
 as vanishing  of  the  second  invariant  of
 electrodynamics, $F^\ast_{\mu\nu}F^{\mu\nu}=0$.
 When this condition is  fulfilled,
 it becomes possible to find such  reference  frames  in
 which the only non-zero observables are  either  the  electric
 field strength (if $F_{\mu\nu}F^{\mu\nu}<0$),
 or  the  magnetic  displacement (if $F_{\mu\nu}F^{\mu\nu}>0$),
 or we are dealing with a pure  radiation-type
 field (the both invariants are equal  to  zero),  though  this
 last case includes also fields which cannot be  considered  as
 propagating waves ({\em e.g.}, some stationary  fields; {\em cf.}
 (Synge 1974)).

                It is remarkable that in the last (pure  radiation-type)
 case, an analogue of the Doppler effect holds: It  is  possible
 (in non-wave situations too) to completely transform away the
 electromagnetic field by switching to moving reference frames,
 although this can be done only asymptotically ($v\rightarrow c$).

\markboth{CHAPTER 4. THE MAXWELL FIELD EQUATIONS}{4.3. MONAD FORM OF
MAXWELL'S EQUATIONS}

\section{Monad representation of Maxwell's equations}    

                First let us  write  down  the  action  of  differential
 operators on $E$, the frame-spatial orientation of the resulting
 quantities being conserved, if these quantities are non-scalar
 ones. The differential operations will be:
 The Lie derivative,
$$
\pounds_\tau E=E_{\sigma;\lambda}\tau^\lambda b^\sigma+
E_\lambda D^\lambda_\sigma b^\sigma+E^\lambda A_{\sigma\lambda}b^\sigma,
     \eqno{(4.3.1)}
$$
 divergence,
$$
{\rm div}\; E=\delta E- G\bullet E={E^\alpha}_{;\alpha}-G\bullet E,
        \eqno{(4.3.2)}
$$
 and curl,
$$
{\rm curl}\; E=\ast(\tau\wedge dE).
         \eqno{(4.3.3)}
$$
 These operations act similarly on the vector $B$.

                Consider now the Maxwell equations (4.1.4a),
 $\delta F=4\pi j$, taking into account (3.1.15):
$$
\delta F\equiv\delta[(E\wedge\tau)+\ast(B\wedge\tau)]=4\pi j.
        \eqno{(4.3.4)}
$$
 Differentiation in the first term yields
$$
\delta(E\wedge\tau)={E^\alpha}_{;\alpha}\tau-E_{\sigma;\lambda}
\tau^\lambda dx^\sigma+E_\alpha D^\alpha_\beta dx^\beta-D^\alpha_\alpha E
+E^\alpha A_{\alpha\beta}dx^\beta
$$
$$
=({\rm div}\; E)\tau-\pounds_\tau E+2E_\alpha D^\alpha_\beta dx^\beta-
D^\alpha_\alpha E,
      \eqno{(4.3.5)}
$$
 and in the second term,
$$
\delta\ast(B\wedge\tau)=\ast d(B\wedge\tau)={\rm curl}\; B+G\times B-
2(B\bullet\omega)\tau.
     \eqno{(4.3.6)}
$$

                In the right-hand side of (4.3.4) we have to make use of
 the decomposition of the four-current (3.2.8),
 $j=\stackrel{(\tau)}{u}\rho(\tau+v)$,
 and to introduce the standard notations for  charge  and
 three-current densities,
$$
\stackrel{(\tau)}{\rho}=\stackrel{(\tau)}{u}\rho, ~ ~
\stackrel{(3)}{j}=\stackrel{(\tau)}{u}\rho v,
         \eqno{(4.3.7)}
$$
 which are non-invariant under  transitions  between  different
 reference frames. We obtain then readily the Maxwell equations
 with sources, the scalar
$$
{\rm div}\; E=4\pi\stackrel{(\tau)}{\rho}+2\omega\bullet B
        \eqno{(4.3.8)}
$$
 and the vector one,
$$
{\rm curl}\; B+G\times B=(\pounds_\tau E-2E_\nu D^\nu_\mu
dx^\mu+D^\alpha_\alpha E)+4\pi\stackrel{(3)}{j}.
     \eqno{(4.3.9)}
$$

                The Maxwell equations without sources (4.1.4b), $dF=0$
(or, equivalently, $\delta\ast F=0$),
split relative to a reference frame in a complete  analogy  to
the equations (4.1.4a), if  we  take  into  account  that  the
 expression (3.1.19) differs from $F$ (3.1.16) in (4.3.4)
 by an exchange of $E$ by
 $(-B)$ and $B$ by $E$. Hence the scalar equation takes the form
$$
{\rm div}\; B=-2\omega\bullet E,       \eqno{(4.3.10)}
$$
 and the vector one,
$$
{\rm curl}\; E+G\times E=-(\pounds_\tau B-2B_\nu D^\nu_\mu dx^\mu+
D^\alpha_\alpha B).     \eqno{(4.3.11)}
$$

                We see that in the presence of  a  gravitational  field,
 and  in  a  non-inertial  reference  frame   characterized  by
 acceleration, rotation, and rate-of-strain tensor  (thus  also
 in the flat Minkowski world in such a  reference  frame),  the
 Maxwell  equations  differ  from   the   customary   Maxwell's
 equations written for an inertial reference frame in the flat
 world, by the presence of additional terms. Such terms can  be
 interpreted as supplementary effective charges  (including  an
 effective  ``magnetic  monopole''  charge),   as   well   as   the
 corresponding currents, {\em cf.} (Mitskievich 1969, Mitskievich  and
 Kalev 1975, Mitskievich, Yefremov and Nesterov 1985). This  is
 in a complete agreement with  the  known  theoretical  results
 claiming  appearance  of  effective  electric   and   magnetic
 (monopole)  charges  in  rotating  reference  frames  in   the
 presence of magnetic and electric fields correspondingly  (see
 (Dehnen, H{\"o}nl and Westpfahl 1961) and (H\"onl and Soergel-Fabricius
 1961) for an approximate approach). Such charges and currents have
 of course a purely kinematic nature,  though  their  objective
 existence is confirmed by the characteristic structure of  the
 corresponding electric and magnetic fields revealed  by  their
 action on charged particles when considered in these reference
 frames. Obvious examples of this interpretation are  presented
 in the following pages where we consider  exact  solutions  of
 self-consistent systems of gravitational  and  electromagnetic
 fields together with charged perfect fluid.

                It is however  worth  considering  the  analogy  between
 classical mechanics and electrodynamics in more general terms.
 In the former theory, the  electromagnetic  forces  acting  on
 charged particles, contain characteristics of the  latters  at
 most  algebraically  (in  the  three-dimensional  description,
 electric field even enters  the  equations  of  motion  as  an
 inhomogeneity). In the field equations the analogous terms are
 the sources (mostly  inhomogeneity,  but  the  presence  of  a
 conducting medium is revealed by a term  proportional  to  the
 electric  field, {\em i.e.}  we  encounter   the   same   algebraic
 dependence, now of the field variable). This  analogy  between
 forces and sources can be traced more strictly and formally in
 the Lagrange-Euler form of equations of motion (for  the  both
 mechanics   and   field   theory):    they    correspond    to
 differentiation of the Lagrangian with  respect  to  canonical
 coordinates  (in   the   field   case,   the   electromagnetic
 four-potential). Moreover, both the  forces  and  the  sources
 stem from one and the same terms in the interaction Lagrangian
 density  (which becomes the mechanical interaction  Lagrangian
 when point-like sources are  considered).  In  fact,  this  is
 intimately related to an obvious generalization of  the  third
 Newtonian  law  of  mechanics  to   include   interaction   of
 mechanical  particles  and  fields.  Now,   in   mechanics   a
 non-inertial frame is revealed by appearance of the forces  of
 inertia  (the  rotational  case  is  usually  considered:  the
 centrifugal and Coriolis forces). One has also  to  expect  an
 appearance of some (let us say) ``sources  of  inertia'',  these
 however never having been considered in a far-reaching analogy
 with the forces of inertia of mechanics.  But  after  we  have
 come to the equations (4.3.8)---(4.3.11),  this  analogy  became
 not a mere possibility, but an inevitable necessity.

                The Maxwell field equations seem to  have  a  form  much
 distinct from that of the equations of  motion  in  mechanics,
 but it is now obvious that the terms connected  with  rotation
 of the reference frame (the vector $\omega$) in (4.3.8) and (4.3.10),
 are direct analogs of the Coriolis  force  in mechanics.
 Equations (4.3.9) and (4.3.11) contain the kinematic ``sources
 of inertia''  due  to  the  translational  acceleration  and  to
 deformations (dilatation and shear) of  the  reference  frame;
 similar terms are known (though less discussed)  in  mechanics
 too. The problem now is to find out the role and importance of
 the corresponding non-inertial effects in  the  field  theory,
 primarily in electrodynamics. The following pages contain some
 examples of how to handle this problem.

\pagestyle{headings}

\section{A charged fluid without electric field}      

                We consider here exact solutions of the
 Einstein-Maxwell  equations  and  equations  for  a   charged
 perfect fluid, which were obtained as a generalization of  the
 G\"odel space-time though having a more restricted isometry  group
 (Mitskievich and Tsalakou 1991). These
 two families of solutions are described by a general metric
$$
ds^2=e^{2\alpha}(dt+fdx)^2-e^{2\beta}dx^2-e^{2\gamma}dy^2-e^{2\delta}dz^2
     \eqno{(4.4.1)}
$$
 For the first of the families, (A), one has
$e^\alpha=a, ~ e^{2\beta}=E\exp(2Bz+C)+\lambda z+\nu, ~ 2\gamma=2Bz+C, ~
\delta=(\gamma-\beta)\ln a, ~ f=bz/a^2,$
 while the invariant energy density of the fluid is
$$
\mu=(\kappa a^2)^{-1}[(b^2/a^2-\kappa a^2k^2/2\pi)
\exp(-2Bz-C)-3B^2E],
$$
 its pressure $p=B^2E/(\kappa a^2)$, the invariant charge density
$$
\rho=-\sqrt{2}bk/(4\pi a^2)\exp(-2Bz-C),
$$
 where $\lambda=(\kappa A^2k^2/\pi-b^2/a^2)/2B$, $\nu=(\lambda+D)/2B$,
  $B$, $C$, $D$ and $E$ being integration constants and $k$ a  parameter
  which controls  the  switching  on  and  off  the  charge  and
  electromagnetic field.

                For the second family, ~ (B), ~
$2\alpha=2Bz+C$, ~  $e^{2\beta}=(b/2B)^2e^{-2\alpha}-\kappa(akz)^2/2\pi
+2Dz+E$, ~  $e^{2\gamma}=F^2$, ~
$\delta=\alpha-\beta+\gamma$, ~  $f=-(b/2B)e^{-2\alpha}$,
  while the invariant energy density and pressure of  the  fluid
  are
$$
\mu=(2\kappa F^2)^{-1}e^{-2\alpha}[2BD+\kappa a^2k^2(1-2Bz)/2\pi],
$$
$$
p=(2\kappa F^2)^{-1}e^{-2\alpha}[2BD+\kappa a^2k^2(1-2Bz)/2\pi],
$$
 and the invariant charge density of the fluid
$$
\rho=-\sqrt{2}abk(4\pi F^2)^{-1}e^{-3\alpha}.
$$
  Here $B$, $C$, $D$, $E$ and $F$ are  integration  constants.  The  both
  families of solutions contain also arbitrary constants $a$ and
  $b$. For a special choice of the constants, the  electromagnetic
  field is switched off, and the G\"odel (1949) cosmological model
  is recovered.

                We shall not discuss here the  method  of  obtaining
  these solutions, but their hydrodynamical and electromagnetic properties
  are of considerable interest. It is natural to employ
  in their study  the  co-moving  (with  respect  to  the  fluid)
  reference frame described by the monad field
$$
\tau=u=e^\alpha(dt+fdx).        \eqno{(4.4.2)}
$$

                Since the problem is a  stationary  one,
the  rate-of-strain tensor vanishes for the both  families  of  solutions,
  but acceleration and rotation in general survive,
$$
G=-\alpha'dz, ~ ~ \omega=\frac{1}{2}f'dy.       \eqno{(4.4.3)}
$$
  Substituting  into  these  expressions the   functions
  entering the families (A) and (B), immediately shows  for  the
  family (A) that fluid is in a geodesic motion ($G=0$),  while its
  rotation is homogeneous in the whole of space ($\omega=(b/2a^2)dy$
  being an exact form), the rotation  axis  directed  along  the $y$
  coordinate line. The family (A) fluid may be  considered  then
  as  a  superposition  of  two  distinct  fluids,  a   standard
  electrically neutral G\"odel fluid (if no cosmological  term  is
  introduced, it is a stiff matter), and a charged  non-coherent
  dust. In contrast to this case, the fluid in the family (B) not
  only  rotates  (its  rotation  is  inhomogeneous   everywhere,
  although  the  rotation  axis  is  still  oriented  along  the $y$
  coordinate line), but moreover its particles are accelerated
  (non-geodesic motion). This is connected with  non-homogeneity
  of the pressure (as opposite to the family (A) case) and  by  no
  means with any interaction with the electromagnetic field ({\em cf.}
  Section 3.2). The acceleration is directed along $z$ axis,
  and it behaves exponentially with respect to this coordinate.

                For the both families (A) and (B), the electromagnetic field
  is described (in the coordinated basis) by one  and  the  same
  potential 1-form
$$
A=ka(dt+\sqrt{2}z~dx).  \eqno{(4.4.4)}
$$
  It is remarkable that this field is of magnetic type since
$$
F_{\mu\nu}F^{\mu\nu}=4a^2k^2exp(-2\beta-2\delta)~>~0,   \eqno{(4.4.5)}
$$
  and moreover, it belongs to the pure type
  ($F^{\;\ast}_{\mu\nu}F^{\mu\nu}=0$); in  the
  reference frame co-moving with the fluid,  electric  field  is
  absent completely, in spite of the  fact  that  the  fluid  is
  electrically charged (and the sign  of  the  charge  does  not
  change anywhere). The magnetic field in  the  reference  frame
  under consideration is
$$
B=\sqrt{2}ak~exp(\gamma-\beta-\delta)dy.        \eqno{(4.4.6)}
$$
  We have thus come to an apparently  paradoxical  situation  in
  which a charged  fluid  does  not  generate  (in  a  co-moving
  reference frame) any electric field at all,  so  that  on  its
  particles no lines of force of this field do neither begin nor
  end, although when these families  of solutions were obtained,
  the Maxwell equations with a non-zero four-current density  in
  the right-hand side were considered (and they  were  satisfied
  indeed); for some other examples see Islam (1985).
  It seems that this fluid generates a magnetic field only, and
  this does not influence the motion of the  particles  of  this
  fluid, since the Lorentz force is identically  equal  to  zero
  (in the co-moving reference frame the particles are at rest by
  definition). The apparent paradox is resolved very simply (see
  Mitskievich and Tsalakou 1991) when it is taken  into  account
  that Maxwell's equations  should  be  considered   here  in  a
  rotating reference frame, taking  the  form  (4.3.8)---(4.3.11).
  Then, a direct substitution of the  quantities  characterizing
  the  fluid  and  electromagnetic  field, into {\em e.g.} equation
  (4.3.8), shows that  there  occurs  at  all  points  an  exact
  compensation of the charge density by the  scalar  product  of
  the angular velocity of rotation of the  reference  frame  and
  the magnetic displacement vector $B$, this justifying  vanishing
  of the left-hand side of (4.3.8) which takes place  simply  by
  virtue  of  absence  of  the  very  electric  field.   Similar
  considerations hold concerning all other  Maxwell's  equations
  in this reference frame.

                In this problem, the crucial fact is that we consider an
  exact solution of a self-consistent system  of  equations  for
  the fields and fluid, where the reference frame is objectively
  determined by the very statement of the problem, so  that  the
  final result asserting the  possibility  of  a  charged  fluid
  which effectively generates no electric field, becomes obvious from the
  fact of no electromagnetic interaction between the  (non-test)
  particles of this fluid. And the solution is an exact one! The
  particles  of  this  fluid  interact  with  each  other   only
  gravitationally (in the solution B, through pressure too).
  This effect has however a nature much simpler
  that the general relativistic one: it may arise in special relativistic
  and even non-relativistic circumstances. In such a highly non-linear
  theory as general relativity, it is difficult to immediately conclude
  whether the fields (both gravitational and electromagnetic ones ---
  though Maxwell's equations remain linear in general relativity too)
  are generated by the system of charges (and masses) under consideration,
  or there is a superposition of such a generated field with a free
  one. This is a case when general relativistic and reference frame
  considerations cannot help, thus it is worth treating this problem
  on the Minkowski flat space-time background. A readily obtainable
  conclusion is that in fact the both fields are present, that
  generated dynamically by the charged fluid, and a free (sourceless)
  field which is essentially magnetic and orthogonal to the three-velocity
  of the fluid particles. It is easy to see that even in the case of
  slowly moving fluid, {\em i.e.} when it is non-relativistic (though the
  electromagnetic field always represents a relativistic object), such
  systems of charges and electromagnetic fields are realizable for
  which the Lorentz force completely vanishes. In the particular case
  of our generalization of G\"odel's solution, we would characterize it
  as a superposition of the charges G\"odel and Bonnor--Melvin
  universes.

\pagestyle{myheadings}
\markboth{CHAPTER 4. THE MAXWELL FIELD EQUATIONS}{4.5. KINEMATIC
MAGNETIC CHARGES}

\section{An  Einstein-Maxwell  field  with   kinematic   magnetic
 charges}      

                Let us consider  now  another  example  of  non-inertial
  effects in exact solutions of Einstein-Maxwell equations. This
  will be a vacuum solution without  any  material  carriers  of
  charges (including magnetic charges, or monopoles), but  in  a
  rotating reference frame there will appear a  distribution  of
  purely kinematic monopoles, {\em i.e.} we shall have ${\rm div}B\neq0$.

                To this end we consider the same form of metric  as  in
  the previous section, (4.4.1), but employ another  Ansatz  for
  the electromagnetic field,
$$
A=kt~dy,        \eqno{(4.5.1)}
$$
  which  automatically  leads  to  satisfaction   of   Maxwell's
  equations without sources. Thus we may consider  the  case  of
  absence of any matter in our new space-time. The  electromagnetic
  invariants are now
$$
{\rm and} ~ ~ ~ \left.\begin{array}{l}
F_{\mu\nu}F^{\mu\nu}=2ke^{-2\gamma}(f^2e^{-2\beta}-e^{-2\alpha}), \\
~~ ~ ~ ~ \\
F^{\;\ast}_{\mu\nu}F^{\mu\nu}=0.
        \end{array}\right\}      \eqno{(4.5.2)}
$$
  so that electric  and  magnetic  fields   should  be  mutually
  orthogonal. For simplicity it is natural  to  admit  that  the
  first invariant is also  equal  to  zero, {\em i.e.}  we  would  be
  dealing with a null electromagnetic field.  Thus we postulate
$$
f^2=e^{2(\beta-\alpha)}.        \eqno{(4.5.3)}
$$
  The corresponding stress-energy tensor is
$$
T_{\rm em}=\frac{k^2}{4\pi}e^{-2(\alpha-\gamma)}(\theta^{(0)}-
\theta^{(1)})\otimes(\theta^{(0)}-\theta^{(1)});      \eqno{(4.5.4)}
$$
  although a null tetrad is more appropriate in  this  case,  we
  shall use the  same  orthonormal  basis  as  in  the  previous
  section.

                Since $\mu=0$ (no perfect fluid is present),  we  consider
  the case $\alpha+\beta=0$ (this  is  obvious  from  the  00  and  11
  components of Einstein's equations). A suitable choice of  the
  coordinate $z$ yields $\gamma=\delta$, and we come easily to the following
  solution:
$$
e^{2\alpha}=e^{-2\beta}=f^{-1}=\frac{\kappa k^2}{4\pi}z^2+Az+B, ~
\gamma=\delta=az+b.     \eqno{(4.5.5)}
$$
  Then the squared interval takes a remarkably simple form  (let
  $a=0=b$),
$$
ds^2=e^{2\alpha}dt^2+2dt~dx-dy^2-dz^2,  \eqno{(4.5.6)}
$$
  with null $x$ coordinate lines (or, which is here the same, null
  hypersurfaces $t=$ const), --- a special case  of  the  $pp$-waves
  (see Kramer, Stephani, MacCallum and Herlt (1980), p. 233ff.).

If we take $\tau=\vartheta^{(0)}=e^\alpha dt+e^{-\alpha}dx$, we get
$$
G=-k\alpha'e^{-\delta}\vartheta^{(3)}, ~
\omega=\frac{1}{2}f'e^{-\gamma}\vartheta^{(2)}, ~ D=0,
      \eqno{(4.5.6)}
$$
  as in the previous section, {\em cf.} (4.4.3), and
$$
E=-ke^{-\alpha-\gamma}\vartheta^{(2)},
B=-ke^{-\alpha-\gamma}\vartheta^{(3)}.  \eqno{(4.5.7)}
$$

                Turning  now  to  Maxwell's  equations,  we  find  that
  $\omega\bullet B=0$,  so that ${\rm div}E=0$,
  but $\omega\bullet E=\frac{1}{2}kf'e^{-\alpha-2\gamma}\neq0$,
  hence ${\rm div}B\neq0$ too, and we come to a distribution of  effective
  magnetic charges as kinematic manifestation of the reference
  frame rotation. The same  conclusions  immediately  follow
  from the observations concerning orientation and $z$-dependence
  of the (co)vectors $E$ and $B$. As to the observability  of  such
  purely classical kinematic monopoles,
  we insist that it is perfectly possible,  although
  there is still some  deficiency  in  the  definition  of  the
  physical three-space of a  rotating  reference  frame,  since
  this submanifold is only local (non-holonomic), so that  it  is
  impossible to speak unequivocally about integration over  its
  parts (in order to produce electric and magnetic fluxes). But
  if infinitesimal regions are considered (and for a continuous
  distribution of magnetic charges this is just the case), it is quite
  plausible that such kinematic charges have to be as real as,
  {\em e.g.}, the Coriolis force is.

        Thus we conclude that when experiments on detection of
  dynamical magnetic monopoles are performed, it is important to clearly
  distinguish them from kinematic magnetic charges. However when
  decay processes are considered, no such a problem could arise,
  though at a classical level influences of the both seem to
  be largely identical (in the kinematic case, the very rotation
  of the Earth may simulate a presence of monopoles).

\newpage

\pagestyle{headings}

\chapter{The Einstein field equations}

\section{The four-dimensional Einstein equations} 

         Let us begin with some remarks on the structure of
 four-dimensional (priory to performing the monad splitting)
 Einstein's equations. In chapter 4 we mentioned the Palatini
 method when the Maxwell equations were deduced using the
 electromagnetic Lagrangian. In the gravitation theory, due to
 the principle of equivalence, there exists only a relative
 gravitational force acting on test particles, to which a
 relative field strength corresponds (we would rather prefer to
 use the expression ``gravitational inhomogeneity field'', since
 observable quantities are in any case determined relative to a
 reference frame). In order to come to this concept, one may
 use the geodesic deviation equation,
$$
\nabla_u\nabla_uw={\sf R}(u.v)u={R^\alpha}_{\beta\gamma\delta}
u^\beta u^\gamma w^\delta\partial_\alpha,       \eqno{(5.1.1)}
$$
 which is deduced from the assumption of geodesic motion of
 free test particles, $\nabla_uu=0$, and includes a vector field $w$
commuting with the four-velocity field $u$ of the particles, so
 that $w$ describes closeness of world lines in such a time-like
 congruence (hence, this equation is related to the focusing
 effects of the motion of particles, and to the concept of
 space-time singularities). From (5.1.1) it is clear that the
 role of the gravitational inhomogeneity field is played by the
 Riemann--Christoffel curvature tensor (this is the only
 ingredient in the geodesic deviation equation which does not
 depend on characteristics of motion of the particles
 themselves). If one applies the electromagnetic Lagrangian
 (4.1.1) while throwing off a divergence term according to
 (4.1.2), it is natural to find an analogy between the
 quantities
$$
A\Leftrightarrow g^{\alpha\beta}, ~ ~
\delta F\Leftrightarrow {R^\sigma}_{\alpha\beta\sigma}\equiv
R_{\alpha\beta}.        \eqno{(5.1.2)}
$$
 Thus we come to the gravitational Lagrangian density
$$
\frak L\;_g=\frac{\sqrt{-g}}{16\pi\gamma}g^{\alpha\beta}R_{\alpha\beta}
        \eqno{(5.1.3)}
$$
 in full conformity to the electromagnetic theory, while
 the constant coefficient involves the Newtonian gravitational
 constant $\gamma$ (the Einsteinian constant is also widely used,
 $\kappa=8\pi\gamma$). Every variational method, including that of
 Palatini (1919) ({\em cf.} for other approaches (Landau and
 Lifshitz 1971, Fock 1964) and (DeWitt 1965) where an
 important concept of the second variational (functional)
 derivative is also considered), now yields the well known
 Einstein gravitational field equations,
$$
R_{\alpha\beta}-\frac{1}{2}g_{\alpha\beta}R=-\kappa T_{\alpha\beta}.
   \eqno{(5.1.4)}
$$
 Hence it is logical to expect that the monad splitting of
 these equations (supplemented also by the sourceless equations
 in analogy with the Maxwell theory), should yield scalar and
 vector field equations which would exhibit some analogy with
 Maxwell's equations under the similar splitting (there arise
 also three-tensor gravitational field equations which are of
 course {\em sui generis}).

         Any physical field has certainly its own characteristic
 features (otherwise is would not deserve the status of a
 separate field); this is true with respect to the gravitational
 field too whose equations have their own peculiarities.
 We shall disclose the latters in course of the following
 consideration, but the problem is so extensive and complicated
 that it is worth being a subject of separate study in
 another publication: here we are interested mainly in
 applications of the monad formalism.

\pagestyle{myheadings}
\markboth{CHAPTER 5. THE EINSTEIN FIELD EQUATIONS}{5.2. MONAD FORM
OF EINSTEIN'S EQUATIONS}

\section{Monad representation of Einstein's equations}  

         First we consider equations containing no source terms
 which are similar to Maxwell's equations (4.3.11) and
 (4.3.12). These turned to be identities when the electric
 field strength and magnetic displacement were expressed
 through the four-potential (the reader may do this easily in
 the monad language as an exercise); the new gravitational
 equations are identities in the same sense. Remember
 expressions for divergence and curl of section 2.4, in
 particular (2.4.4). Inserting them into (2.4.7) and (2.4.9)
 (the first of these relations is repeatedly used in the
 second one), we obtain
$$
{\rm div}\; \omega=G\bullet\omega
        \eqno{(5.2.1)}
$$
 and
$$
{\rm curl}\; G=2\left(\pounds_\tau\omega-
2\omega^\alpha D_{\alpha\beta}dx^\beta
+\Theta\omega\right). \eqno{(5.2.2)}
$$
 When these equations are compared with Maxwell's equations
 (4.3.10) and (4.3.11), one sees that they are in fact the
 same up to a substitution of $G$ instead of $E$ and $(-2\omega)$
 instead of $B$; this substitution is exactly the same which can
 be observed in (3.3.11) and the term analogous to $G\times E$ of
 (4.3.11) naturally vanishes.

         As to Einstein's equations proper, they enter the monad
 constructions {\em via} $R_{\alpha\beta}\tau^\alpha\tau^\beta$,
 $R_{\alpha\beta}\tau^\alpha b^\beta$,
 $R_{\alpha\beta}b^\alpha\otimes b^\beta$,
 while it is
 the best to denote
 $T_{\alpha\beta}\tau^\alpha\tau^\beta=:\epsilon$,
 $T_{\alpha\beta}\tau^\alpha b^\beta=:\sigma$,
 $T_{\alpha\beta}b^\alpha\otimes b^\beta=:\eta$,
 where $\epsilon$ is energy density scalar,
 $\sigma$  energy flow density (or equivalently, momentum density)
 covector, and $\eta$ tensor of the purely spatial
 stresses (or, momentum flow density), all taken with respect
 to the reference frame $\tau$ (so that for perfect fluid in a
 co-moving frame, $\tau=u$, one has $\epsilon=\mu$,
 $\sigma=0$, $\eta=p(g - u\otimes u$),
 {\em cf.} (3.2.6); for perfect fluid in an arbitrary frame, $\epsilon$,
 $\sigma$,
 and $\eta$ can be read off from (3.2.7)). Then energy-momentum
 tensor takes the form
$$
T=\epsilon\;\tau\otimes\tau+\tau\otimes\sigma+\sigma\otimes\tau
+\eta,  \eqno{(5.2.3)}
$$
 while $\tau\cdot\sigma\equiv0$, so that the contracted
 energy-momentum tensor is  ${\rm tr}T=\epsilon+{\rm tr}\eta$.

         The obvious identity
$$
R_{\alpha\beta}\tau^\alpha\tau^\beta=({\tau^\alpha}_{;\beta;\alpha}-
{\tau^\alpha}_{;\alpha;\beta})\tau^\beta
$$
 can be easily brought into the form
$$
R_{\alpha\beta}\tau^\alpha\tau^\beta=2\Theta_{,\alpha}\tau^\alpha
-{G^\alpha}_{;\alpha}+D_{\alpha\beta}D^{\alpha\beta}
-A_{\alpha\beta}A^{\alpha\beta},        \eqno{(5.2.4)}
$$
 and it yields finally a scalar equation (its special case for
 $G=0$ is the Raychaudhuri equation, see (Ryan and Shepley
 1975)):
$$
{\rm div}\; G=\frac{\kappa}{2}(\epsilon-{\rm tr}\eta)+2\Theta_{,\alpha}
\tau^\alpha-G\bullet G-2\omega\bullet\omega+
D_{\alpha\beta}D^{\alpha\beta}. \eqno{(5.2.5)}
$$
 Here we used definition of divergence, (2.4.5), and the fact
 that $A_{\alpha\beta}A^{\alpha\beta}\equiv2\omega\bullet\omega$.
  If now the relation $G^\alpha A_{\alpha\beta}dx^\beta=\omega\times G$
  is inserted into the generalized Codazzi equations (2.4.12), one
 obtains the vector equation
$$
{\rm curl}\; \omega=-\kappa\sigma+2\omega\times G+(D^\mu_{\alpha;\nu}b^\nu_\mu
-2\Theta_{,\alpha})b^\alpha.    \eqno{(5.2.6)}
$$

         The tensor equation can be obtained without difficulty
 on the basis of the generalized Gauss equations (2.4.13),
 using as well relation
 $$
 R_{\alpha\beta}\equiv{R^\kappa}_{\alpha\beta\lambda}\tau^\lambda
 \tau_\kappa+{R^\kappa}_{\alpha\beta\lambda}b^\lambda_\kappa
 $$
 where the
 left-hand side (Ricci tensor) is to be expressed through the
 energy-momentum tensor (sources of Einstein's equations),
 while
 $$
 {R^\kappa}_{\alpha\beta\lambda}\tau^\lambda\tau_\kappa
 =(\tau_{\alpha;\beta;\lambda}-\tau_{\alpha;\lambda;\beta})\tau^\lambda,
 $$
 so that this purely three-spatial quantity is easily expressible
 through the reference frame characteristics only:
$$
R_{\alpha\beta\gamma\delta}\tau^\beta\tau^\delta=
\frac{1}{2}G_{\mu;\nu}(b^\mu_\alpha b^\nu_\gamma+b^\mu_\gamma b^\nu_\alpha)
-\pounds_\tau D_{\gamma\alpha}-G_\alpha G_\alpha+D_{\gamma\nu}D^\nu_\alpha
$$
$$
+\omega_\alpha\omega_\gamma+\omega\bullet\omega\;b_{\alpha\gamma}
+2e_{\mu\delta(\alpha}D^\delta_{\gamma)}\omega^\mu.
$$
 At the same time, the crafty identities (1.2.9) yield
$$
R_{\alpha\beta\gamma\delta}b^\alpha_\kappa b^\beta_\lambda
b^\gamma_\mu b^\delta_\nu=-R_{\alpha\beta\gamma\delta}
\tau^\beta\tau^\delta {e_{\kappa\lambda}}^\alpha {e_{\mu\nu}}^\gamma
-\frac{1}{2}R(b_{\kappa\mu}b_{\lambda\nu}-b_{\kappa\nu}b_{\lambda\mu})
$$
$$
-R^{\alpha\gamma}e_{\kappa\lambda\alpha}e_{\mu\nu\gamma}+
R^\delta_\beta\tau^\beta\tau_\delta(b_{\kappa\mu}b_{\lambda\nu}
-b_{\kappa\nu}b_{\lambda\mu}),
$$
 but this is exactly the quantity which should be substituted
 into the right-hand part of the Gauss equations (2.4.13), or
 in their contracted form (2.4.14), which makes no difference
 whatever since for a three-dimensional case the Riemann--Christoffel
 and Ricci tensors are equivalent (they both
 possess one and the same number of independent non-trivial
 components). The subsequent calculations represent a mere
 routine, yielding finally
$$
r_{\lambda\mu}=\frac{1}{2}G_{\kappa;\nu}(b^\kappa_\lambda b^\nu_\mu
+b^\kappa_\mu b^\nu_\lambda)-\pounds_\tau D_{\lambda\mu}-
G_\lambda G_\mu+2D_{\lambda\nu}D^\nu_\mu-2\Theta D_{\lambda\mu}
$$
$$
+2e_{\beta\delta(\lambda}D^\delta_{\mu)}\omega^\beta+
\frac{\kappa}{2}\epsilon b_{\lambda\mu}-\kappa(\eta_{\lambda\mu}
-\frac{1}{2}b_{\lambda\mu}{\rm tr}\eta).        \eqno{(5.2.7)}
$$
 This tensor equation can be found in literature (however, in
 forms differing much one from the other, not only concerning
 notations; {\em cf.} (Vladimirov 1982, Zel'manov and Agakov 1989)).
 It has no counterpart in Maxwell's electrodynamics indeed,
 its characteristic feature being presence of Ricci tensor of
 the three-dimensional space of the reference frame, so that
 (5.2.7) may be interpreted also as means for obtaining this
 latter quantity. We shall not discuss here the important role
 of obtained equations from the point of view of the Cauchy
 problem approach.

         In the dynamical equations which include sources of the
 field, specific character of gravitation manifests itself
 more clearly, although there are certain common terms
 existing in electromagnetic equations too. The most striking
 feature of (5.2.6) is absence of time derivative of the
 acceleration vector ({\em cf.} $\pounds_\tau E$ in (4.3.9)). Dehnen (1962)
 succeeded in eliminating this distinction but at the cost of
 a highly artificial approach, and a coordinate representation
 was essential ({\em cf.} also our approach (Mitskievich 1969) which
 is probably more transparent). One knows however from the
 formulation of the Cauchy problem for Einstein's gravitational
 field that the noted distinction of Einstein's and
 Maxwell's theories is a crucial one (see {\em e.g.} (Mitskievich,
 Yefremov and Nesterov 1985)).

          Thus a question arises, how far-reaching can be the
 natural analogy between gravitation and electromagnetism
 (without abandoning the standard context of Einstein's theory
 of gravity)? The study of the equations of motion of electrically
 charged test particles in chapter 3 has already shown
 that this analogy is very profound, starting with the
 four-dimensional form of these equations, (3.1.4), and being
 revealed more comprehensively in the monad splitted equations
 (3.3.9) -- (3.3.11). In this connection there was even
 extinguished the basic distinction of one theory from
 another, the uniqueness of the gravitational equivalence
 principle, although the latter is manifested primarily by
 equations of motion of the particles. In equation (3.1.4)
 this enters as the repeated participation of the four-velocity
 (or the absence of a similar repetition for the
 four-potential). Equations of the very gravitational field
 are however much more involved than one would expect: if the
 set of equations, being in fact identities when acceleration,
 rotation and rate-of-strain are expressed through the monad
 field, still remains quasi-Maxwellian in its form, the more
 profound dynamical equations of gravity, (5.2.4) and (5.2.5),
 differ from the corresponding Maxwell equations, (4.3.8) and
 (4.3.9), radically enough.

         Does there exist any roundabout way which could outflank
 this difference without revising the already existing theory
 (for such a revision there exist no reasons at all according
 to the profound belief of the most of the specialists)? This
 way really does exist, but it is not as straightforward one
 as the approach just considered above. This alternative
 representation of the gravitation theory is based on the
 Riemann--Christoffel curvature tensor instead of that of Ricci
 used in Einstein's equations; moreover, field equations now
 are first-order differential ones with respect to this tensor
 (third-order to the metric tensor), while the Ricci tensor
 enters Einstein's equations purely algebraically. Nevertheless,
 the physical contents of this higher order theory is the
 same as it was for Einstein's theory. The gravitational field
 equations take then the form of contracted Bianchi identities
 with a substitution of the right-hand side of Einstein's
 equations instead of the Ricci tensor; then the four-dimensional
 equations (without introduction of any reference frame
 approach) show analogy with the Maxwell equations:
$$
{R^\sigma}_{\tau\mu\nu;\sigma}=\kappa[T_{\tau\nu;\mu}-
T_{\tau\mu;\nu}-\frac{1}{2}(g_{\tau\nu}T_{,\mu}-
g_{\tau\mu}T_{,\nu})]   \eqno{(5.2.8)}
$$
 (old quasi-Maxwellian gravitational equations; the new ones
 are (5.4.3)). A detailed study of these equations see in
 (Mitskievich 1969) where also the variational formalism (with
 the corresponding Lagrangian density) was proposed. The first
 attempt of splitting the gravitational inhomogeneity field
 $R_{\alpha\beta\gamma\delta}$ in a reference frame sense
 (not exactly the monad one)
 was made by Matte (1953) and discussed in a modernized form by
 Zakharov (1972) (see also Zel'manov and Agakov (1989)). Hereto
 the notations were introduced,
$$
\overline X_{\alpha\beta}=R_{\mu\alpha\nu\beta}\tau^\mu\tau^\nu, ~ ~
\overline Y_{\alpha\beta}=-R^{~ \ast}_{\mu\alpha\nu\beta}\tau^\mu\tau^\nu,
~ ~    \overline Z_{\alpha\beta}=
-R^{~ \ast~\;\ast}_{\mu\alpha\nu\beta}\tau^\mu\tau^\nu,
        \eqno{(5.2.9)}
$$
 which are clearly similar to (3.1.9) and (3.1.15) for the
 electromagnetic case (though they are quadratic with respect
 to projections onto $\tau$). Dashes were written here since we
 shall use below somewhat different quantities $X$ and $Y$:
$$
X_{\alpha\beta}=C_{\mu\alpha\nu\beta}\tau^\mu\tau^\nu, ~ ~ ~
Y_{\alpha\beta}=-C^{\;\ast}_{\mu\alpha\nu\beta}\tau^\mu\tau^\nu.
        \eqno{(5.2.10)}
$$

\pagestyle{myheadings}
\markboth{CHAPTER 5. THE EINSTEIN FIELD EQUATIONS}{5.3. THE GEODESIC
DEVIATION EQUATION}

\section{The geodesic deviation equation and a new level of 
analogy between gravitation and electromagnetism}

         As it was pointed out in section 5.1, the existing
 exact analogy between gravitation and electromagnetism
 enables one to formulate the gravitational Lagrangian along
 the same lines as it was for the electromagnetic one. The
 directing principle was the use of equations of motion of a
 particle interacting with the corresponding field in order to
 determine the field strength concept; for gravitational field
 one has to consider the geodesic deviation equation (5.1.1).
 Obviously, it is worth making use of this equation also for
 finding hints of how to introduce analogues of the electric
 and magnetic field vectors connected with the gravitational
 inhomogeneity field.

         Let us express the full Riemann--Christoffel tensor in
 (5.1.1) through the Weyl conformal curvature tensor, Ricci
 tensor and the scalar curvature. Consider now  the term
 $C_{\alpha\beta\gamma\delta}u^\beta u^\gamma w^\delta$ which
 may be written also as
 $C_{\alpha\beta\gamma\delta}dx^{\alpha\beta}dx^{\gamma\delta}
 (\cdot,u,u,w)$. The Weyl tensor may be considered as
 the representative of a free gravitational field, the other
 terms entering the Riemann--Christoffel curvature being
 expressed through the gravitational field sources ({\em cf.} the
 ideas of Pirani and Schild (1961) and Wheeler (1993)).
 The quantities $X$ and
 $Y$ which enter the Weyl tensor, are not only symmetric in their
 indices, but also traceless, so that they are the only
 irreducible representatives of free (intrinsic) gravitational
 inhomogeneity field. We apply here the expression (2.2.21)
 for the basis 2-forms and note that the necessary
 contractions are
$$
(\tau\wedge b^\mu)(\cdot,u)=\frac{1}{2}\stackrel{(\tau)}{u}
(\tau v^\mu-b^\mu),     \eqno{(5.3.1)}
$$
$$
\ast(\tau\wedge b^\mu)(\cdot,u)=-\frac{1}{2}\stackrel{(\tau)}{u}
v\times b^\mu,           \eqno{(5.3.2)}
$$
$$
\ast(\tau\wedge b^\mu)(u,w)=-\frac{1}{2}\stackrel{(\tau)}{u}
(v\times w)^\mu.        \eqno{(5.3.3)}
$$

          Recall now that in the electromagnetic case one has
$$
F(\cdot,u)=\frac{1}{2}\stackrel{(\tau)}{u}[(E\bullet v)\tau+E
+(v\times B)]           \eqno{(5.3.4)}
$$
 which gives a combination of the Lorentz force with the work
 of the field {\em pro} unit time (and unit charge of the particle);
 the absence of the second vector, $w$, is here but natural.

        In the case of the Weyl tensor we have
$$
C_{\alpha\beta\gamma\delta}dx^{\alpha\beta}dx^{\gamma\delta}
(\cdot,u,u,w)=\left(\stackrel{(\tau)}{u}\right)^2
[(M\bullet v)\tau+M+(v\times N)]        \eqno{(5.3.5)}
$$
 where
$$
M:=\left[X_{\alpha\beta}\left(\stackrel{(\tau)}{w}v^\beta-
w^\beta\right)+\frac{1}{2}Y_{\alpha\beta}(v\times w)^\beta\right]
b^\alpha        \eqno{(5.3.6)}
$$
 and
$$
N:=\left[-X_{\alpha\beta}(v\times w)^\beta
+\frac{1}{2}Y_{\alpha\beta}\left(\stackrel{(\tau)}{w}v^\beta-
w^\beta\right)\right]b^\alpha.  \eqno{(5.3.7)}
$$
 For $X$ and $Y$ see (5.2.10); it is clear that the analogue
 of $\overline Z$ but
 without a dash would merely coincide with $X$, so it is
 superfluous. The formulae (5.3.4) and (5.3.5) are strikingly
 similar to each other; the factor $\frac{1}{2}$ in (5.3.4) is due to the
 definition of the 2-form $F$, and the second power of the
 standard relativistic factor $\stackrel{(\tau)}{u}$
 results from the repeated
 appearance of the four-velocity $u$ in the left-hand side of
 (5.3.5). It is however impossible to completely identify $M$ and
 $N$ as exact analogues of the electromagnetic quantities $E$ and $B$
 since the formers include a dependence on $u$, the velocity of
 particle, and on another vector field $w$ (density of world
 lines of a cloud of test particles), so they cannot be
 interpreted as pure field quantities. Consider instead an
 equation generalizing the geodesic deviation equation (5.1.1)
 to the case of electrically charged test masses (with equal
 charge/mass ratio $e/m$),
$$
\nabla_u\nabla_u w=\frac{e}{m}(F_{\alpha\beta}u^\beta)_{;\gamma}
w^\gamma dx^\alpha+{\sf R}(u,w)u.       \eqno{(5.3.8)}
$$
 We see that there exists certain difference in the structure
 of the electromagnetic and gravitational (curvature) terms:
 the latter is quadratic in $u$. When expressed with respect to a
 reference frame, equation (5.3.8) contains alongside with the
 term independent of the three-velocity $v$ (quasi-electric
 vector) and that linear in the velocity (in electrodynamics,
 the vector product containing the magnetic displacement vector
 $B$), also a term quadratic in the three-velocity of the
 particle, which cannot have any counterpart in the electromagnetic
 theory. As to the other vector field, $w$, it is specific
 for the deviation type equations, and it is present to the
 same extent also in the electromagnetic part of (5.3.8).

         The only reasonable way out of this situation seems to
 be an introduction of the already well known symmetric
 three-tensors, $X$ and $Y$, present in the motion-of-the-particle
 dependent quantities $M$ and $N$, (5.3.6) and (5.3.7). The first
 is clearly most closely related to the electric type and the
 other to the magnetic type field, but the intrinsic gravitational
 inhomogeneity field tensor (the Weyl tensor) exhibits
 certain peculiarities when entering the geodesic deviation
 equation. It remains to interprete these as manifestations of
 the higher tensor rank of the gravitational field.

\section{New quasi-Maxwellian equations of the gravitational field} 

\markboth{CHAPTER 5. THE EINSTEIN FIELD EQUATIONS}{5.4. NEW
QUASI-MAXWELLIAN EQUATIONS}

         Once contracted Bianchi identities (1.2.23) read
$$
{R^{\nu\lambda}_{~ \ast}}_{\kappa\nu;\lambda}=0     \eqno{(5.4.1)}
$$
 or equivalently,
 $$
 {R^\alpha}_{\beta\gamma\delta;\alpha}=R_{\beta\gamma;\delta}
 -R_{\beta\delta;\gamma}.       \eqno{(5.4.2)}
 $$
 When the right-hand side of Einstein's equations is substituted
 into the right-hand part of (5.4.2), this yields the old
 quasi-Maxwellian equations (5.2.8). It is however better to
 expess the Riemann--Christoffel tensor in the left-hand side of
 (5.4.2) (or (5.2.8)) in terms of the Weyl conformal curvature,
 Ricci tensor and scalar curvature (see (1.2.24)), and
 further express the latters in terms of the stress-energy
 tensor and its trace using Einstein's equations.

         It is easy to find that (5.4.2) and (5.2.8) yield in
 this way new equations
 $$
 {C_{\sigma\tau\mu\nu}}^{;\sigma}=\frac{\kappa}{2}[T_{\tau\nu;\mu}
 -T_{\tau\mu;\nu}-\frac{1}{3}(T_{,\mu}g_{\tau\nu}-
T_{,\nu}g_{\tau\mu})]          \eqno{(5.4.3)}
$$
 (in fact, one can find these equations already in the book by
 Eisenhart (1926), though without their physical interpretation).
 They hold for arbitrary solutions of Einstein's equations;
 moreover (see Lichnerowicz (1960)), if a solution of
 Einstein's equations is taken as initial boundary values, it
 is reproduced in the whole space-time as a solution of equations
 (5.2.8) (so of course also of (5.4.3)). These equations
 determine (when considered in analogy with Maxwell's
 equations for the field tensor $F$) the intrinsic gravitational
 inhomogeneity field tensor $C$ (the other parts of the
 Riemann--Christoffel tensor describe another --- extrinsic ---
 kind of the
 gravitational inhomogeneity field which is algebraically
 reducible to its sources, the stress-energy tensor $T$).

         It is important that non-contracted Bianchi identities
 yield exactly the same equations (the contraction does not
 lead to any information loss in this case): they can be
 written with help of the Weyl tensor, as it already was the
 case for the contracted identities ({\em cf.} (5.4.2) and then
 (5.4.3)). Thus the dual conjugated Weyl tensor now enters the
 left-hand side as ${C^{\;\ast}_{\sigma\tau\mu\nu}}^{;\sigma}$. But
 $C^{\;\ast}_{\sigma\tau\mu\nu}\equiv C^{~ \; ~ \;\ast}_{\sigma\tau\mu\nu}$,
 so that the
 resulting quasi-Maxwellian equations are nothing else than
 the equation (5.4.3) with dual conjugation in the indices $\mu$
 and $\nu$! As a matter of fact, this is equivalent to
 representation of the Bianchi identities as vanishing of
 divergence of the right-hand side of (1.2.9).

         Now we have to express the left-hand side of (5.4.3)
 through the quasi-electric and quasi-magnetic space-like
 symmetric tensors $X$ and $Y$. To this end we first substitute
 relation (2.2.21) into the coordinates-free expression of
 Weyl's tensor and then take account of the definitions
 (5.2.10). The resulting decomposition is
$$
\frac{1}{4}C_{\alpha\beta\gamma\delta}dx^{\alpha\beta}\otimes
dx^{\gamma\delta}=X_{\beta\delta}[(\tau\wedge b^\beta)\otimes
(\tau\wedge b^\delta)-\ast(\tau\wedge b^\beta)\otimes
\ast(\tau\wedge b^\delta)]
$$
$$
+Y_{\beta\delta}[(\tau\wedge b^\beta)\otimes(\tau\wedge b^\delta)
+\ast(\tau\wedge b^\beta)\otimes\ast(\tau\wedge b^\delta)].
        \eqno{(5.4.4)}
$$
 An analogy with the definition (3.1.15) is here obvious,
 although summations over the indices of $X$ and $Y$ components
 are shared among the both two-forms of the tensor (double
 bivector) basis. The pure component form of the same
 expression is
$$
C_{\alpha\beta\gamma\delta}=X_{\beta\delta}\tau_\alpha\tau_\gamma
+X_{\alpha\gamma}\tau_\beta\tau_\delta-X_{\alpha\delta}\tau_\beta\tau_\gamma
-X_{\beta\gamma}\tau_\alpha\tau_\delta
-X^{\kappa\lambda}E_{\alpha\beta\mu\kappa}E_{\gamma\delta\nu\lambda}
\tau^\mu\tau^\nu
$$
$$
+Y_{\beta\lambda}{E_{\gamma\delta}}^{\nu\lambda}\tau_\alpha\tau_\nu-
Y_{\alpha\lambda}{E_{\gamma\delta}}^{\nu\lambda}\tau_\beta\tau_\nu+
Y_{\lambda\delta}{E_{\alpha\beta}}^{\nu\lambda}\tau_\gamma\tau_\nu -
Y_{\lambda\gamma}{E_{\alpha\beta}}^{\nu\lambda}\tau_\delta\tau_\nu.
        \eqno{(5.4.5)}
$$

         Consider now the left-hand side of the quasi-Maxwellian
 equations (5.4.3), {\em i.e.} four-dimensional divergence of the
 Weyl curvature. A somewhat lengthy calculation yields the
 following decomposition with respect to the monad $\tau$:
$$
{C_{\alpha\beta\gamma\delta}}^{;\alpha}=
-\tau_\beta(\tau_\gamma{\rm I}_\delta-\tau_\delta{\rm I}_\gamma)-
\tau_\beta{\rm II}^\lambda e_{\gamma\delta\lambda}
+\tau_\gamma{\rm III}_{\beta\delta}-\tau_\delta{\rm III}_{\beta\gamma}
+{{\rm IV}_\beta}^\lambda e_{\gamma\delta\lambda}
        \eqno{(5.4.6)}
$$
 with
$$
\left.
\begin{array}{rr}
{\rm III}_{\beta\gamma}e^{\beta\gamma\lambda}-{\rm II}^\lambda=0, ~
& {\rm III}_{\alpha\beta}g^{\alpha\beta}=0, \\
{\rm IV}_{\beta\gamma}e^{\beta\gamma\lambda}+{\rm I}^\lambda=0,  ~
& {\rm IV}_{\alpha\beta}g^{\alpha\beta}=0 ~
\end{array}
\right\}
        \eqno{(5.4.7)}
$$
 (consequences of the four-dimensional Ricci identities for $C$
 as well as of its tracelessness). Here
$$
{\rm I}_\delta={X_{\alpha\sigma}}^{;\alpha}b^\sigma_\delta+
G^\alpha X_{\alpha\delta}+3\omega^\alpha Y_{\alpha\delta}
+D^{\alpha\nu}Y^\lambda_\alpha e_{\delta\nu\lambda},
        \eqno{(5.4.8)}
$$
$$
{\rm II}_\delta={Y_{\alpha\sigma}}^{;\alpha}b^\sigma_\delta+
G^\alpha Y_{\alpha\delta}-3\omega^\alpha X_{\alpha\delta}
-D^{\alpha\nu}X^\lambda_\alpha e_{\delta\nu\lambda},
        \eqno{(5.4.9)}
$$
$$
{\rm III}_{(\beta\delta)}=
\pounds_\tau X_{\beta\delta}-\frac{5}{2}(X^\alpha_\beta D_{\alpha\delta}
+X^\alpha_\delta D_{\alpha\beta})+\frac{1}{2}\omega^\gamma
(X^\alpha_\beta e_{\alpha\delta\gamma}+
X^\alpha_\delta e_{\alpha\beta\gamma})
$$
$$
+4\Theta X_{\beta\delta}+X_{\mu\nu}D^{\mu\nu}b_{\beta\delta}
-G^\nu Y^\lambda_\beta e_{\delta\nu\lambda}
-G^\nu Y^\lambda_\delta e_{\beta\nu\lambda}
+\frac{1}{2}Y^{\kappa\lambda;\alpha}(
e_{\alpha\beta\kappa}b_{\lambda\delta}+
e_{\alpha\delta\kappa}b_{\lambda\beta}),         \eqno{(5.4.10)}
$$
$$
{\rm IV}_{(\beta\delta)}=
\pounds_\tau Y_{\beta\delta}-\frac{5}{2}(Y^\alpha_\beta D_{\alpha\delta}
+Y^\alpha_\delta D_{\alpha\beta})+\frac{1}{2}\omega^\gamma
(Y^\alpha_\beta e_{\alpha\delta\gamma}+
Y^\alpha_\delta e_{\alpha\beta\gamma})
$$
$$
+4\Theta Y_{\beta\delta}+Y_{\mu\nu}D^{\mu\nu}b_{\beta\delta}
-G^\nu X^\lambda_\beta e_{\delta\nu\lambda}
-G^\nu X^\lambda_\delta e_{\beta\nu\lambda}
+\frac{1}{2}X^{\kappa\lambda;\alpha}(
e_{\alpha\beta\kappa}b_{\lambda\delta}+
e_{\alpha\delta\kappa}b_{\lambda\beta})         \eqno{(5.4.11)}
$$
 (it is clear from (5.4.7) that only the symmetric parts of
 III and IV are independent constructions). Since the both
 quasi-electric and quasi-magnetic gravitational inhomogeneity
 fields are two-index tensors and not vectors, as it was the
 case for electromagnetic field, it is inappropriate to
 introduce here operations as curl or div ({\em cf.} Section 4.3),
 as well as to make use of three-dimensional scalar and vector
 products.

         In a vacuum, the field equations of gravitational field
 (the higher-order counterpart of Einstein's equations) reduce
 to vanishing of the quantities (5.4.8) -- (5.4.11). A straightforward
 comparison with the corresponding parts of Maxwell's
 equations (4.3.8) -- (4.3.12) shows existence of every
 parallels to electrodynamics, with natural differences in the
 coefficients, while the new specific terms in the gravitational
 field equations are those which cannot in principle arise
 when three-dimensional representation of fields is realized
 with vectors ($E$, $B$) and not tensors ($X$, $Y$). The only
 exclusion is appearance of terms with $\omega^\gamma$, the reference frame
 rotation vector, in (5.4.10) and (5.4.11), but this can be
 easily understood since this comes from the expression of the
 time-derivative, {\em e.g.}
$$
\pounds_\tau X_{\beta\delta}=X_{\beta\delta;\alpha}\tau^\alpha
+X_{\alpha\delta}{\tau^\alpha}_{;\beta}
+X_{\beta\alpha}{\tau^\alpha}_{;\delta},
$$
 which contains as compared with $\pounds_\tau E_\beta$,
 an extra term  with ${\tau^\alpha}_{;\delta}$,
 and the complete compensation (as it was in (4.3.5) and
 (4.3.1)) is now impossible. In order to facilitate such a
 comparison, we give here expressions for Maxwell's equations
 of Chapter 4 in the component form:
$$
{E^\alpha}_{;\alpha}+G^\alpha E_\alpha+2\omega^\alpha B_\alpha
=4\pi\stackrel{(\tau)}{\rho},   \eqno{(5.4.12)}
$$
$$
{B^\alpha}_{;\alpha}+G^\alpha B_\alpha-2\omega^\alpha E_\alpha=0,
        \eqno{(5.4.13)}
$$
$$
\pounds_\tau E_\delta-2E^\alpha D_{\alpha\delta}+2\Theta E_\delta
+G^\nu B^\lambda e_{\delta\nu\lambda}+
B^{\kappa;\alpha}e_{\alpha\delta\kappa}
=-4\pi\stackrel{(3)}{j}_\delta,        \eqno{(5.4.14)}
$$
$$
\pounds_\tau B_\delta-2B^\alpha D_{\alpha\delta}+2\Theta B_\delta
-G^\nu E^\lambda e_{\delta\nu\lambda}-
E^{\kappa;\alpha}e_{\alpha\delta\kappa}=0.      \eqno{(5.4.15)}
$$
 It is obvious that this representation of the gravitational
 field equations is completely in line with the form of
 Maxwell's equations, in contrast with the direct decomposition
 of Einstein's equations: in particular, the time-derivatives
 always enter together with the analogues of curls of the
 counterpart fields (this was exactly the stumbling-block in
 section 5.2).

         As to the right-hand side of the quasi-Maxwellian
 equations, we shall restrict our consideration to a perfect
 fluid in a co-moving reference frame, with the energy-momentum
 tensor (3.2.6). Then the quasi-Maxwellian equations (5.4.3)
 take form
$$
{\rm I}_\delta=\frac{\kappa}{3}\mu_{,\nu}b^\nu_\delta,
        \eqno{(5.4.16)}
$$
$$
{\rm II}_\delta=\kappa(\mu+p)\omega_\delta,     \eqno{(5.4.17)}
$$
$$
{\rm III}_{(\beta\delta)}=\kappa(\mu+p)\left[\frac{1}{3}\Theta
b_{\beta\delta}-\frac{1}{2}D_{\beta\delta}\right],
        \eqno{(5.4.18)}
$$
$$
{\rm IV}_{(\beta\delta)}=0,     \eqno{(5.4.19)}
$$
 where for the left-hand sides, expressions (5.4.8) -- (5.4.11)
 should be substituted. The right-hand side expressions are now
 remarkably simple, and it is easy to check that they satisfy
 conditions (5.4.7) (since in (5.4.16) -- (5.4.19) equations of
 motion (3.2.10) and (3.2.11) were taken into account, they
 should be used in this check also).

\markboth{CHAPTER 5. THE EINSTEIN FIELD EQUATIONS}{5.5. CLASSIFICATION
OF GRAVITATIONAL FIELDS}

\section{Remarks on classification of intrinsic gravitational   
 fields}

         Now it is clear that gravitational inhomogeneity fields
 can be classified in analogy to the very simple
 classification of electromagnetic fields which divides the
 latters into the fields of electric, magnetic and null types
 (see {\em e.g.} Synge (1956)). Very naturally, this classification
 is completely in line with the well known algebraic
 classification of Petrov (1966) (see also (G\'eh\'eniau 1957,
 Pirani 1957, Bel 1959, Debever 1959, Penrose 1960; Kramer, Stephani,
 MacCallum and Herlt 1980)).

         There exist only four independent invariants of
 the Weyl tensor $C$:
$$
\left.
\begin{array}{ll}
I_1=C_{\alpha\beta\gamma\delta}C^{\alpha\beta\gamma\delta},~ &
I_2=C^{\;\ast}_{\alpha\beta\gamma\delta}C^{\alpha\beta\gamma\delta}, \\
I_3=C_{\alpha\beta\gamma\delta}C^{\alpha\beta\epsilon\eta}
{C_{\epsilon\eta}}^{\gamma\delta},~ &
I_4=C^{\;\ast}_{\alpha\beta\gamma\delta}C^{\alpha\beta\epsilon\eta}
{C_{\epsilon\eta}}^{\gamma\delta}
\end{array}     \right\}        \eqno{(5.5.1)}
$$
 (they can be found as the identically non-vanishing constructions
 among the fourteen invariants of the Riemann--Christoffel
 tensor when the latter is substituted by the Weyl
 conformal curvature tensor, see {\em e.g.} (Mitskievich, Yefremov and
 Nesterov 1985), p. 76). When one expresses the Weyl tensor
 {\em via} (5.4.4) or (5.4.5), a representation of the invariants in
 terms of the quasi-electric and quasi-magnetic gravitational
 fields emerges:
$$
\left.
\begin{array}{ll}
I_1=8(X_{\beta\delta}X^{\beta\delta}-Y_{\beta\delta}Y^{\beta\delta}),~ &
I_3=16X^\alpha_\beta(X^\beta_\gamma X^\gamma_\alpha-
3Y^\beta_\gamma Y^\gamma_\alpha), \\
I_2=-16X_{\beta\delta}Y^{\beta\delta},  &
I_4=16Y^\alpha_\beta(Y^\beta_\gamma Y^\gamma_\alpha-
3X^\beta_\gamma X^\gamma_\alpha).
\end{array}    \right\}        \eqno{(5.5.2)}
$$
 Note that these invariants do not change under the both independent
 infinite-parameter groups of transformations, those
 of systems of coordinates and of reference frames. They are
 closely connected with (anti-)self-dual fields ({\em cf.} the
 electromagnetic case, section 3.1). To this end observe that
$$
\stackrel{\pm~~~~}{C^{\;\ast}_{\alpha\beta\gamma\delta}}=
\pm i\stackrel{\pm}{C}_{\alpha\beta\gamma\delta}
        \eqno{(5.5.3)}
$$
 where
$$
\stackrel{\pm}{C}_{\alpha\beta\gamma\delta}:=
C_{\alpha\beta\gamma\delta}\mp iC^{\;\ast}_{\alpha\beta\gamma\delta}.
        \eqno{(5.5.4)}
$$
 Then
$$
\stackrel{\pm}{X}_{\alpha\beta}:=
\stackrel{\pm}{C}_{\alpha\mu\beta\nu}\tau^\mu\tau^\nu=
X_{\alpha\beta}\pm iY_{\alpha\beta}
        \eqno{(5.5.5)}
$$
 and finally
$$
I_3\mp iI_4=\stackrel{\pm~~~~}{{C_{\alpha\beta}}^{\gamma\delta}}
\stackrel{\pm~~~~}{{C_{\gamma\delta}}^{\epsilon\eta}}
\stackrel{\pm~~~~}{{C_{\epsilon\eta}}^{\alpha\beta}}=
16\stackrel{\pm~}{X_\alpha^\beta}\stackrel{\pm~}{X_\beta^\gamma}
\stackrel{\pm}{X_\gamma^\alpha}.
        \eqno{(5.5.6)}
$$
 Similarly,
$$
I_1\mp iI_2=8\stackrel{\pm~}{X_\alpha^\beta}
\stackrel{\pm~}{X_\beta^\alpha}.       \eqno{(5.5.7)}
$$
 Since the quadratic constructions formed exclusively of the
 real (not (anti-)self-dual) $X$ or $Y$ are positive definite, it
 is clear that the sign of $I_1$ determines the electric (plus) or
 magnetic (minus) type of the gravitational inhomogeneity
 field, while $I_1=0$ corresponds to the semi-null type field.
 The gravitational inhomogeneity field is purely quasi-electric,
 purely quasi-magnetic, or null when all the other
 invariants vanish (in the case of electromagnetic field the
 situation is more simple since there exist only two corresponding
 invariants, and vanishing of the only one second invariant
 provides the condition of a pure field). When we conclude
 that the field is purely (quasi-)electric (or magnetic), this
 does not mean that the alternative field is absent identically:
 it simply can be transformed away by a choice of the reference
 frame. The gravitational null case has the same propery
 as its electromagnetic counterpart: the null field merely
 changes its amplitude under transitions between different
 reference frames (a manifestation of the Doppler effect in the
 non-wave representation), with a possibility to transform away
 (at least, locally) the whole of the field in an asymptotical
 sense (when the velocity of a new reference frame is tending
 to the speed of light with respect to the initial frame).

        This amazingly complete and simple analogy between
 electromagnetism and gravitation is well known indeed, but
 its existence is worth being mentioned since it is widely
 underestimated. We shall not go into more detail concerning
 the Petrov classification in connection with the properties
 of the Weyl tensor invariants, as well as the corresponding
 properties of $X$ and $Y$: these aspects may be considered as not
 so closely related to the reference frames problems.

\pagestyle{headings}

\section{Example of the Taub--NUT field} 

         We consider here first the Taub--NUT (in fact, especially
 NUT) field since it incorporates the both quasi-electric and
 quasi-magnetic properties which are already very well known
 (for references see {\em e.g.} (Kramer, Stephani, MacCallum and
 Herlt 1980)). Its metric reads
$$
ds^2=e^{2\alpha}(dt+2l\cos\vartheta d\phi)^2-e^{-2\alpha}dr^2
-(r^2+l^2)(d\vartheta^2+\sin^2\vartheta d\phi^2),
$$
 so that the natural orthogonal tetrad has to be chosen as
$$
\left.
\begin{array}{lr}
\theta^{(0)}=e^\alpha(dt+2l\cos\vartheta d\phi), ~ &
\theta^{(1)}=e^{-\alpha}dr, \\
\theta^{(2)}=(r^2+l^2)^{1/2}d\vartheta, ~ &
\theta^{(3)}=(r^2+l^2)^{1/2}\sin\vartheta d\phi.
\end{array}     \right\}        \eqno{(5.6.1)}
$$
 Here $e^{2\alpha}=(r^2-2mr-l^2)/(r^2+l^2)$. The only non-vanishing
 independent curvature components in this basis are
$$
R_{(0)(1)(0)(1)}=-R_{(2)(3)(2)(3)}=2U,
$$
$$
R_{(0)(2)(0)(2)}=R_{(0)(3)(0)(3)}=-R_{(1)(2)(1)(2)}=-R_{(1)(3)(1)(3)}=U,
$$
$$
R_{(0)(1)(2)(3)}=-2R_{(0)(3)(1)(2)}=-2R_{(0)(2)(3)(1)}=2V,
$$
 where
$$
\left.
\begin{array}{l}
U=(l^4+3ml^2r-3l^2r^2-mr^3)(r^2+l^2)^{-3}, \\
V=l(ml^2-3l^2r-3mr^2+r^3)(r^2+l^2)^{-2}.
\end{array}     \right\}        \eqno{(5.6.2)}
$$

  The most convenient choice of the monad field is
$$
\tau=\theta^{(0)}=e^\alpha(dt+2l\cos\vartheta d\phi)
        \eqno{(5.6.3)}
$$
 which describes an accelerated and rotating (but without
 deformations) reference frame,
$$
G=-\alpha'e^\alpha\theta^{(1)}, ~ ~ \omega=
-le^\alpha(r^2+l^2)^{-1}\theta^{(1)}    \eqno{(5.6.4)}
$$
 (the both vectors have radial orientation, and an effective
 ``spherical symmetry'' takes place). Thus from the point of view
 of section 5.2, we can treat the NUT field as possessing
 analogues of the both electric ($G$) and magnetic ($\omega$) fields,
 their radial orientation corresponding to a similarity to the
 magnetic monopole field superimposed on a point charge field.

         The latter interpretation depends however crucially on
 choice of the reference frame, since it is always possible to
 consider a non-rotating frame, {\em e.g.} with
$$
\overline\tau=N(dt+2ld\phi),    \eqno{(5.6.5)}
$$
$N^2=(r^2-2mr-l^2)(r^2+l^2)[(r^2+l^2)^2-4l^2(r^2-2mr-l^2)
\tan^2(\vartheta/2)]$,
 thus showing analogy with an electric field only. We do not
 present here the corresponding acceleration vector which has
 components along axes $r$ and $\vartheta$, since it is ruther unsightly.

         In the rotating reference frame (5.6.3), the tensors of
 quasi-electric (gravitoelectric) and quasi-magnetic (gravitomagnetic)
 gravitational inhomogeneity
 fields (5.2.9) are diagonal with the components
$$
\left.
\begin{array}{lr}
X_{(1)(1)}=-2U, ~ ~ & X_{(2)(2)}=X_{(3)(3)}=U, \\
Y_{(1)(1)}=2V, ~ ~ & Y_{(2)(2)}=Y_{(3)(3)}=-V.
\end{array}     \right\}        \eqno{(5.6.6)}
$$
 Then the Weyl tensor invariants read
$$
\left.
\begin{array}{ll}
I_1=48(U^2-V^2), ~~ & I_2=96UV, \\
I_3=96U(U^2-3V^2), ~~ & I_4=-96V(V^2-3U^2).
\end{array}     \right\}        \eqno{(5.6.7)}
$$
 Here the above remarks on the classification of gravitational
 inhomogeneity fields (section 5.5) are clearly applicable.

        For the (anti-)self-dual Weyl tensor (5.5.4) and
 corresponding space-like complex tensors (5.5.5), it is easy
 to find that
$$
U\pm iV=-(r^2+l^2)^{-3}(m\mp il)(r\pm il)^3     \eqno{(5.6.8)}
$$
 ({\em cf.} a superposition of electric point charge and magnetic
 monopole fields in Maxwell's theory). Then
$$
\stackrel{\pm}{X}=\left(
\begin{tabular}{rrr}
$-2$ & 0 & 0 \\
0  & 1 & 0 \\
0  & 0 & 1
\end{tabular}   \right)(U\mp iV).       \eqno{(5.6.9)}
$$
 It is worth mentioning that this type of solution may be
 obtained in a purely linear theory using cohomology and
 twistor methods, {\em cf.} (Hughston 1979).

         As to the description of quasi-electric and quasi-magnetic
 gravitational inhomogeneity fields in a non-rotating
 reference frame, {\em e.g.} that of (5.6.5), it involves them both
 in the Taub--NUT space-time, although they enter not so
 symmetric as it is the case for the rotating frame (see
 (5.6.6)). The transition between the two frames is given by
$$
\tau=K\tilde \tau+L\tilde \theta^{(3)}, ~ ~ \theta^{(3)}=
L\tilde \tau+K\tilde \theta^{(3)},
$$
 the other two basis covectors remaining unchanged. Here
$$
K=[1-4l^2e^{2\alpha}(r^2+l^2)^{-1}\tan^2(\vartheta/2)]^{-1/1},
$$
$$
L=2Kle^\alpha(r^2+l^2)^{-1/2}\tan(\vartheta/2),
$$
 so that $K^2-L^2=1$, the standard Minkowski relation which holds
 in general relativity for an orthonormal basis. Now,
$$
\tilde X_{(1)(1)}=-(2K^2+L^2)U,~ ~ \tilde X_{(2)(2)}=(K^2+2L^2)U,~ ~
\tilde X_{(3)(3)}=U,
$$
$$
\tilde Y_{(1)(1)}=(2K^2+L^2)V,~ ~ \tilde Y_{(2)(2)}=-(K^2+2L^2)V,~ ~
\tilde Y_{(3)(3)}=-V,
$$
 the tensors ramaining diagonal and traceless, while neither of
 the fields may be transformed away by these means. This
 situation shows of course more parallels with the electromagnetic
 field case than the description of gravitation using
 only $G$, $\omega$, and $D$ (see section 5.2), and the general
 transformation laws for $X$ and $G$ can be obtained in strict
 analogy with electromagnetism (see, {\em e.g.}, Landau and Lifshitz
 (1971), although in gravitational case the transformation
 coefficients enter qua\-dratically, as it is above).

\pagestyle{myheadings}
\markboth{CHAPTER 5. THE EINSTEIN FIELD EQUATIONS}{5.7. THE SPINNING
PENCIL-OF-LIGHT FIELD}

\section{Example of the spinning pencil-of-light field} 

         As a pencil of light we understand an infinitesimally
 thin rectilinear energy distribution moving along itself with
 the velocity of light. Thus no electromagnetic field is to be
 considered, and all expressions will be taken outside the
 singular line of the source. In general the intensity of the
 source may vary along the line, these variations propagating
 with the luminal velocity in one and the same direction. The
 first (approximate) study of such a situation was undertaken
 by Tolman, Ehrenfest and Podolsky (1931) (see also (Tolman
 1934)); exact solutions having been found by Peres (1959) and
 Bonnor (1969) (the latter has considered also the case of a
 spinning source, and he first gave a proper physical
 interpretation to the results). Later this subject was
 revisited by Mitskievich (1981) and Mitskievich and Kumaradtya
 (1989) who have studied dragging phenomenon in the pencil-of-light
 field (see also (Mitskievich 1990)), as well as
 additivity properties of such fields. From these papers it is
 clear that the pencil-of-light field possesses both quasi-electric
 and quasi-magnetic properties. Formally it may be
 considered as a gravitational wave field, but, as this is the
 case in the Einstein--Maxwell pp-wave in section 4.5, this
 field can be in particular a stationary one, thus being
 definitively no wave at all.

         The simplest description of the spinning pencil-of-light
 field is connected with a semi-null tetrad,
$$
ds^2=2\theta^{(0)}\theta^{(1)}-\theta^{(2)}\theta^{(2)}
-\theta^{(3)}\theta^{(3)},      \eqno{(5.7.1)}
$$
 where
$$
\left.
\begin{array}{ll}
\stackrel{\rm N~~}{\theta^{(0)}}=dv,~~ &
\stackrel{\rm N~~}{\theta^{(1)}}=du+Fdv+Hd\phi, \\
\stackrel{\rm N~~}{\theta^{(2)}}=d\rho, ~~ &
\stackrel{\rm N~~}{\theta^{(3)}}=\rho d\phi.
\end{array}     \right\}   \eqno{(5.7.2)}
$$
 Here
$$
F=k\ln(\sigma\rho), ~~~ H=H(v),         \eqno{(5.7.3)}
$$
 $H$ being an arbitrary function of $v$ (as also $k$ and $\sigma$ may be).
 The solution is a vacuum one (outside the singular line $\rho=0$)
 so that the only non-trivial curvature components
$$
\stackrel{N}{R}_{(0)(2)(0)(2)}=-\stackrel{N}{R}_{(0)(3)(0)(3)}=
\frac{k}{\rho^2}, ~~~
\stackrel{N}{R}_{(0)(2)(0)(3)}=-\frac{\dot H}{\rho^2}
$$
 are at the same time components of the Weyl tensor.

         For our aims it is better to use an orthonormal tetrad
 which reads
$$
\left.
\begin{array}{ll}
\theta^{(0)}=2^{-1/2}[(1+F)dv+du+Hd\phi],~~ & \theta^{(2)}=d\rho, \\
\theta^{(1)}=2^{-1/2}[(1-F)dv-du-Hd\phi],~~ & \theta^{(3)}=\rho d\phi,
\end{array}     \right\}        \eqno{(5.7.4)}
$$
 and the non-trivial curvature components are
$$
R_{(0)(2)(0)(2)}=R_{(1)(2)(1)(2)}=R_{(0)(2)(1)(2)}=-R_{(0)(3)(0)(3)}
=-R_{(1)(3)(1)(3)}
$$
$$
=-R_{(0)(3)(1)(3)}=k/(2\rho^2),
$$
$$
R_{(0)(2)(0)(3)}=R_{(1)(2)(1)(3)}=R_{(0)(2)(1)(3)}
=R_{(1)(2)(0)(3)}=-\dot H/(2\rho^2).
$$
 Thus in the reference frame with $\tau=\theta^{(0)}$,
$$
\left.
\begin{array}{l}
X{(2)(2)}=-X{(3)(3)}=Y{(2)(3)}=k/(2\rho^2), \\
Y{(2)(2)}=-Y{(3)(3)}=-X{(2)(3)}=\dot H/(2\rho^2),
\end{array}     \right\}
        \eqno{(5.7.5)}
$$
 or for (anti-)self-dual representation,
$$
\stackrel{\pm}{X}=\left(
\begin{tabular}{rrr}
0 & 0 & 0 \\
0 & 1 & $\pm 1$ \\
0 & $\pm i$ & $-1$
\end{tabular}
\right)(k\pm i\dot H)/(2\rho^2)
        \eqno{(5.7.6)}
$$
 ({\em cf.} expression (5.6.9) for the NUT field). Here the numerator
 is an arbitrary complex function of the retarded time $v$. The
 field resembles that of a charged rectilinear wire with a
 current in electrodynamics, but the gravitational case is
 much richer, incorporating in particular such a property as
 angular momentum ("polarization") of the source (in
 electrodynamics this would be probably an effect of the
 Aharonov--Bohm type).

         The monad field $\tau=\theta^{(0)}$ (5.7.4) corresponds
 not only to an acceleration, but also to a rotation,
$$
G=(-k\theta^{(2)}+\dot H\theta^{(3)})/(2\rho),~~~
\omega=-(\dot H\theta^{(2)}+k\theta^{(3)})/(4\rho),
        \eqno{(5.7.7)}
$$
 which are here orthogonal to each other. We see that in this
 representation, as it was the case for (5.7.5) too, the both
 linear and angular momenta of the source (neither of which is
 transformable away by switching to frames in translational
 and/or rotational motion), contribute to the both quasi-electric
 and quasi-magnetic parts of the total gravitational
 field. But in this approach characteristic to section 5.2, the
 principle of equivalence permits to get rid of rotation or
 acceleration, and moreover, not only locally, but also in a
 finite region (we speak then correspondingly on either a
 normal, or geodesic congruence), so that only in the
 gravitational inhomogeneity field approach, there is guaranteed
 the invariant property of a field to be of quasi-electric,
 quasi-magnetic, or null type, independent of a concrete choice
 of reference frame. This shows once more that for gravitational
 fields the feature of inhomogeneity plays a crucial role.
 We may thus choose a non-rotating monad congruence, {\em e.g.} as
 $\tau=N(dv+du)$, $N^{-2}=2(1-k\ln(\sigma\rho))-(H/\rho)^2$, but this only
 makes the expressions for $X$ and $Y$ unpleasant, so we shall
 not consider here this subject in more detail. It is however
 worth mentioning that a non-rotating frame can be simultaneously
 introduced in the space-time of a pencil of light only
 in a limited band of the radial coordinate $\rho$. This is the so
 called local stationarity property known also for the
 ergosphere region of a Kerr black hole ({\em cf.} Hawking and Ellis
 (1973)). This limited band can be moved arbitrarily, but it
 cannot be expanded to fill the whole space-time of a pencil of
 light at once.

\pagestyle{headings}

\section{Gravitational fields of the G\"odel universe}        

         Finally we come to the G\"odel space-time (G\"odel 1949)
 which has certain remarkable properties. We consider its
 metric in the form
$$
ds^2=a^2[(dt+\sqrt{2}zdx)^2-z^2dx^2-dy^2-z^{-2}dz^2],  \eqno{(5.8.1)}
$$
 with the corresponding natural orthonormal basis. It is clear
 that the co-moving tetrad is $\tau=a(dt+\sqrt{2}zdx)$ which is
 rotating but geodesic ($G=0$, $\omega=2^{-1/2}dy$), while the
 stationarity yields absence of deformations ($D_{\alpha\beta}=0$). We
 assume here that the cosmological constant is equal to zero,
 so that $\mu=p=(2\kappa a^2)^{-1}$ (see, {\em e.g.},
 (Stephani 1990)). Then
 the intrinsic gravitational inhomogeneity field is very
 simple,
$$
-2X_{(1)(1)}=X_{(2)(2)}=-2X_{(3)(3)}=(6a^2)^{-1},  ~~~
Y_{(\alpha)(\beta)}=0.  \eqno{(5.8.2)}
$$
 Thus there exists no quasi-magnetic gravitational field in the
 G\"odel universe, although this is rotating (in a sharp contrast
 with the electromagnetic field in generalizations of this
 universe where a charged fluid does not produce any electric
 field in its co-moving frame, see section 4.4).

         As to the gravitational inhomogeneity field sources, they
 may be seen from the quasi-Maxwellian equations (5.4.3) with a
 substitution of (5.4.8)---(5.4.11) and (5.4.16)---(5.4.19):
$$
{X_{\alpha\sigma}}^{;\alpha}b^\sigma_\delta=
(\kappa/3)\mu_{,\sigma}b^\sigma_\delta, \eqno{(5.8.3)}
$$
$$
-3\omega^\alpha X_{\alpha\delta}=\kappa(\mu+p)\omega_\delta,
        \eqno{(5.8.4)}
$$
$$
\frac{1}{2}\omega^\gamma(X^\alpha_\beta e_{\alpha\delta\gamma}
+X^\alpha_\delta e_{\alpha\beta\gamma})=0,      \eqno{(5.8.5)}
$$
 while the equation corresponding to (5.4.11),
 reduces to $0\equiv 0$. Equations (5.8.3)---(5.8.5) are satisfied
 also indeed. We have already excluded from them terms being
 identically equal to zero, this is the reason why (5.8.4) and
 (5.8.5) are algebraic with respect to $X$. In fact (5.8.4)
 contains an analogue of ${\rm div}\; B$ in Maxwell's equations, and it
 shows how the other terms mutually compensate to permit $Y$ to
 vanish completely. Similarly, (5.8.5) corresponds to
 Maxwell's equation with ${\rm curl}\; B$.

\newpage

\pagestyle{myheadings}
\markboth{CHAPTER 6. CONCLUDING REMARKS}{CHAPTER 6. CONCLUDING REMARKS}

\chapter{Concluding remarks}  

          In this paper it was of course impossible  to  sensibly
characterize the whole variety of use of reference frames  in
general relativity, uncluding their new development tendencies.
 We have deliberately ignored  here  the  case  of  normal
congruences (theory of holonomic hypersurfaces) which  is  so
important in building  the  canonical  formalism  in  general
relativity to the end  of  gravity  quantization.  Elementary
ideas of quantization are discussed in (Mitskievich 1969) and
 fundamentals of the canonical formalism, in  (Misner, Thorne and
 Wheeler 1973, Mitskievich, Yefremov and Nesterov 1985).
 It is however worth mentioning  here  some  new  works  which
 could be helpful  in  understanding  the  recent  trends  and
 developments in these areas.

         Much difficulty in realization of  canonical  formalism
 in field theory is due to the problem of equations of constraints
 (see Mitskievich, Yefremov and Nesterov 1985).  With
 help of gauge theory ideas, and in particular  those  of  the
 Yang--Mills theory, Ashtekar (see Ashtekar 1988a)  succeeded
 in making exceptionally good choice of field canonical  variables
 and applying them to different types of problems  ({\em e.g.}
 Ashtekar 1986, Ashtekar 1988b, Ashtekar 1989). One of trends
 in the development of this theory consisted in a study of spaces
 with self-dual Weyl tensor (Koshti and Dadhich 1990)  for
 which a set of general theorems were obtained (in a more conventional
 approach, such spaces are often called ``heavenly spaces'' (or
 ${\cal H}$-spaces); they were studied and applied to solving
 Einstein's equation, especially by E. Newman, J. Pleba\'nski, and
 D. Finley). The very foundations
 of the gravitational field theory were also considered,
 and among them the action  principle  with  the  Palatini
 approach (Capovilla, Dell, Jacobson and Mason 1991,
 Floreanini and Percacci 1990)
 and the covariant action  formulation  (Jacobson  and  Smolin
 1988). Sometimes new fruitful developments were  achieved  in
 the theory of initial data using conformally flat spaces  and
 space-time complexification (Wagh and Saraykar 1989).

         The attention was focused  in  this  direction  however
 mainly to the canonical formalism in gravitation theory where
 vacuum constraints were brought  to  a  solvable  formulation
 (Robinson and Soteriou 1990)  and  canonical  transformations
 were introduced together with the corresponding generators in
 the field  theory  (Dolan  1989).  Ashtekar's  formalism  was
 connected with the  tetrad  approach  (Henneaux, Nelson and Schomblond
 1989), then a first order tetrad form was introduced
 for description of the action principle  and  formulation
 of the canonical theory (Kamimura  and  Fukayama  1990),  and
 elegant results were obtained in  a  further  use  of  three-spinors
 in Ashtekar's formalism (Perj\'es 1990).

         Several papers, including the initial works of Ashtekar
 himself, were dedicated to development  of  methods  of
 3+1-splitting of space-time (Ashtekar 1988b,  Ashtekar, Jacobson
 and Smolin 1988, Wallner 1990). On  basis  of  the  canonical
 formalism were also, quite naturally, considered  the  energy
 problems ({\em e.g.}, its localizability) (Nester 1991).

         Another  branch  of  these  studies  was  directed   to
 obtaining new solutions of the gravitational field  equations
 with help of specific choices of Ashtekar's canonical variables.
 Hereby solutions for gravitational instantons were found
 (Samuel  1988),  spherically  symmetric  cosmological  models
 (Koshti and  Dadhich  1989)  as  well  as  a  wide  class  of
 spherically symmetric problems of general relativity  studied
 (Bengtsson  1990),  while  in  the  latter  paper  a  general
 discussion of further prospects in applications of Ashtekar's
 formalism  was  given  being  of   considerable   independent
 interest. The formalism was applied  to  the  Kantowski-Sachs
 type solutions with a perfect fluid  (Bombelli  and  Torrence
 1990) as well as Bianchi types  cosmological  models  (Kodama
 1988) (here quantization of these models was  also  performed
 using Ashtekar's canonical variables). It is significant that
 Ashtakar's formalism does not in general imply non-degeneracy
 of the  metric  tensor,  so  that  perfectly  new  solutions,
 inadmissible from the standpoint of the  traditional  theory,
 enter scope of the study. For example, several solutions with
 massive scalar field were thus obtained without explicit  use
 of a metric (Peld\'an 1990).

         As we see, even a single, though explosion-like evolving
 in the last years trend of application of  reference  frames,
 has produced by itself a great number of fine  results.  This
 is not always the reference frame trend proper, of  the  type
 which was discussed below in our paper. The authors go  often
 beyond the limits of physical reference frames while considering
 either complexified spaces (such as self-dual ones which
 may have both coordinated, tetrad or spinor representations),
 or null congruences, or else. But  in  all  the  cases,
 far-reaching analogies between the theories  of  gravitation  and
 electromagnetism  (or  Yang-Mills  theory)  emerged  clearly,
 which were often used by the very authors  for  determination
 of their own  research  directions  and  formulation  of  new
 approaches. This sets our hopes on the  fruitfulness  of  the
 parallel  description  of  electromagnetism  and  gravitation,
 however fragmentary and concise it could
 be done here. It is worth repeating that we tried to give here a review
 of a part of the reference frames  formalism  from  the  side
 being used, successfully enough, for description of observables
 and interpretation of physical effects which are calculated
 initially, as a rule, in a four-dimensional form  without
 any relation to specific reference frames, so that  further
 interpretation efforts are unavoidable.

        Returning to the presentation of reference frames formalism
 given in this paper, we feel it necessary to emphasize that
 it differs from previous approaches mainly in a more consistent
 use of modern differential geometry techniques. At the same
 time, new and more convenient notations were introduced which
 are as near to the conventional ones of the old three-dimensional
 vector calculus as it was possible to achieve. It is even possible
 that the reader could first take our notations for those conventional
 ones and consider them to not be general covariant. As a matter of
 fact, they however {\em are} covariant under arbitrary transformations of
 coordinates, and the distinction between different reference frames
 is to be seen in the use of a concrete monad $\tau$ which is of course
 a real four-vector (or covector: the coincident notations of the both
 cannot create any confusion). The advantage of such notations (as
 div or curl) is in greatly facilitated interpretation possibilities
 when one considers equations and effects of relativistic physics in
 arbitrary reference frames.

\newpage

\pagestyle{myheadings}
\markboth{REFERENCES}{REFERENCES}

\addcontentsline{toc}{chapter}{References}
\chapter*{References}
\begin{itemize}
\item[~~] Anderson, J.L. (1958a) {\it Phys. Rev.} {\bf  110}, 1197.
\item[~~] Anderson, J.L. (1958b) {\it Phys. Rev.} {\bf 111}, 965.
\item[~~] Anderson, J.L. (1959) {\it Phys. Rev.} {\bf 114}, 1182.
\item[~~] Ashby, Neil, and Shahid-Saless, Bahman (1990) {\it Phys. Rev. D}
{\bf 42}, 1118.
\item[~~] Ashtekar, A. (1986) {\it Phys. Rev. Lett.} {\bf 57}, 2244.
\item[~~] Ashtekar,  A.  (1988a)  {\it New  Perspectives  in  Canonical
Gravity} (Napoli: Bibliopolis).
\item[~~] Ashtekar, A. (1988b) {\it Contemporary Mathematics} {\bf 71}, 39.
\item[~~] Ashtekar, A. (1989) In: {\it Proc.  9th  Int.  Congr.  Math.
Phys., Swansea, 17-27 July, 1988} (Bristol, New York: AMS), p. 286.
\item[~~] Ashtekar, A.,  Jacobson,  T.,  and  Smolin,  L.  (1988)
{\it Commun. Math. Phys.}  {\bf 115}, 631.
\item[~~] Bel, L. (1959) {\it Compt. Rend. Paris} {\bf 248}, 2161.
\item[~~] Bengtsson, I. (1990) {\it Class. Quantum Grav.} {\bf 7}, 27.
\item[~~] Bohr,  N.,  and  Rosenfeld,  L.  (1933)   {\it Kgl.   Danske
Videnskab. Selskab., Mat.-fys. Medd.} {\bf 12}, 8.
\item[~~] Bombelli, L., and Torrence, R.J. (1990) {\it Class.  Quantum
Grav.} {\bf 7}, 1747.
\item[~~] Bonnor, W.B. (1969) {\it Commun. Math. Phys.} {\bf 13}, 163.
\item[~~] Bowler, M.G. (1976) {\it Gravitation and Relativity} (Oxford:
Pergamon).
\item[~~] Boyer, T.H. (1980) In: {\it Foundations of Radiation  Theory
and Quantum Electrodynamics} (New York, London: Plenum), p. 49.
\item[~~] Brill, D.  (1972)  In:  {\it Methods  of  Local  and  Global
Differential  Geometry   in  General   Relativity}   (Berlin:
Springer), p. 45.
\item[~~] Brillouin, L. (1970) {\it Relativity Reexamined}  (New  York:
Academic Press).
\item[~~] Capovilla, R., Dell, J., Jacobson, T.,  and  Mason,  L.
(1991) {\it Class. Quantum  Grav.} {\bf 8}, 41.
\item[~~] Cattaneo, C. (1958) {\it Nuovo Cim.} {\bf 10}, 318.
\item[~~] Cattaneo, C. (1959) {\it Ann. mat. pura e appl.} {\bf 48}, 361.
\item[~~] Cattaneo, C. (1961) {\it Rend. mat. e appl.} {\bf 20}, 18.
\item[~~] Cattaneo, C. (1962) {\it Rend. mat. e appl.} {\bf 21}, 373.
\item[~~] Choquet-Bruhat, Y., DeWitt-Morette,  C.,  and  Dillard-Bleick,  M.
(1982)  {\it Analysis,   Manifolds,   and   Physics}
(Amsterdam: North-Holland).
\item[~~] Debever, R. (1959) {\it Comptes Rendus Paris} {\bf 249}, 1744.
\item[~~] Dehnen, H. (1962) {\it Zs. Naturforsch.} {\bf 17a}, 18.
\item[~~] Dehnen, H., H\"onl, H.,  and  Westpfahl,  K.  (1961)  {\it Zs.
Physik} {\bf 164}, 483.
\item[~~] DeWitt, B.S. (1965)  {\it Dynamical  Theory  of  Groups  and
Fields} (New York:   Gordon and Breach).
\item[~~] DeWitt, B.S. (1966) {\it Phys. Rev. Lett.} {\bf 16}, 1092.
\item[~~] Dolan, B.P. (1989) {\it Phys. Lett.} {\bf B 233}, 89.
\item[~~] Eckart, C. (1940) {\it Phys. Rev.} {\bf 58}, 919.
\item[~~] Eguchi, T., Gilkey, P.B., and Hanson, A.J. (1980) {\it Phys.
Reports} {\bf 66}, 213.
\item[~~] Ehlers, J. (1961) {\it Akad. Wiss. Lit. Mainz, Abhandl.
Math.-Nat. Kl.} {\bf 11}
\item[~~] Ehlers, J. (1993) {\it Gen. Relat. and Grav.} {\bf 25},
1225.
\item[~~] Eisenhart, L.P. (1926) {\it Riemannian  Geometry}  (Princeton,
N.J.: Princeton   University Press).
\item[~~] Eisenhart,   L.P.   (1933)   {\it Continuous    Groups    of
Transformations}  (Princeton,  N.J.:   Princeton   University
Press).
\item[~~] Eisenhart,   L.P.   (1972)    {\it Non-Riemannian    Geometry}
(Providence, RI: American Mathematical Society).
\item[~~] Elst, H. van, and Ellis, G.F.R. (1996) {Class. Quantum Grav.}
{\bf 13}, 1099.
\item[~~] Epikhin, E.N., Pulido, I., and Mitskievich, N.V. (1972) In:
{\it Abstracts of the reports presented at the 3rd Soviet
Gravitational Conference} (Erevan: University Press), p. 380.
\item[~~] Floreanini, R., and Percacci, R. (1990) {\it Class.  Quantum
Grav.} {\bf 7}, 1805.
\item[~~] Fock, V.A.  (1964)  {\it The  Theory  of  Space,  Time,  and
Gravitation} (New York:   MacMillan).
\item[~~] Fock, V.A. (1971) {\it Voprosy  filisofii},  No.  3,  46.  In
Russian.
\item[~~] G\'eh\'eniau, J. (1957) {\it Comptes Rendus Paris} {\bf 244}, 723.
\item[~~] G\"odel, K. (1949) {\it Revs. Mod. Phys.} {\bf 21}, 447.
\item[~~] Goldstein,  H.  (1965)  {\it Classical  Mechanics}  (Reading,
London: Addison--Wes\-ley).
\item[~~] Goldstein,  S.  (1976)  {\it Lectures  in  Fluid   Mechanics}
(Providence, R.I.: AMS).
\item[~~] Gorbatsievich, A.K. (1985) {\it Quantum Mechanics in General
Relativity} (Minsk:   Universitetskoie). In Russian.
\item[~~] Hawking, S.W., and Ellis, G.F.R. (1973) {\it The Large Scale
Structure  of  Space-Time}  (Cambridge:  Cambridge  University
Press).
\item[~~] Henneaux, M., Nelson, J.E., and Schomblond,  C.  (1989)
{\it Phys. Rev.} {\bf D 39}, 434.
\item[~~] H\"onl, H., and Soergel-Fabricius, Chr. (1961) {\it Zs.  Phys.}
{\bf 163}, 571.
\item[~~] Hughston, L.P. (1979) {\it Twistors and  Particles}  (Berlin,
Heidelberg, New York:   Springer-Verlag).
\item[~~] Islam,  J.N.  (1985)   {\it Rotating   fields   in   general
relativity} (Cambridge, Cambridge University Press).
\item[~~] Israel, W. (1970) {\it Commun. Dublin Inst. Adv.  Studies}  {\bf A
19}, 1.
\item[~~] Ivanitskaya,   O.S.    (1969)    {\it Generalized    Lorentz
Transformations  and  Their  Applications}  (Minsk:  Nauka   i
tekhnika). In Russian.
\item[~~] Ivanitskaya,   O.S.   (1979)   {\it Lorentzian   Basis   and
Gravitational  Effects  in  Einstein's   Gravitation   Theory}
(Minsk: Nauka i tekhnika) In Russian.
\item[~~] Ivanitskaya, O.S., Mitskievich, N.V.,  and  Vladimirov,
Yu.S. (1985) {\it Reference Frames in General Relativity.} Preprint
No. 374, Inst. of Physics,  Acad. Sci. BSSR, Minsk.
\item[~~] Ivanitskaya, O.S., Mitskievich, N.V.,  and  Vladimirov,
Yu.S.  (1986)  In:  {\it Relativity in  Celestial  Mechanics  and
Astrometry} (Dordrecht: Reidel). Pp.  177-186.
\item[~~] Jacobson, T., and  Smolin,  L.  (1988)  {\it Class.  Quantum
Grav.} {\bf 5}, 583.
\item[~~] Jantzen, R.T., Carini, P., and Bini, D. (1992)
{\it Ann. Phys. (USA)} {\bf 219}, 1.
\item[~~] Kamimura, K., and Fukayama, T. (1990) {\it Phys. Rev.} {\bf D  41},
1885.
\item[~~] Khaikin,  S.E.  (1947)  {\it Mechanics,  2nd  ed.}   (Moscow,
Leningrad: GITTL). In   Russian.
\item[~~] Kodama, H. (1988) {\it Prog. Theoret. Phys.} {\bf 80}, 1024.
\item[~~] Koshti, S., and Dadhich, N. (1989) {\it Class. Quantum Grav.}
{\bf 6}, L223.
\item[~~] Koshti, S., and Dadhich, N. (1990) {\it Class. Quantum Grav.}
{\bf 7}, L5.
\item[~~] Kramer, D., Stephani, H., MacCallum M., and  Herlt,  E.
(1980) {\it Exact Solutions of Einstein's Field Equations} (Berlin:
Deutscher Verlag der Wissenschaften; Cambridge, UK: Cambridge
University Press).
\item[~~] Lanczos, C. (1938) {\it Ann. Math.} {\bf 39}, 842.
\item[~~] Lanczos, C. (1962) {\it Revs. Mod. Phys.} {\bf 34}, 379.
\item[~~] Landau, L.D., and Lifshitz, E.M. (1971) {\it Classical Field
Theory} (Reading,  Mass.: Addison-Wesley).
\item[~~] Lichnerowicz, A. (1955)  {\it Th\'eories  relativistes  de  la
gravitation et de l'\'elec\-tro\-mag\-n\'e\-tisme} (Paris: Masson).
\item[~~] Lichnerowicz, A. (1960) {\it Ann. di Mat. Pura e Appl.} {\bf 50}, 1.
\item[~~] Lief, B. (1951) {\it Phys. Rev.} {\bf 84}, 345.
\item[~~] Lovelock, D. (1971) {\it J. Math. Phys.} {\bf 12}, 498.
\item[~~] MacCallum, M.A.H. (1979) In: {\it General Relativity: An
Einstein Centenary Survey} ed. S.W. Hawking and W. Israel (Cambridge:
Cambridge University Press).
\item[~~] Macdonald,  D., and Thorne, K.S.  (1982)  {\it Monthly  Not.
Roy. Astr. Soc.} {\bf 198},  345.
\item[~~] Marck, J.-A. (1983) {\it Proc. Roy. Soc. London A} {\bf 385}, 431.
\item[~~] Massa, E. (1974) {\it Gen. Relat. and Grav.} {\bf 5}, 555.
\item[~~] Massa, E. (1974) {\it Gen. Relat. and Grav.} {\bf 5}, 573.
\item[~~] Massa, E. (1974) {\it Gen. Relat. and Grav.} {\bf 5}, 715.
\item[~~] Matte, A. (1953) {\it Canad. J. Math.} {\bf 5}, 1.
\item[~~] Meier, W., and Sali\'e, N. (1979) {\it Theoret. and Math. Phys.}
{\bf 38}, 408. In Russian.
\item[~~] Misner, C.W., Thorne, K.S., and  Wheeler,  J.A.  (1973)
{\it Gravitation} (San   Francisco: W.H. Freeman).
\item[~~] Mitskievich, N.V. (1969)  {\it Physical  Fields  in  General
Relativity} (Moscow:  Nauka). In Russian.
\item[~~] Mitskievich, N.V.  (1972)  In:  {\it Einsteinian  Collection
1971} (Moscow: Nauka),  p. 67. In Russian.
\item[~~] Mitskievich, N.V. (1975) In:  {\it Problems  of  Gravitation
Theory} (Erevan: University Press), p. 104. In Russian.
\item[~~] Mitskievich, N.V. (1976)  In:  {\it Problems  of  Theory  of
Gravitation and  Elementary  Particles}  (Moscow:  Atomizdat).
Fasc. 7, p. 15. In Russian.
\item[~~] Mitskievich, N.V. (1979) {\it Supplementary Chapter}  to  the
Russian Translation  (Moscow: Mir) of the Book: Bowler (1976).
In Russian.
\item[~~] Mitskievich, N.V. (1981a) {\it Experim. Technik d. Phys.}  {\bf 29},
213.
\item[~~] Mitskievich, N.V. (1981b) In: {\it Abstracts of the reports
presented
at the 5th Soviet Gravitational Conference} (Moscow: Moscow University
Press), p. 43. In Russian.
\item[~~] Mitskievich,  N.V.   (1983)   {\it Proc.   Einstein   Found.
Internat.} {\bf 1}, 137.
\item[~~] Mitskievich, N.V. (1989)  In:  {\it Gravitation  and  Waves.}
Transactions of Inst. of Physics, Estonian Acad. Sci.,  Tartu
{\bf 65}, p. 104.
\item[~~] Mitskievich, N.V. (1990) In: {\it Differential Geometry  and
Its Applications} (Singapore: World Scientific), p. 297.
\item[~~] Mitskievich, N.V., and Cindra, J.L. (1988) In: {\it Current problems of
Quantum Mechanics and Statistical Physics} (Moscow: Peoples' Friendship
University Press), p. 89. In Russian.
\item[~~] Mitskievich, N.V., and Gupta, Salil (1980) In: {\it Abstracts of
contributed papers, 9th International Conference of GRG} (Jena:
Friedrich Schiller University Press), p. 190.
\item[~~] Mitskievich,  N.V.,  and  Kalev,  D.A.  (1975)  {\it Comptes
Rendus Acad. Bulg. Sci.} {\bf 28}, 735. In Russian.
\item[~~] Mitskievich, N.V., and Kumaradtya, K.K. (1989) {\it J. Math.
Phys.} {\bf 30}, 1095.
\item[~~] Mitskievich, N.V., and  Merkulov,  S.A.  (1985)  {\it Tensor
Calculus  in  Field  Theory}  (Moscow:   Peoples'   Friendship
University Press). In Russian.
\item[~~] Mitskievich, N.V.,  and  Mujica,  J.D.  (1968)  {\it Doklady
Akad. Nauk SSSR} {\bf 176}, 809. In Russian.
\item[~~] Mitskievich, N.V., and Nesterov,  A.I.  ((1981)  {\it Exper.
Techn. der Physik} {\bf 29}, 333.
\item[~~] Mitskievich, N.V., and  Nesterov,  A.I.  (1991)  {\it Class.
Quantum Grav.} {\bf 8}, L45.
\item[~~] Mitskievich, N.V., and Pulido, I. (1970) {\it Doklady  Akad.  Nauk
SSSR}  {\bf 192}, 1263. In Russian.
\item[~~] Mitskievich, N.V., and Ribeiro Teodoro, M. (1969)  {\it Sov.
Phys. JETP} {\bf 29},   515.
\item[~~] Mitskievich, N.V., and  Tsalakou,  G.A.  (1991)  {\it Class.
Quantum Grav.} {\bf 8}, 209.
\item[~~] Mitskievich, N.V., and Uldin, A.V. (1983) {\it A gravitational lens
in the Kerr field, Deponent No. 2654-83} (Moscow: Institute of
Scientific and Technical Information, Academy of Sciences of USSR)
In Russian.
\item[~~] Mitskievich, N.V., Yefremov, A.P., and  Nesterov,  A.I.
(1985). {\it Dynamics of Fields  in  General  Relativity}  (Moscow:
Energoatomizdat) In Russian.
\item[~~] Mitskievich, N.V., and  Zaharow,  V.N.  (1970)  {\it Doklady
Akad. Nauk SSSR}  {\bf 195}, 321. In Russian.
\item[~~] Nester, J.M. (1991) {\it Class. Quantum Grav.} {\bf 8}, L19.
\item[~~] Newton, Sir Isaac  (1962)  {\it Mathematical  Principles  of
Natural Philosophy.} (Ed. and comm. by Florian Cajori.) Vol. 1
\& 2. (Berkeley and  Los Angeles,  University  of  California
Press).
\item[~~] Noether, E (1918) {\it G\"otting. Nachr.}, 235.
\item[~~] Nordtvedt, Ken (1988) {\it Int. J. Theoret. Phys.} {\bf 27}, 1395.
\item[~~] Palatini, A. (1919) {\it Rend. circolo mat. Palermo} {\bf 43}, 203.
\item[~~] Papini, G. (1966) {\it Physics Letters} {\bf 23}, 418.
\item[~~] Papini, G. (1969) {\it Nuovo Cim.} {\bf 63 B}, 549.
\item[~~] Peld\'an, P. (1990) {\it Phys. Lett.} {\bf B 248}, 62.
\item[~~] Penrose, R. (1960) {\it Ann. Phys. (USA)} {\bf 10}, 171.
\item[~~] Penrose,  R.,  and  Rindler,  W.  (1984a)  {\it Spinors  and
Space-Time. Vol.  1.  Two-Spinor  Calculus  and  Relativistic
Fields}  (Cambridge: Cambridge University Press).
\item[~~] Penrose,  R.,  and  Rindler,  W.  (1984b)  {\it Spinors  and
Space-Time. Vol. 2. Spinor and Twistor Methods in  Space-Time
Geometry.} (Cambridge: Cambridge University Press).
\item[~~] Peres, A. (1959) {\it Phys. Rev. Lett.} {\bf 3}, 571.
\item[~~] Perj\'es, Z. (1990) {\it The Parametric  Manifold  Picture  of
Space-Time.} Preprint, Central  Research  Inst.  for  Physics,
Budapest, Hungary.
\item[~~] Petrov, A.Z. (1966) {\it New Methods in General
Theory of Relativity} (Moscow: Nauka). In Russian.
\item[~~] Pirani, F.A.E. (1957) {\it Phys. Rev.} {\bf 105}, 1089.
\item[~~] Pirani, F.A.E, and Schild, A. (1961) {\it Bulletin de l'Acad\'emie
Polonaise des Sciences, S\'erie de sciences math., astr. et phys.}
{\bf IX}, No. 7, 543.
\item[~~] Regge, T. (1958) {\it Nuovo Cimento} {\bf 7}, 215.
\item[~~] Robinson, D.C., and Soteriou, C. (1990) {\it Class.  Quantum
Grav.} {\bf 7}, L247.
\item[~~] Rodichev, V.I. (1972) In: {\it Einsteinian  Collection  1971}
(Moscow: Nauka). In  Russian.
\item[~~] Rodichev,  V.I.  (1974)  {\it Theory   of   Gravitation   in
Orthonormal Tetrad Frames}  (Moscow: Nauka). In Russian.
\item[~~] Ryan,  M.P.,  and  Shepley,  L.C.  (1975)   {\it Homogeneous
Relativistic   Cosmologies}   (Princeton,   N.J.:    Princeton
University Press).
\item[~~] Sachs, R.K., and Wu, H. (1977) {\it General  Relativity  for
Mathematicians} (New  York, Heidelberg, Berlin: Springer).
\item[~~] Sakina, K., and Chiba, J. (1980) {\it Lett. N. Cim.} {\bf 27}, 184.
\item[~~] Sali\'e, N. (1986) {\it Astron. Nachr.} {\bf 307}, 335.
\item[~~] Samuel, J. (1988) {\it Class. Quantum Grav.} {\bf 5}, L123.
\item[~~] Schmutzer, E. (1968) {\it Relativistische Physik (klassi\-sche
Theorie)} (Leipzig: Teubner).
\item[~~] Schmutzer, E. (1975) {\it Scr. Fac. Sci. Nat. UJEP Brunensis,
Physica} {\bf 3--4, 5}, 279.
\item[~~] Schmutzer, E. (1989) {\it Grundlagen der Theoretischen
Physik} (Mannheim, Wi\-en, Z\"urich:
BI-Wis\-sen\-schafts\-ver\-lag), Teile 1 \& 2.
\item[~~] Schmutzer, E., and Pleba\'nski, J.F. (1977) {\it Fortschr.  d.
Physik} {\bf 25}, 37.
\item[~~] Schmutzer, E., und Sch\"utz, W. (1983)  {\it Galileo  Galilei},
5. Auflage (Leipzig:  Teubner).
\item[~~] Schouten, I.A., and Struik, D.J. (1935)  {\it Einf\"uhrung  in
die  neueren  Methoden  der  Differentialgeometrie.}  Vol.   1
(Gronongen: Noordhoff).
\item[~~] Schouten, I.A., and Struik, D.J. (1938)  {\it Einf\"uhrung  in
die  neueren  Methoden  der  Differentialgeometrie.}  Vol.   2
(Gronongen: Noordhoff).
\item[~~] Schr\"odinger, E. (1956) {\it Expanding Universes}  (Cambridge:
Cambridge University Press).
\item[~~] Shteingrad, Z.A. (1974) {\it Doklady Akad.  Nauk  SSSR}  {\bf 217},
1296. In Russian.
\item[~~] Stephani,  H.  (1990)  {\it General  Relativity}  (Cambridge:
Cambridge University Press).
\item[~~] Synge,  J.L.  (1960)  {\it Relativity:  The  General  Theory}
(Amsterdam: North-Hol\-land).
\item[~~] Synge,  J.L.  (1965)  {\it Relativity:  The  Special  Theory}
(Amsterdam: North-Hol\-land).
\item[~~] Synge, J.L.  (1974)  {\it Hermathena  (a  Dublin  University
review)}. No. cxvii (without pagination).
\item[~~] Taub, A.H. (1978) {\it Ann. Rev. Fluid Mech.} {\bf 10}, 301.
\item[~~] Thorne, K.S., and Macdonald,  D.  (1982)  {\it Monthly  Not.
Roy. Astr. Soc.} {\bf 198}, 339 (Microfiche MN 198/1).
\item[~~] Thorne, K.S., Price, R.H., and Macdonald,  D.A.  (1986)
editors, {\it Black Holes: The Membrane Paradigm} (New  Haven,  CT:
Yale University Press).
\item[~~] Tolman, R.C.  (1934) {\it  Relativity,  Thermodynamics,  and
    Cosmology} (Oxford: Clarendon Press).
\item[~~] Tolman, R.C., Ehrenfest, P., and Podolsky, B. (1931)
{\it Phys. Rev} {\bf 37}, 602.
\item[~~] Torres del Castillo, G.F. (1992) {\it Rev. Mex. F\'\i s.} {\bf 38}, 484.
\item[~~] Trautman, A.(1956) {\it Bulletin de l'Acad\'emie
Polonaise des Sciences, S\'erie de sciences math., astr. et phys.}
{\bf IV}, 665 \& 671.
\item[~~] Trautman, A. (1957) {\it Bulletin de l'Acad\'emie
Polonaise des Sciences, S\'erie de sciences math., astr. et phys.}
{\bf V}, 721.
\item[~~] Tsoubelis, D., and Economou, A. (1988) {\it Gen. Relat. and Gravit.}
{\bf 20}, 37.
\item[~~] Tsoubelis, D., Economou, A., and Stoghianidis, E. (1987)
{\it Phys. Rev. D} {\bf 36}, 1045.
\item[~~] Vladimirov,   Yu.S.   (1982)   {\it Reference   Frames    in
Gravitation Theory} (Moscow: Energoizdat). In Russian.
\item[~~] Vladimirov, Yu.S., Mitskievich, N.V.,  and  Horsky,  J.
(1987)  {\it Space,  Time, Gravitation}  (Moscow:  Mir).   Revised
English translation of a Russian edition  of  1984  (Moscow:
Nauka).
\item[~~] Wagh, S.M., and Saraykar, R.V. (1989) {\it Phys. Rev.} {\bf D  39},
670.
\item[~~] Wallner, R.P. (1990) {\it Phys. Rev.} {\bf D 42}, 441.
\item[~~] Westenholz, C. von (1986)  {\it Differential  Forms   in
Mathematical Physics} (Amsterdam: North-Holland).
\item[~~] Wheeler,  J.A.  (1962)  {\it Geometrodynamics}   (New   York:
Academic Press).
\item[~~] Wheeler, J.A. (1993) {\it Private communication}.
\item[~~] Yano, K. (1955) {\it The Theory of Lie Derivatives  and  Its
Applications} (Amsterdam: North-Holland).
\item[~~] Zakharov, V.D. (1973) {\it Gravitational Waves in Einstein's
Theory} (New York: Halsted Press).
\item[~~] Zel'manov, A.L. (1956) {\it Doklady  Akad.  Nauk  SSSR}  {\bf  107},
815. In Russian.
\item[~~] Zel'manov, A.L. (1959) {\it Doklady  Akad.  Nauk  SSSR}  {\bf 124},
1030.  In Russian.
\item[~~] Zel'manov, A.L. (1973) {\it Doklady  Akad.  Nauk  SSSR}  {\bf 209},
822.  In Russian.
\item[~~] Zel'manov, A.L. (1976) {\it Doklady Akad. Nauk SSSR} {\bf 227}, 78.
In Russian.
\item[~~] Zel'manov, A.L., and Agakov, V.G.  (1989)  {\it Elements  of
General Theory of Relativity} (Moscow: Nauka). In Russian.
\end{itemize}

\newpage


\begin{theindex}
\addcontentsline{toc}{chapter}{Index}

\item acceleration of reference frame  32, 33
\item ADM formalism  22
\item Aharonov-Bohm type effect 106
\item analogy between gravitation and electromagnetism 50, 91, 101
\item Ashtekar's formalism 110ff
\indexspace
\item Bach brackets 5
\item Bach--Lanczos invariant 8
\item basis of a $p$-form 5
\item Bianchi identity 12, 13
\item black-body radiation  25
\indexspace
\item canonical
	\subitem formalism  26, 109
	\subitem variables 109
\item Cartan's forms  12
\item Cauchy problem  39, 91
\item centrifugal force 24, 81
\item charge density 79
	\subitem in a frame  47
	\subitem invariant  46
\item charges, effective (kinematic) 80, 86
\item Christoffel symbols  9
\item chronometric invariants  21, 30, 31
\item classification
	\subitem of electromagnetic fields 77
	\subitem of gravitational fields 101
\item Codazzi equations, generalized  39, 90
\item coefficients $T_a|^\tau_\sigma$ 10
\item connection
	\subitem coefficients  9
	\subitem forms  12
\item constraints 109
\item cosmological red shift 54
\item covariant
	\subitem derivative (;) 10
	\subitem differentiation ($\nabla$) 8
\item co-moving frame  45
\item Coriolis force 24, 81
\item crafty identities  7, 75, 92
\item crossing a horizon 56
\item curl 36, 78
\item curl of acceleration  37
\item curvature
	\subitem operator  11
	\subitem tensor  11
\indexspace
\item de-Rhamian operator  16
\item differentiation operator $\delta$ 16
\item dilatation  33
\item div curl  37
\item divergence  16, 78
\item Doppler
	\subitem effect analogue 78
	\subitem shift  52
\item dragging
	\subitem generalized manifestations of 72
	\subitem in the Kerr field 60
	\subitem in the NUT field 61
	\subitem in the pencil-of-light field 62ff
	\subitem of electromagnetic fields in conductors and
			superconductors 70ff
	\subitem of dragging 64
\item dual conjugation  6
\indexspace
\item Einstein's gravitational field equations  88
\item electric
	\subitem field strength  44
	\subitem type fields 77
\item electromagnetic
	\subitem field
		\subsubitem invariants  44
		\subsubitem strength  42
	\subitem energy density 76
	\subitem four-potential  42
\item energy (mass)
	\subitem  density 47
	\subitem  scalar  45
\item energy-momentum
	\subitem tensor  46, 89
		\subsubitem density 75
	\subitem vector  45
\item equation of motion, for a charged test particle  41
\item equivalence of gravity and acceleration  50
\item expansion  33
\item exterior
	\subitem differentiation  12, 16
	\subitem differential of monad  34
\indexspace
\item field strength 42
\item first integrals of energy and angular momentum 55
\item forces of inertia 50, 81
\item four-current
	\subitem decomposition 79
	\subitem  density  46
\item frequency shift  51, 56
\item Friedmann world  53
\item FW (Fermi--Walker)
	\subitem differentiation  14
	\subitem differentiation, generalized  15
	\subitem curvature  14, 38
\indexspace
\item Gauss equations, generalized  39, 90
\item Gauss--Bonnet invariant 8
\item geodesic deviation equation  42, 93
	\subitem generalized for charged masses 95
\item geodesic equation  14
\item G\"odel space-time 107
	\subitem generalized 81ff
\item gradient of monad vector 32
\item gravitational instantons 90
\item gravitational inhomogeneity field tensors
		(intrinsic and extrinsic) 96
\indexspace
\item ``heavenly spaces'' 109
\item Hodge star  6
\item horizon  56
\item hydrodynamics  31
\indexspace
\item indices, collective and individual  5, 6
\item infinitesimal transformations  10
\item invariants
	\subitem electromagnetic
	\subitem gravitational 100
\item isometry  15
\indexspace
\item Kerr
	\subitem gravitational lens 61
	\subitem space-time 59
\item Kerr--Newman field  55
\item Killing
	\subitem vector field  15
	\subitem tensor  55
\item Killingian reference frame 57
\item kinemetric invariants  21
\item Kronecker symbol with collective indices  6
\indexspace
\item Lagrangian density
	\subitem electromagnetic 73
	\subitem gravitational 88
\item Lanczos identities  8
\item Leibniz property 8
	\subitem generalized 16
\item Levi-Civit\`a axial tensor 6, 7, 29
\item Lie derivative  10, 15, 33, 78
\item local stationarity 106
\item Lorentz force  42
	\subitem equal to zero 65
\item Lorenz gauge condition 74
\item Lovelock Lagrangians 8
\indexspace
\item Mach's principle 1
\item magnetic
	\subitem charges, effective  80, 86
	\subitem displacement vector  44
	\subitem monopole
		\subsubitem dynamical 86
		\subsubitem effective 80
	\subitem type fields 77
\item Maxwell's equations 73, 74, 78, 79
\item mixed triple product 30
\item momentum three-vector  45
\item monad 3
	\subitem basis  34
	\subitem field 27
	\subitem formalism  30
\indexspace
\item Noether theorem  25, 75
\item non-geodesic motion of a fluid  48
\item non-inertial frame effects  50, 81
\item normal congruence  38
\item null type fields 77
\indexspace
\item Palatini approach 73, 74, 88, 110
\item pencil of light 104
\item perfect fluid  46
\item ``polarization'' 106
\item Poynting vector  76
\item {\em pp}-waves  67
\item precession of a gyroscope 68
\item pressure  46
\item product
	\subitem three-scalar 28
	\subitem vector 29
\item projector  27
\indexspace
\item quantization 26, 109
\item quantum physics  25
\item quasi-electric tensor 97, 98, 100
\item quasi-magnetic tensor 97, 98, 100
\item quasi-Maxwellian (gravitational field) equations
	\subitem for perfect fluid 99
	\subitem new  96
	\subitem old 93
\indexspace
\item radiation-type fields  78
\item rate-of-strain tensor  32, 33
\item Raychaudhuri equation  90
\item red shift  51, 52, 56
\item reference
	\subitem body  2, 31
	\subitem frame,
		\subsubitem congruence of  24
		\subsubitem global  24, 51
		\subsubitem in quantum physics 26
\item rest mass  44
\item Ricci
	\subitem identities 12, 13
	\subitem three-tensor 39
\item Riemann-Christoffel tensor  7
\item Rindler vacuum  25
\item rotation of reference frame   33
	\subitem and dragging 67, 68
\indexspace
\item scalar
	\subitem mass  46
	\subitem multiplication of Cartan forms 8
	\subitem product, three-dimensional 28
\item Schr\"odinger-Brill formula  53
\item Schwarzschild space-time 51
\item self-dual quantities  44,
\item shear tensor  33
\item signature 5
\item skew tensor $b^a_b$ 29
\item ``sources of inertia'' 81
\item speed of light  28
\item spin-spin and spin-orbital interactions 69
\item {\em Stress}  77
\item stress-energy tensor  46
	\subitem electromagnetic 75, 76
\item structure
	\subitem coefficients  9
	\subitem equations
		\subsubitem 1st  12
		\subsubitem 2nd  13
\indexspace
\item Taub-NUT field 101
\item $\tau$-curvature  38
\item $\tau$-differentiation
\item tetrad formalism  22
\item three-current density 79
\item three-divergence  36
\item three-metric tensor 27, 30
\item three-space 3
	\subitem curvature tensor 38
\item three-velocity  29
\item transport
	\subitem Fermi-Walker 14
	\subitem parallel  13
\item triad  22
\indexspace
\item uncertainty relation for time and energy  46
\indexspace
\item vector product 29
\indexspace
\item Weyl tensor 7, 12, 93, 94
	\subitem decomposition 97
\indexspace
\item Yang-Mills theory 109
\indexspace
\item zero-point radiation 25

\end{theindex}

\end{document}